\documentclass[aps,showpacs,preprintnumbers,amsmath,amssymb,superscriptaddress,nofootinbib]{revtex4}
\usepackage{graphicx}
\usepackage{dcolumn}
\usepackage{here}
\usepackage{epsfig}
\usepackage{graphicx}
\def\lsim{\mathrel{\raise.3ex\hbox{$<$\kern-.75em\lower1ex\hbox{$\sim$}}}}
\def\gsim{\mathrel{\raise.3ex\hbox{$>$\kern-.75em\lower1ex\hbox{$\sim$}}}}

\newcommand{\cp}{{\cal CP}}
\newcommand{\At}{A_t}
\newcommand{\Ab}{A_b}
\newcommand{\mgl}{m_{\tilde{g}}}
\newcommand{\He}{h_1}
\newcommand{\Hz}{h_2}
\newcommand{\Hd}{h_3}
\newcommand{\mHe}{m_{\He}}
\newcommand{\mHz}{m_{\Hz}}
\newcommand{\mHd}{m_{\Hd}}

\newcommand{\BE}{\begin{equation}}
\newcommand{\EE}{\end{equation}}
\newcommand{\BEA}{\begin{eqnarray}}
\newcommand{\EEA}{\end{eqnarray}}
\newcommand{\non}{\nonumber}
\newcommand{\KL}{\left(}
\newcommand{\KR}{\right)}

\def\gfvh{g_{fV\h}}
\def\gfvphi{g_{fV\phi}}
\def\gama{\gam a}
\def\gamb{\gam b}

\def\non{\nonumber}

\def\Hhat{\widehat H}
\def\vo{v_0}
\def\ho{h_0}
\def\mho{m_{\ho}}
\def\phio{\phi_0}
\def\mphio{m_{\phio}}

\def\gamc{$\gam$C}
\def\nsd{N_{SD}}
\def\hsm{h_{SM}}
\def\mhsm{m_{\hsm}}
\def\mw{m_W}
\def\mz{m_Z}

\def\anti{\overline}
\def\gev{~{\rm GeV}}
\def\tev{~{\rm TeV}}

\def\lwh{\widehat\Lambda_W}

\def\lphi{\Lambda_\phi}

\def\mphi{m_\phi}

\def\hbar{\overline h}
\def\lam{\lambda}
\def\mpl{M_{Pl}}

\def\h{h}
\def\mh{m_{\h}}
\def\fbi{~{\rm fb}^{-1}}
\def\tanb{\tan\beta}
\def\lam{\lambda}

\def\hh{H}
\def\hl{h}

\def\ha{A}
\def\mha{m_{\ha}}
\def\gam{\gamma}
\def\epem{e^+e^-}
\def\gfvh{g_{fV\h}}
\def\gfvphi{g_{fV\phi}}
\def\gama{\gam a}
\def\gamb{\gam b}

\def\sig{\sigma}

\def\anti{\overline}
\def\gam{\gamma}
\def\br{B}
\def\ifmath#1{\relax\ifmmode #1\else $#1$\fi}

\def\lsim{\mathrel{\raise.3ex\hbox{$<$\kern-.75em\lower1ex\hbox{$\sim$}}}}
\def\gsim{\mathrel{\raise.3ex\hbox{$>$\kern-.75em\lower1ex\hbox{$\sim$}}}}
\def\half{\ifmath{{\textstyle{1 \over 2}}}}

    \def\fillboxx#1#2{\hbox to #1{\vbox to #2{\vfil}\hfil}    }

\def\ie{{\it i.e.}}

\def\vev#1{\langle #1 \rangle}

\def\gl{\wt g}
\def\mgl{m_{\gl}}

\def\tanb{\tan\beta}

\def\mz{m_Z}
\def\mw{m_W}

\def\h{h}
\def\mh{m_{\h}}
\def\wt{\widetilde}

\def\lam{\lambda}
\def\br{BR}
\def\tauptaum{\tau^+\tau^-}

\def\gam{\gamma}

\def\anti{\overline}
\def\epem{e^+e^-}

\def\rtsee{\sqrt s_{ee}}
\def\ie{{\it i.e.}}

\def\eps{\epsilon}
\def\anti{\overline}

\def\mw{m_W}
\def\mz{m_Z}
\def\h{h}
\def\mh{m_{\h}}

\def\hsm{h_{SM}}
\def\mhsm{m_{\hsm}}

\def\hl{h^0}

\def\ha{A^0}
\def\mha{m_{\ha}}

\def\hh{H^0}

\def\fbi{~{\rm fb}^{-1}}

\def\gev{~{\rm GeV}}
\def\tev{~{\rm TeV}}

\def\MPL #1 #2 #3 {{ Mod.~Phys.~Lett.}~{\bf#1} (#3) #2}
\def\NPB #1 #2 #3 {{ Nucl.~Phys.}~{\bf #1} (#3) #2}
\def\PLB #1 #2 #3 {{ Phys.~Lett.}~{\bf #1} (#3) #2}
\def\PR #1 #2 #3 {{ Phys.~Rep.}~{\bf#1} (#3) #2}
\def\PRD #1 #2 #3 {{ Phys.~Rev.}~{\bf #1} (#3) #2}
\def\PRL #1 #2 #3 {{ Phys.~Rev.~Lett.}~{\bf#1} (#3) #2}
\def\RMP #1 #2 #3 {{ Rev.~Mod.~Phys.}~{\bf#1} (#3) #2}
\def\ZPC #1 #2 #3 {{ Z.~Phys.}~{\bf #1} (#3) #2}
\def\IJMP #1 #2 #3 {{ Int.~J.~Mod.~Phys.}~{\bf#1} (#3) #2}
\def\NIM #1 #2 #3 {{ Nucl.~Inst.~and~Meth.}~{\bf#1} {#3} #2}
\def\JHEP #1 #2 #3 {{ JHEP}~{\bf#1} (#3) #2}

\newcommand{\nc}{\newcommand}
\nc{\beq}{\begin{equation}}   \nc{\eeq}{\end{equation}}
\nc{\bea}{\begin{eqnarray}}   \nc{\eea}{\end{eqnarray}}
\nc{\baa}{\begin{array}}      \nc{\eaa}{\end{array}}
\nc{\bit}{\begin{itemize}}    \nc{\eit}{\end{itemize}}
\nc{\bed}{\begin{description}}    \nc{\eed}{\end{description}}
\nc{\ben}{\begin{enumerate}}  \nc{\een}{\end{enumerate}}
\nc{\bce}{\begin{center}}     \nc{\ece}{\end{center}}
\def\beqa{\begin{eqnarray}}
\def\eeqa{\end{eqnarray}}
\def\lsim{\mathrel{\raise.3ex\hbox{$<$\kern-.75em\lower1ex\hbox{$\sim$}}}}
\def\gsim{\mathrel{\raise.3ex\hbox{$>$\kern-.75em\lower1ex\hbox{$\sim$}}}}
\def\vev#1{\langle #1 \rangle}
\def\lam{\lambda}

\newcommand{\hpm}{\ensuremath{H^{\pm}}}

\begin{document}
\preprint{nuhep-exp/03-003\cr UCD-03-08}

\title{Complementarity of a Low Energy 
Photon Collider and LHC Physics}

\affiliation{University of Pittsburgh, Pittsburgh, Pennsylvania 15260, USA}
\affiliation{Lawrence Livermore  National Laboratory,  Livermore, California 94550, USA}
\affiliation{CERN, CH-1211 Geneva 23, Switzerland}
\affiliation{Institut f\"ur theoretische Elementarteilchenphysik, LMU M\"unchen, Theresienstr.\ 37, D-80333 M\"unchen, Germany}
\affiliation{University of California , Davis, California 95616,    USA}
\affiliation{University of Wisconsin, Madison, Wisconsin, 53706, USA}
\affiliation{Northwestern University, Evanston, Illinois 60201, USA}

\author{David~Asner}
\affiliation{University of Pittsburgh, Pittsburgh, Pennsylvania 15260, USA}

\author{Stephen~Asztalos}
\affiliation{Lawrence Livermore  National Laboratory,  Livermore, California 94550, USA}

\author{Albert~De~Roeck}
\affiliation{CERN, CH-1211 Geneva 23, Switzerland}

\author{Sven~Heinemeyer}
\affiliation{Institut f\"ur theoretische Elementarteilchenphysik, LMU M\"unchen, Theresienstr.\ 37, D-80333 M\"unchen, Germany}

\author{Jeff~Gronberg}
\affiliation{Lawrence Livermore  National Laboratory,  Livermore, California 94550, USA}

\author{John~F.~Gunion}
\affiliation{University of California , Davis, California 95616,    USA}

\author{Heather~E.~Logan}
\affiliation{University of Wisconsin, Madison, Wisconsin, 53706, USA}

\author{Victoria~Martin}
\affiliation{Northwestern University, Evanston, Illinois 60201, USA}

\author{Michal~Szleper}  
\affiliation{Northwestern University, Evanston, Illinois 60201, USA}
\author{Mayda~M.~Velasco}
\affiliation{Northwestern University, Evanston, Illinois 60201, USA}
\begin{abstract}
We discuss the complementarity  between the LHC and a low energy photon 
collider.  We mostly consider the scenario, where the first linear collider 
is a photon collider based on dual beam technology like CLIC.
\end{abstract}

\maketitle
\section{Introduction }

The LHC is scheduled to turn on in the year 2007.
Within the first few years of operation LHC will discover the Higgs boson, 
if it 
exists. When this long awaited new particle is finally found, it 
will no doubt become the most important object to be studied in detail
in high energy physics (HEP) -- unless of course at the same time also many
other new particles, such as SUSY sparticles, are copiously produced
at the LHC --.

The LHC will be able to measure many characteristics
of the Higgs boson rather precisely, such as mass and width. However,
has not yet been demonstrated that 
the couplings to fermions and gauge bosons cannot be measured in a
model independent way, rather ratios of couplings are directly
accessible at LHC.
The measurement of other properties, such as spin and CP quantum 
numbers and Higgs self coupling, will be even more tedious.
Hence data on Higgs measurement  in different reactions such as 
in electron-positron and photon-photon collisions will be needed to
determine the Higgs parameters in greater detail.

If the physics beyond the Standard Model is low energy 
supersymmetry, then  the mass
of the Higgs will be relatively low, e.g. below about 135 GeV~\cite{sven}
 as predicted
in the minimal supersymmetric model (MSSM).
This will put the production of Higgs bosons 
within reach of future lepton or
photon colliders. Such colliders benefit from a much cleaner 
production environment for Higgs particles as compared to 
hadron machines.

A linear $e^+e^-$ collider is seriously considered as 
an option for the next large accelerator 
in HEP~\cite{snowmass} by international consensus. Such a machine 
would certainly be ideal to study the properties of the Higgs particle in 
detail. 
These linear colliders, for which presently 
several proposals 
exist~\cite{nlc,jlc,tesla} are generally huge machines, of a length of 
about 20-40 km, to reach approximately 1 TeV center of mass  energy ($E_{CM}$)
with conventional accelerating techniques.
Regrettably, it is unlikely that the construction of such an accelerator
will start any time before 2007-2009\cite{roadmap}, 
and the construction/commissioning will take of the 
order of 10 years.

Two beam acceleration (TBA) has been proposed as an alternative 
accelerating technique to reach higher accelerating gradients, and is 
presently studied most intensively at CERN through the CLIC 
R\&D project~\cite{c:clic,ctf3a}.
TBA  is still in an experimental stage, but when ready
it will allow to construct $e^+e^-$ colliders with higher $E_{CM}$, and/or, for
a more compact $e^+e^-$ collider to reach 
the energy region of 
interest. That is, one 600\,m accelerating module with the CLIC technology 
will
accelerate electrons by about 70 GeV. The TBA technology  
 has been demonstrated for low currents and small pulses in 
test facilities CTF1 and CTF2 at CERN. 
Presently, the CTF3 test facility\cite{ctf3} is 
under
construction and should demonstrate the feasibility of the machine
parameters for the drive beam for CLIC. When successful 
this will allow to complete a technical design for a machine based 
on TBA.

In this paper we consider physics studies for a Higgs factory which could 
be decided upon and built, based on TBA, soon after
the discovery of a light Higgs at the LHC during the last year of this 
decade. The smallest (but not necessary simplest) collider would be one 
based on two TBA  modules, which accelerate each an electron beam up to 
70-75 GeV. When these beams are converted into photon beams  
via Compton scattering using powerful lasers, Higgs particles with a 
mass of up to 130 GeV can be produced in the s-channel.  
Such photon colliders\,(\gamc) have been 
extensively proposed in Ref.\cite{telnov}, and all $e^+e^-$ linear collider\,(LC)
projects consider such an option as an upgrade of their base-line~\cite{nlc,jlc,tesla}
program.
R\&D projects for \gamc~are presently ongoing. To 
exploit the physics opportunities as discussed in this paper it is  
further imperative that the electron beams can be polarized, which 
is generally 
considered to be technologically feasible.

A low energy Higgs factory driven by a  \gamc~ based on CLIC 
technology, such as the one assumed here, has  already been elaborated 
in Ref.\cite{cliche}, and was coined CLICHE. The basic parameters of such a 
machine and initial physics studies have been presented. As discussed in
Ref.\cite{cliche},
the $ H \rightarrow \gamma \gamma$ vertex is due to loop diagrams making it
sensitive to physics beyond the Standard Model, and as an example, 
they  show how  the precision  obtained on the 
branching ratio measurements made at CLICHE could help us to distinguish
between the Standard Model\,(SM) and its minimal supersymmetric extension, the MSSM.
Further work on the comparison between the SM and the MSSM at a
\gamc\ was discussed in Ref.\cite{korea}.

Here we follow up on this opportunity,
and  we extend the physics arguments in favor of 
vigorously pursuing a \gamc~ collider, either at 
CLICHE,
or possibly  as an adjunct to 
an upcoming LC.  We find that CLICHE by itself 
could provide invaluable complementarity to the LHC.
For example,
we show the crucial role that  CLICHE
could play in: (1) detecting and/or confirming the LHC signal for
a CP-even Higgs boson that decays to two light pseudo-scalar Higgs bosons, 
(2)
providing a precise measurement of the $H\to\gamma\gamma$ partial
width, which 
allows us to test large scales in the
Littlest Higgs model after combining with branching ratio measurements
made at an $e^+e^-$ collider or the LHC, and 
(3) in the presence of Higgs-Radion mixing in the Randall-Sundrum model, 
one could test possible $gg$ and $\gam\gam$ anomalous couplings 
to the $h$ and $\phi$ among many other things, since the LHC will give
access to the  $gg$  coupling, while the \gamc\ will give us the coupling
to $\gam\gam$.


%
\section{Machine and detector design update}

As  discussed in detailed in Ref.\cite{cliche}, the technical
requirements  to produce a \gamc\ with warm accelerating
technology and for TBA are compatible, and therefore their R\&D is in common.
Details about the  parameters for  \gamc\ based on warm, cold and TBA
technology 
can be found in Refs.~\cite{nlc,tesla,cliche}. Here we will  discuss
the recent progress in the R\&D, and detector requirements
due to the environment at the interaction region of a \gamc.

\subsection{R\&D for photon collider technology}

The straw man design for the \gamc\ technology for a warm
machine  that
was presented at Snowmass 2001 by LLNL  has continued to develop.
The MERCURY laser has been commissioning and has reached
it's full repetition rate of 10Hz with 20 Joule pulses.  It is now
undergoing
a major refit to include the second amplifier head. Once that is
completed it will reach its full power of 100 Joules.  The basic layout
of the optics has remained unchanged, but much work is going into the
details of aligning the system and maintaining the laser pulse quality, which
is critical to efficient utilization of the laser power.

A collaboration of DESY and MBI have been exploring a design for a
cavity laser system to reduce the total laser power by exploiting the
larger bunch spacing of TESLA to reuse the laser pulses.  This type of
system cannot be used for the warm machine since the 2.8\,ns bunch
spacing does not allow time to reuse the laser pulse.

Choices in the operating parameters of the electron accelerator can improve
the ratio of delivered luminosity to laser power.  The percentage of
electrons
which Compton scatter is set by the laser photon density, and is independent of the
electron bunch charge.  Maximizing the bunch charge increases the luminosity
at no cost in laser power and should always be pursued.  In fact, since
luminosity
increases as the bunch charge squared, it can be a win to reduce the number
of
bunches, while increasing the bunch charge to keep the total beam current
constant.
This can both increase luminosity, while reducing the total laser power
needed, and therefore reduce cost. 

\subsection{\label{sec:resolved}
Resolved photons and  impact on detector}

The photon is the gauge boson of QED that  can couple  to the electroweak and 
strong interactions  via virtual charged fermion pairs.
As a consequence, photon-photon interactions are more complex than $e^+e^-$
interactions, and a careful simulation of the beam is extremely important.

In a \gamc\ we have 
a $few\times 10^{10}$ primary $e^-$, in addition to 
$\sim\frac{2}{3}\times few \times 10^{10}$ Compton photons. We have used
CAIN~\cite{cainref} to obtain the $ee$, $e\gamma$ and $\gamma\gamma$
luminosity distributions. The  CAIN simulations  include real
particles from beam related backgrounds, such as
$e^+e^-$ from pair production, and real beamstrahlung photons.
Pythia is used to simulate the virtual photon cloud
associated with the electron beam.

\begin{figure}\begin{center}
\resizebox{\textwidth}{!}{
\includegraphics{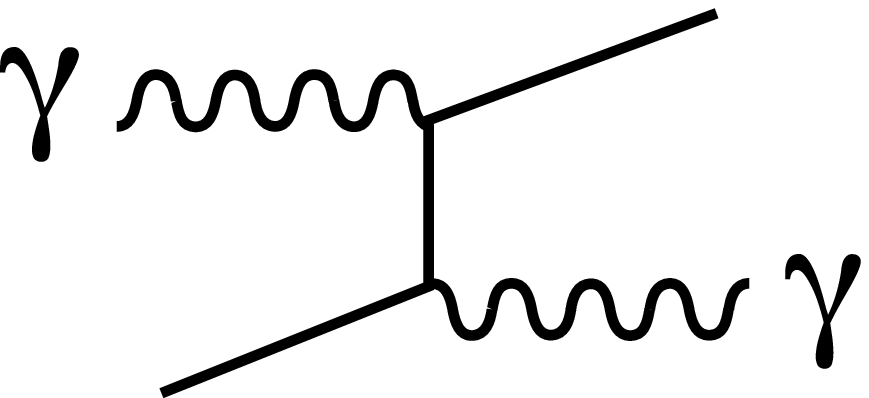}
\hskip 1cm
\includegraphics{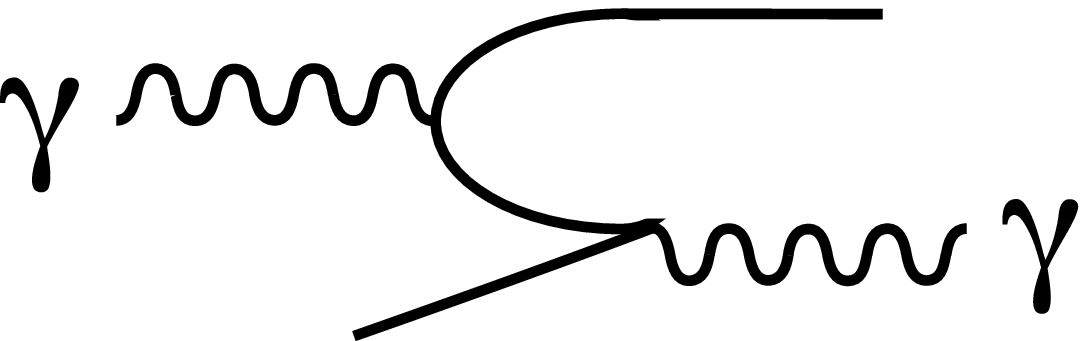}
\hskip 1cm
\includegraphics{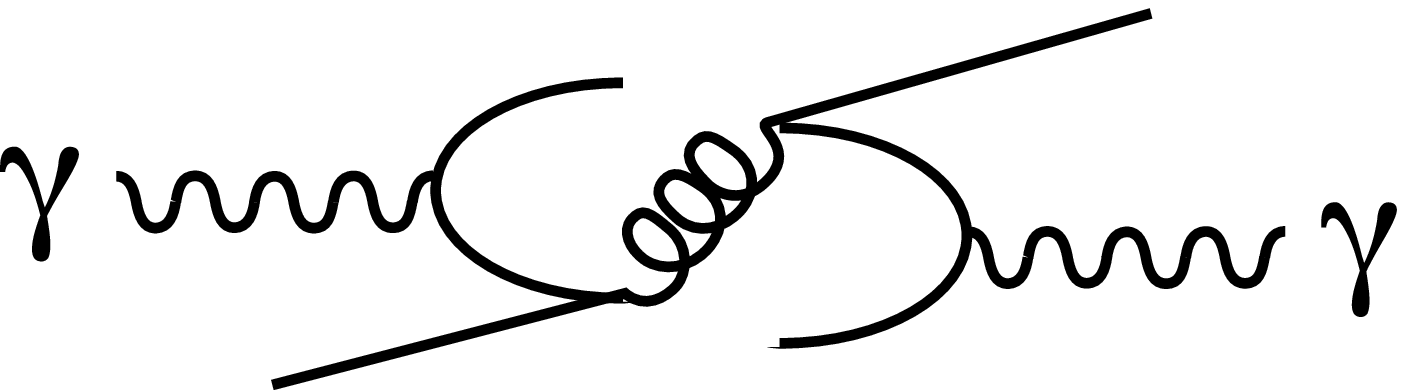}
}
\caption{\label{fig:resphotons}Photon-photon Interactions. 
(left) Direct interactions involve only electroweak couplings.
(center) Once resolved process where one photon probes the parton structure of
the other photon, similar to deep inelastic scattering. (right) Twice resolved
process where the partons of each photon interact, similar to a $\rho-\rho$ collisions.}
\end{center}\end{figure}

As illustrated in Fig.~\ref{fig:resphotons}, photon-photon
interactions can be classified into three types of processes.
Direct interactions that involve only electroweak couplings, the
once resolved process that is similar to deep inelastic scattering
because    one photon probes the parton structure of
the other photon,  and the twice resolved
process that can be thought of as a 
$\rho-\rho$ collisions because the partons of each photon interact.

Similarly,
photon-electron interactions can be classified into direct
interactions involving  only electroweak couplings, and the
once resolved process, where the electron probes the parton structure of
the photon. Additionally, the virtual photon cloud associated with the
electron beam can interact with the Compton photons as described above.

The electron-electron
interact by direct processes, and the virtual electron cloud associated
with each beam, also lead to all combinations of photon-photon and
photon-electron processes.

We use the Pythia~\cite{pythiajetset} Monte Carlo program to simulate the 
resolved photon cross sections, which are potential backgrounds to {\em all}
other two-photon physics processes. These backgrounds, usually referred to as
$\gamma\gamma \to {\rm hadrons}$, are also a concern at $e^+e^-$ and have
been studied in detail~\cite{teslagghh}. However, at a $\gamma\gamma$ collider
the high energy Compton photons provide an additional and dominant source
of $\gam\gam \to {\rm hadron}$ background.

There has been a long standing discrepancy between the resolved photon
backgrounds calculated by Pythia and those computed assuming the
observed LEP cross section for $\gamma\gamma \rightarrow hadrons$
\footnote{At Snowmass 2001, it was noted that the cross sections
predicted by Pythia were much larger than naively expected from
LEP data. At ECFA/DESY in St. Malo, the discrepancy was determined to
be approximately an order of magnitude. A more detailed comparison by
Asner and de Roeck, narrowed the discrepancy to a factor of six. It
was reported at ECFA/DESY in Prague that Pythia predicted the cross
section for $\gamma\gamma \rightarrow hadrons$ to be consistent with
the LEP data after better taking a better choice of parameters.}. 
This problem was  mostly due to the parameters used by  Pythia
to describe this processes,   and it is now understood.
Nominally, we use the default settings for Pythia with the exception of 
the parameters listed in 
Table~\ref{tab:pythiasettings}. The complete Pythia parameter list
used is taken from Ref.~\cite{jetweb}. We find that $\sim 83\%$,$\sim
17\%$ and $\sim 0.4\%$ of
interactions are due to photon-photon, photon-electron, and
electron-electron interactions, respectively, and the photon-photon
and photon-electron interaction cross section is dominated by
resolved photons.
\begin{table}
\caption{\label{tab:pythiasettings} Pythia parameter settings for
resolved photon processes. Other parameters are taken from
Ref.~\cite{jetweb}.}
\begin{center}
\begin{tabular}{lc}
\hline
\hline
Parameter Setting & Explanation\\
\hline
         MSTP(14)=30              & Sets Structure of Photon\\
         MSTP(53)=3               & choice of pion parton
distribution set\\
         MSTP(54)=2               &  choice of pion parton
distribution function library\\
         MSTP(55)=N/A for MSTP(14)=30 & choice of photon parton
distribution set\\
         MSTP(56)=2               &  choice of photon parton
distribution function library\\
        CKIN(3)=1.32               & ptmin  for hard $2\rightarrow$2 interactions\\

\hline
\hline
\end{tabular}
\end{center}
\end{table}

\begin{figure}\begin{center}
\resizebox{\textwidth}{!}{
\includegraphics{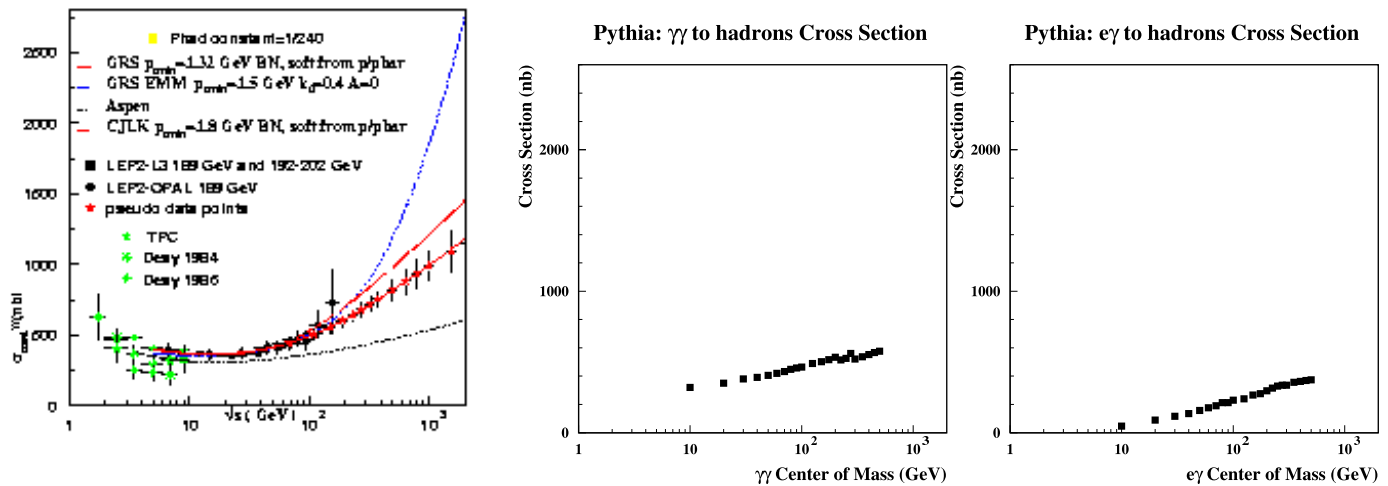}
\includegraphics{gg_res.epsi}
\includegraphics{eg_res.epsi}
}
\caption{\label{fig:2rescs}Photon-Photon and Photon-Electron Cross
Section.
(left) From Ref.~\cite{albert}. Updated photon-photon cross sections
predicted by CJLK, GRS densities and
measurements from LEP. (center) Photon-photon cross section and
(right) Photon-electron cross section from Pythia using parameter
described in the text.} 
\end{center}\end{figure}

Over the energy range, $\sim 10-150$~GeV,
of the LEP data the Pythia and LEP cross sections are in reasonable
agreement, $\sim 400$\,nb, as shown in Fig.~\ref{fig:2rescs}. Also shown
in Fig.~\ref{fig:2rescs} is the cross section for photon-electron
processes, which are not negligible relative to the photon-photon processes.
Beyond the range of the existing data, there are large uncertainties in the
estimated of the 
once and twice resolved photon cross sections. 
These uncertainties at higher energies are a concern for a 500~GeV
linear collider, but not for the lower energy Higgs factory.
Ref.~\cite{butterworthpaper} provides 
an excellent discussion of the theoretical
and experimental challenges of determining the photon structure.
The most recent experimental data is from HERA~\cite{hera} and 
LEP-II~\cite{lep}. Despite the new data, there are 
large uncertaintes at small $x_\gamma$,  and the uncertainties
at  large $x_\gamma$ are also significant.
The QCD stucture function is probed by examining the 
high $E_t$ jets, but extrapolating
to smaller $E_t$ introduces additional uncertainty.
These large uncertainties  could  be reduce by data obtained from 
a proposed 
$\gamma\gamma$ engineering run at the  SLC~\cite{ourproposal}.

\begin{figure}
\resizebox{\textwidth}{!}{
\rotatebox{0}{\includegraphics{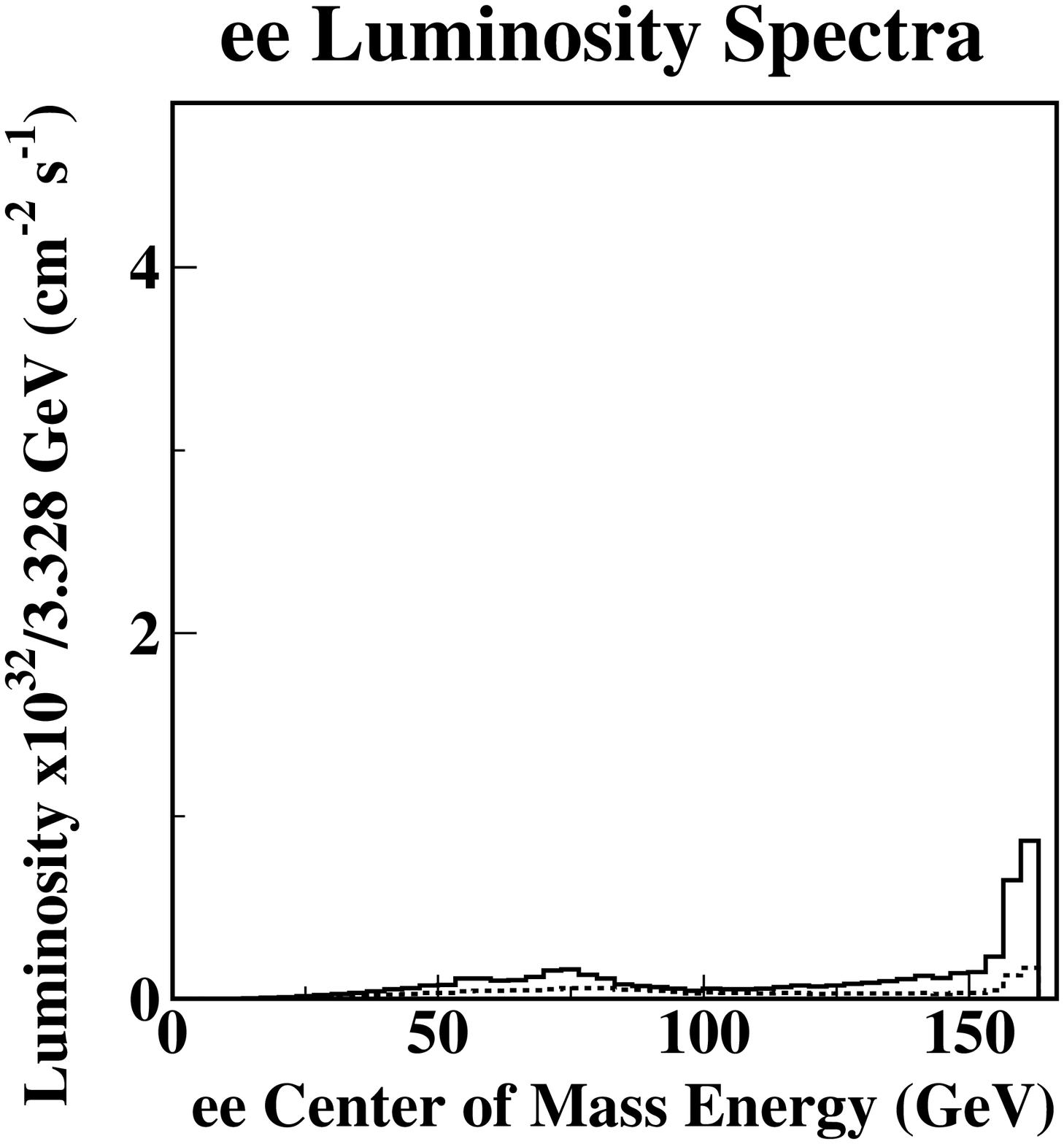}}
\rotatebox{0}{\includegraphics{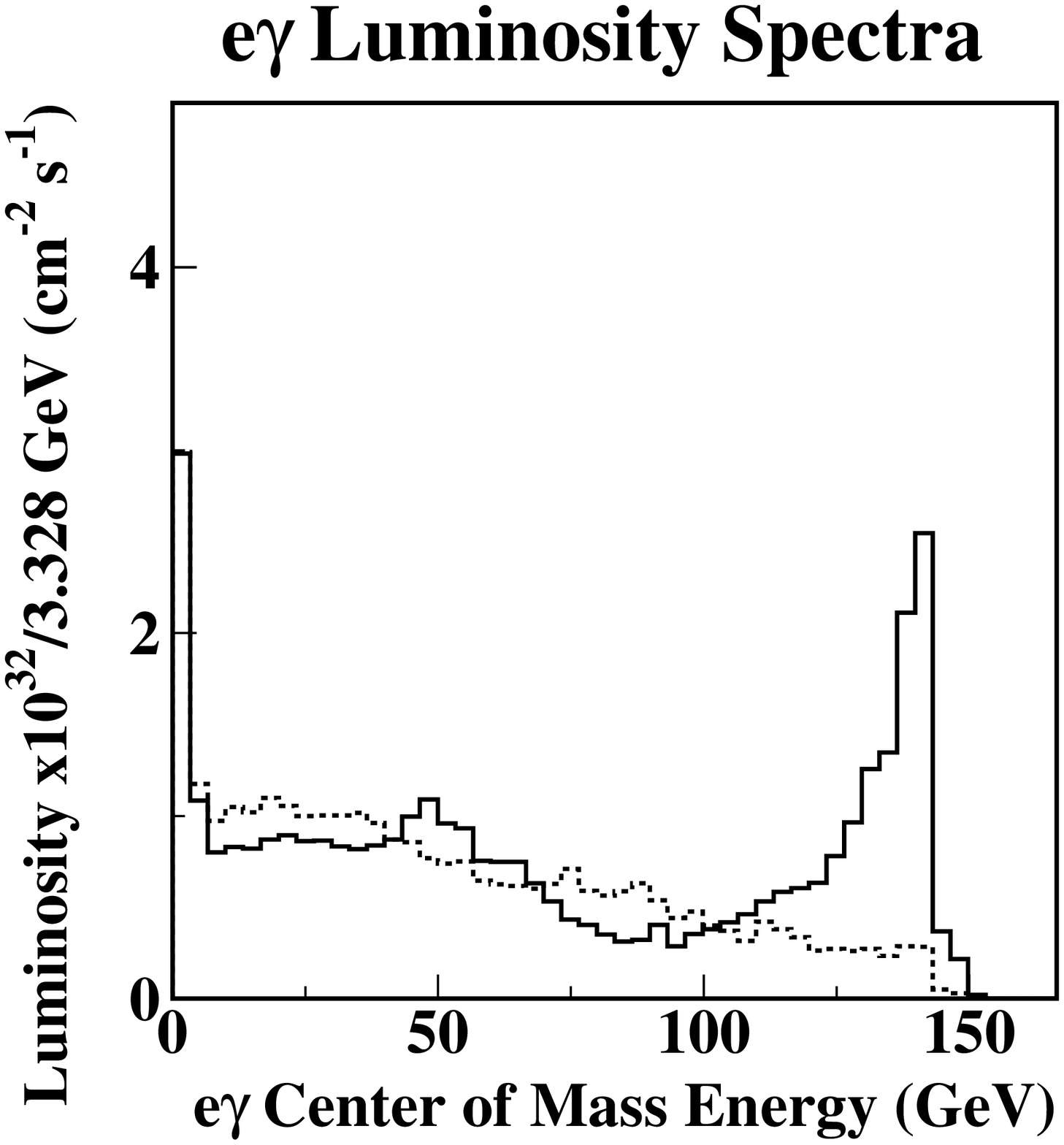}}
\rotatebox{0}{\includegraphics{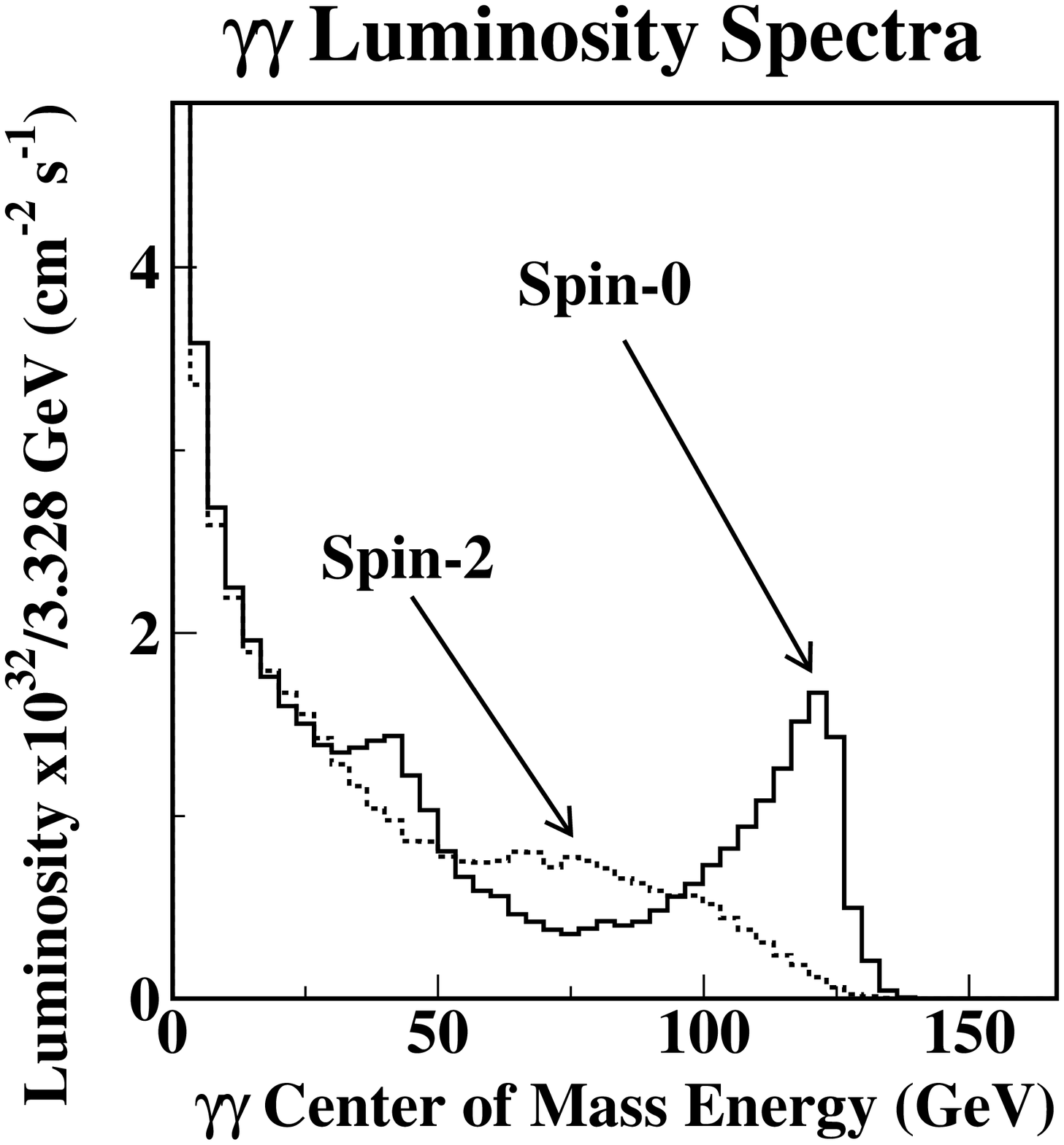}}}
\resizebox{\textwidth}{!}{
\rotatebox{0}{\includegraphics{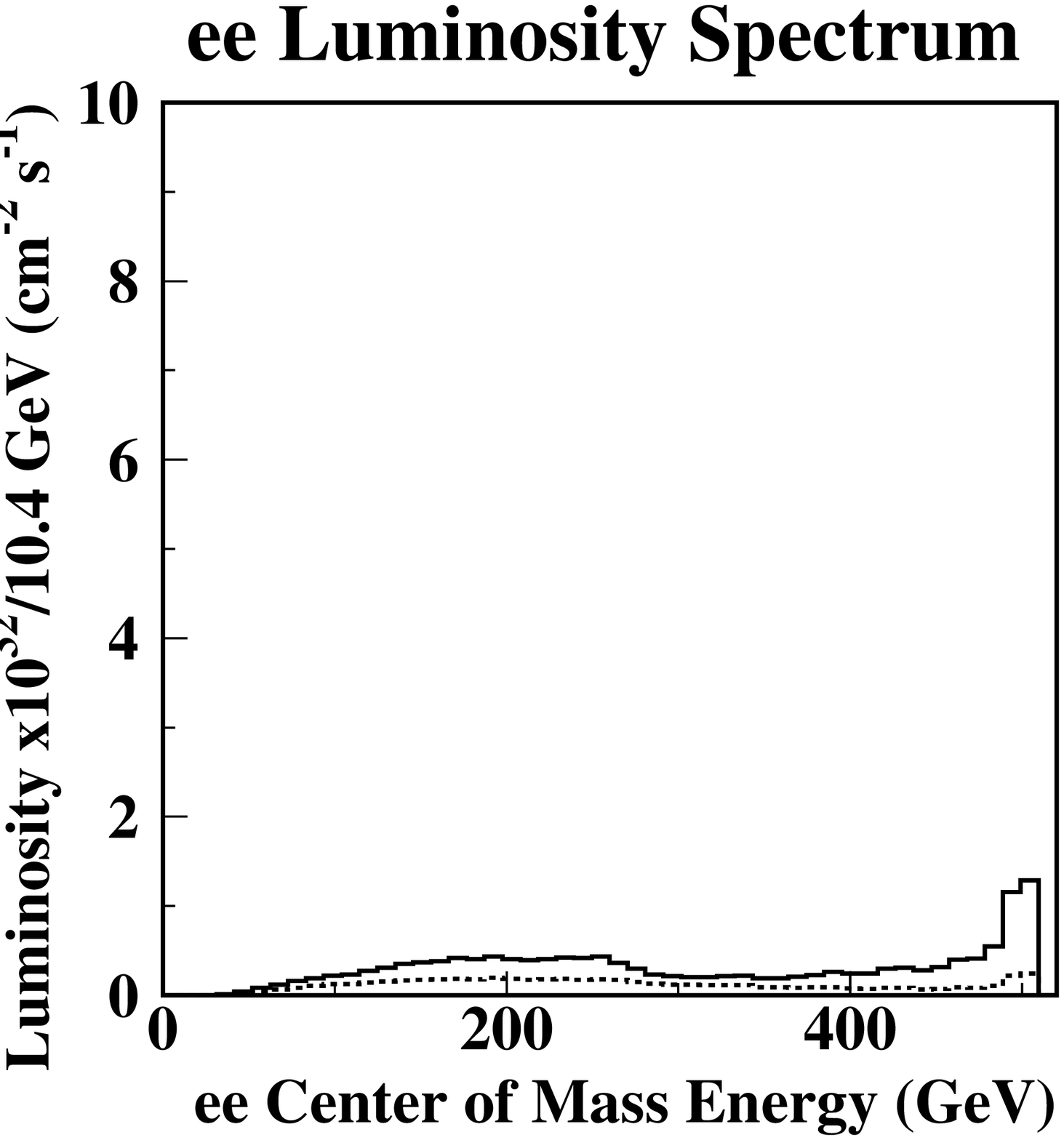}}
\rotatebox{0}{\includegraphics{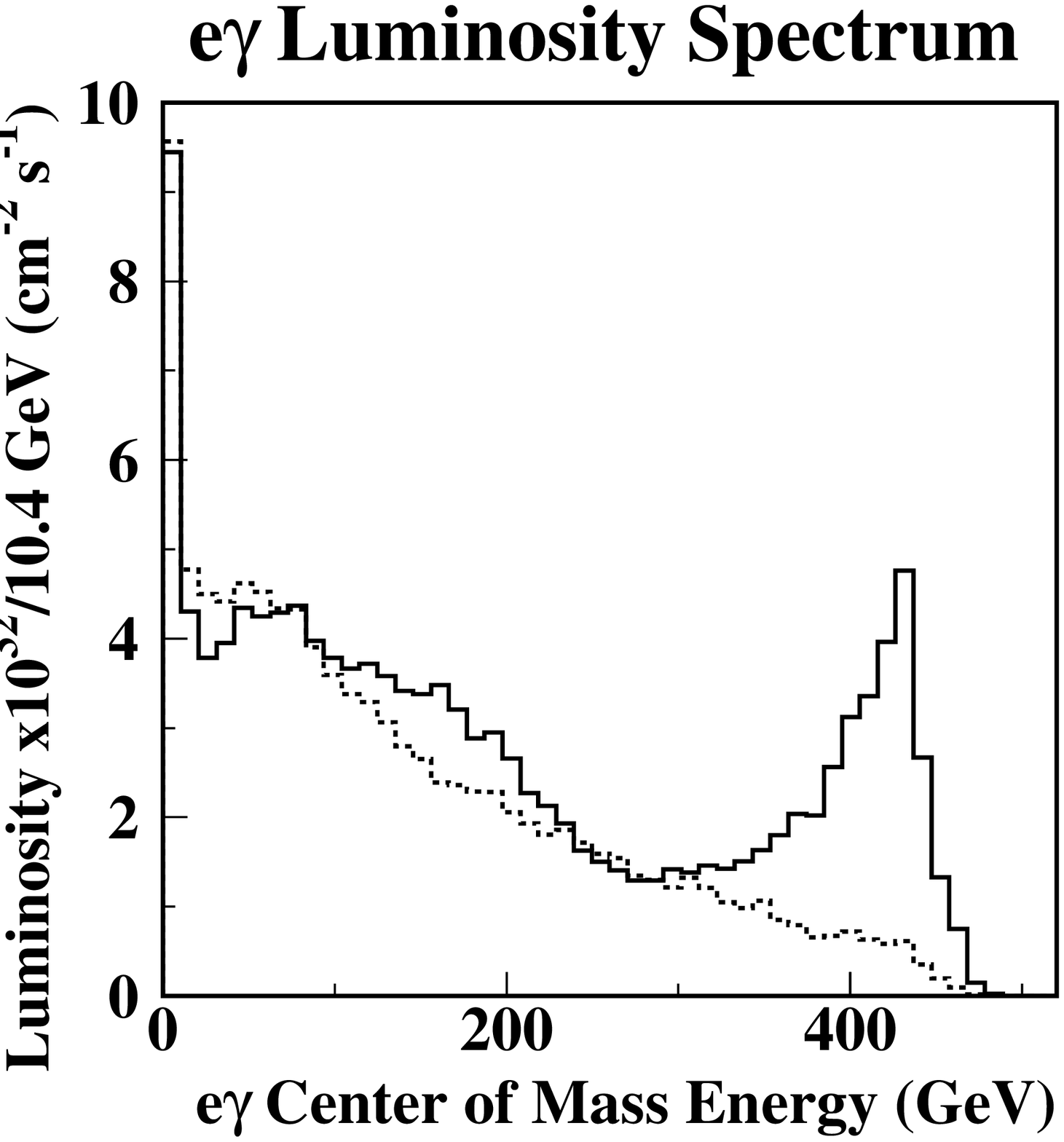}}
\rotatebox{0}{\includegraphics{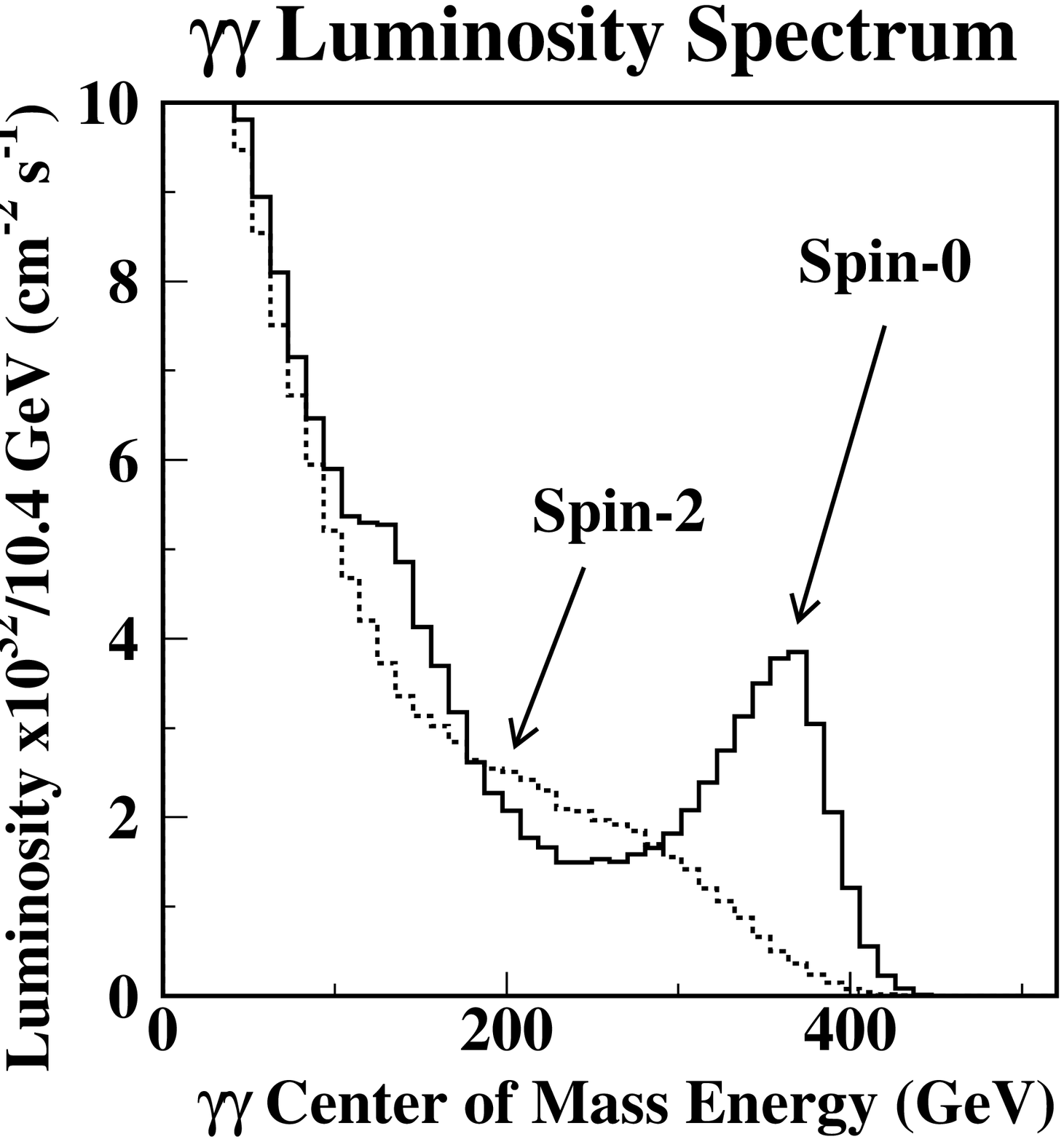}}}
\caption[0]{Luminosity for a $10^7$~sec year and
associated expectation value for the product of photon polarizations, $\vev{\lam\lam'}$, are plotted for $\sqrt s=160$~GeV
($x=4.1$ for $1.054/3~\mu$m laser wavelength),
assuming 80\% electron beam polarizations.}
\label{fig:cainlums}
\end{figure}

Using these new sets of Pythia parameters, we can quatify the effect 
of this `extra' beam related activity in the detector. For this purpose,
we consider the beam parameters for $\rtsee=160$~GeV and $\rtsee=500$~GeV 
given in 
Table~\ref{tab:beampar}, and use the luminosity distributions 
shown in Fig.~\ref{fig:cainlums}.
\begin{table}[h]
\caption{Laser and electron beam parameters for ${\sqrt s} = 160$~GeV
 ${\sqrt s} = 500$~GeV. The beam parameters for ${\sqrt s} = 500$~GeV
differ from the NLC-$e^+e^-$ parameters. The bunch charge has been
doubled to improve luminosity. Consequently, both the vertical emittance, 
$\epsilon_y$, and the bunch length, $\sigma_z$ are increased. Additionally,
the total current is conserved as the rep. rate reduced by a factor of 2.
The optimal laser wavelength decreases as the beam energy decreases.
We assume that non-linear optics are used to triple the laser frequency for
the ${\sqrt s} = 160$~GeV machine and that this procedure is 70\% efficient,
thus more laser power is required.}
\label{tab:beampar}
\begin{center}
\begin{tabular}[c]{ccc}
\hline
\hline
Electron Beam Energy (GeV) & 80 & 250  \\
\hline
$\beta_x/\beta_y$~(mm) & 1.4/0.08 & 4/0.065   \\
$\epsilon_x/\epsilon_y~(\times 10^{-8})$ & 360/7.1 & 360/7.1 \\
$\sigma_x/\sigma_y ~({\rm nm})$ & 179/6 & 172/3.1  \\
$\sigma_z$~(microns) & 156 &  156  \\
$ N~(\times 10^{10})$ & 1.5 & 1.5  \\
$ e^-~{\rm Polarization}~ (\%)$ & 80 & 80 \\
rep. rate (Hz) & 120$\times$95 & 120$\times$95  \\
Laser Pulse Energy (J) & 1.0/70\%=1.4 & 1.0 \\
Laser $\lambda$ (microns) & 1.054/3 = 0.351 &  1.054  \\
CP-IP distance (mm) & 1 & 2  \\
\hline
\hline
\end{tabular}
\end{center}
\end{table}

\begin{figure}\begin{center}
\resizebox{.9\textwidth}{!}{
\includegraphics{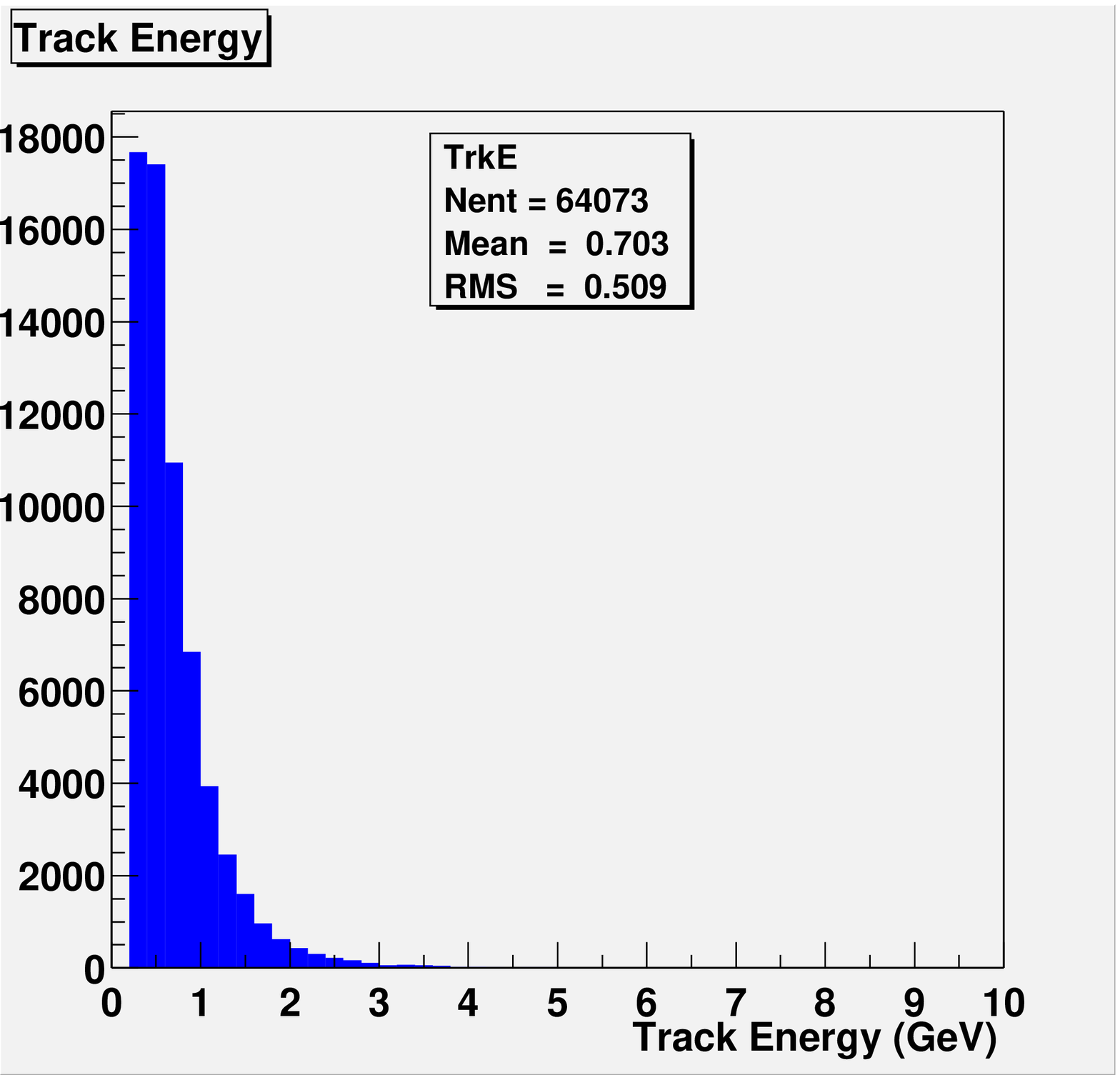} 
\includegraphics{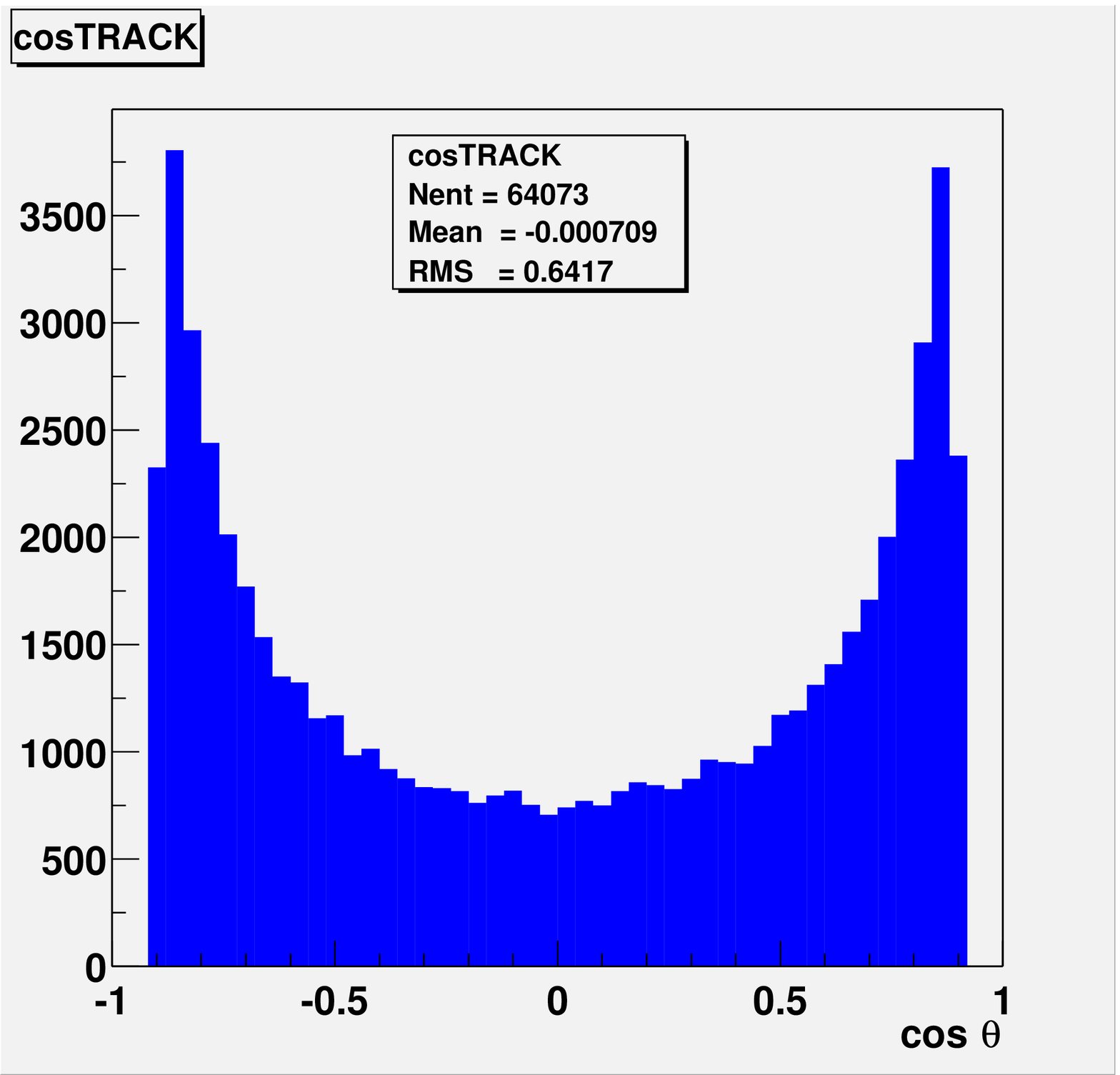}}
\caption{\label{fig:160track}
Tracks contributing to the resolved photon background
for $\rtsee=160$~GeV.
Energy and    $\cos\theta$ distribution for tracks with $p>0.2$~GeV.
The plots corresponds to 6,700 beam crossings.}
\end{center}\end{figure}

\begin{figure}\begin{center}
\resizebox{.9\textwidth}{!}{
\includegraphics{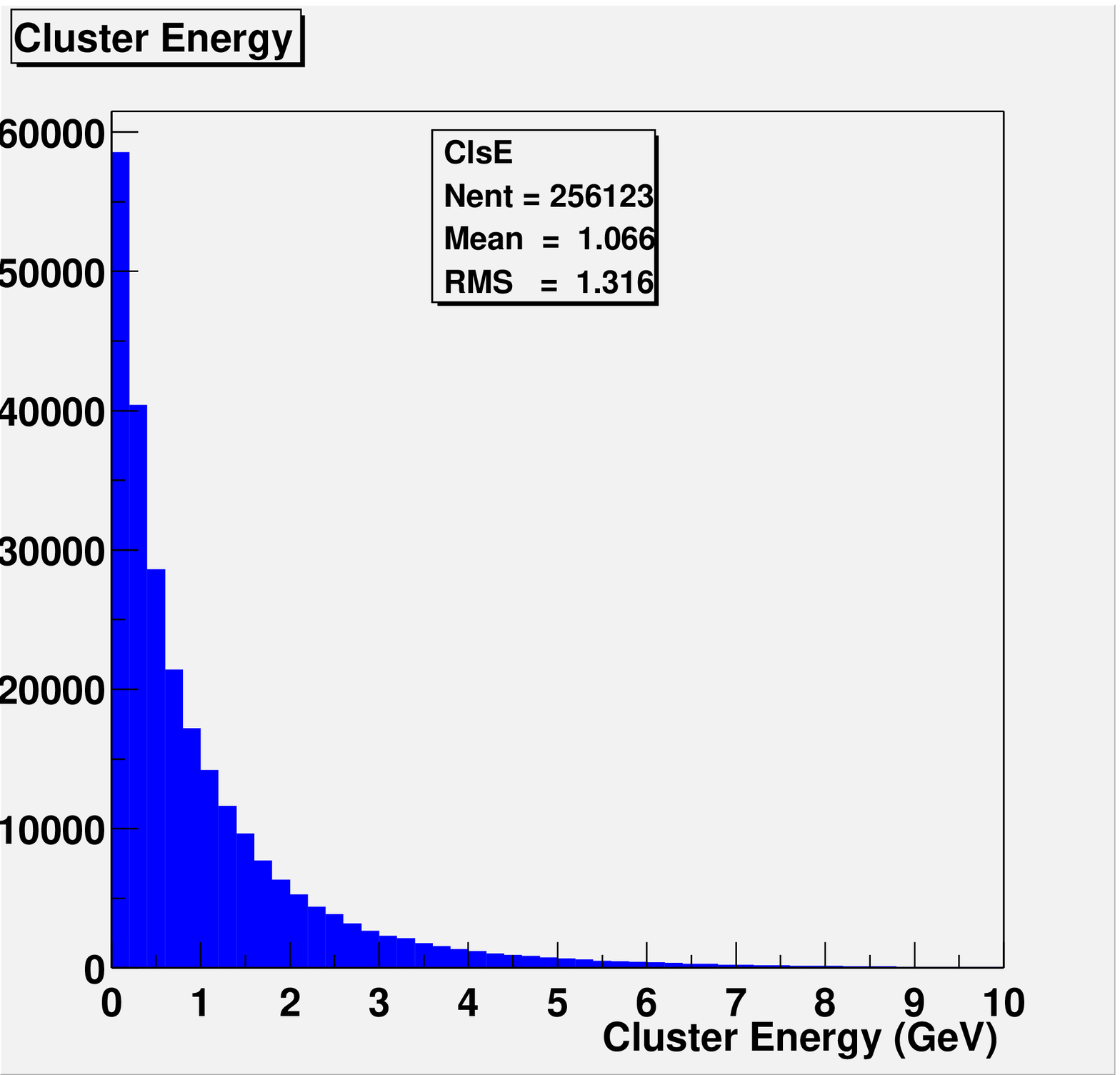} 
\includegraphics{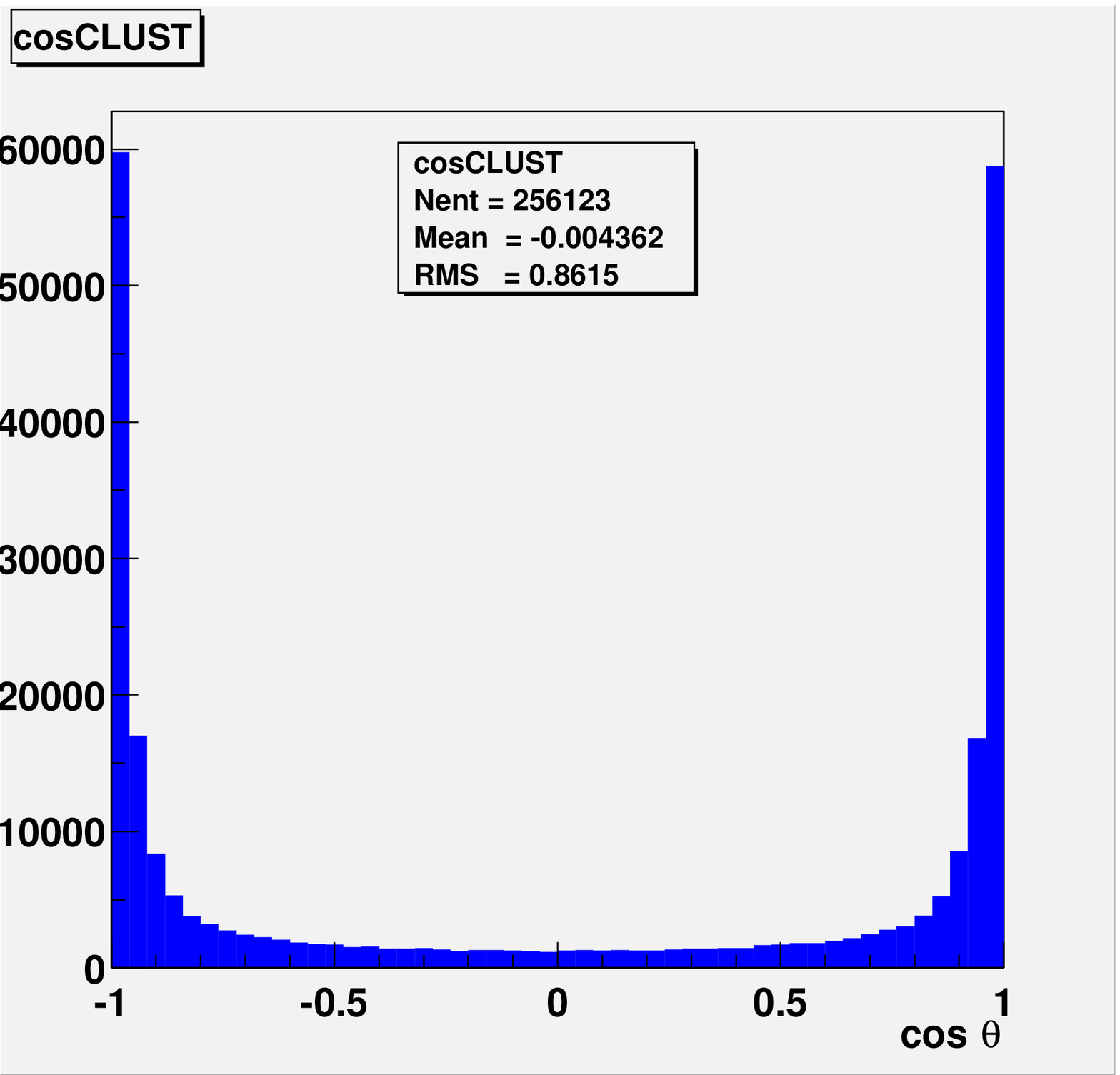}}
\caption{\label{fig:160clust}
Showers contributing to the resolved photon background
for $\rtsee=160$~GeV.
Energy and    $\cos\theta$ distribution for showers with $E>0.1$~GeV.
The plots corresponds to 6,700 beam crossings.}
\end{center}\end{figure}

\begin{figure}\begin{center}
\resizebox{.9\textwidth}{!}{
\includegraphics{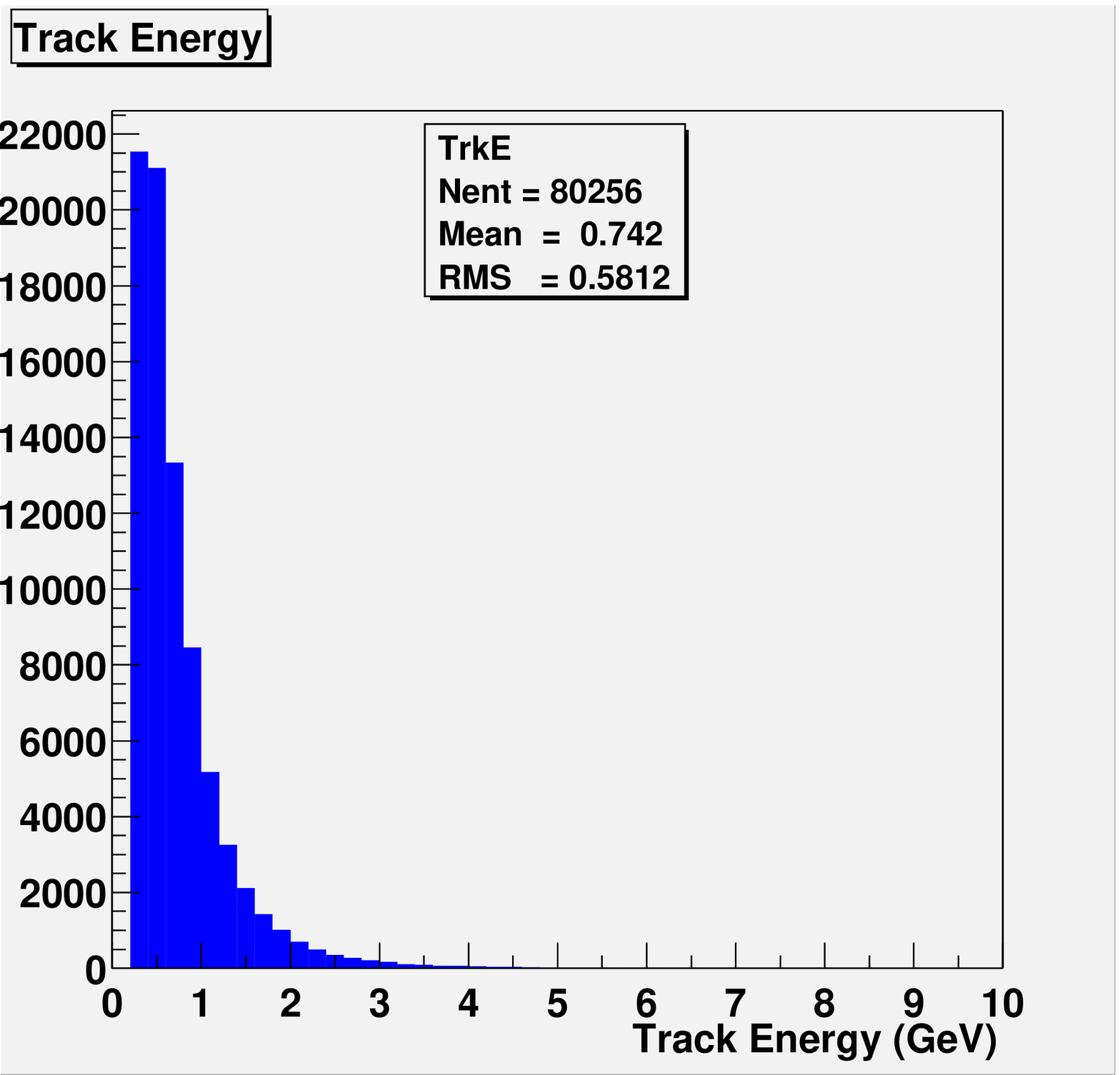} 
\includegraphics{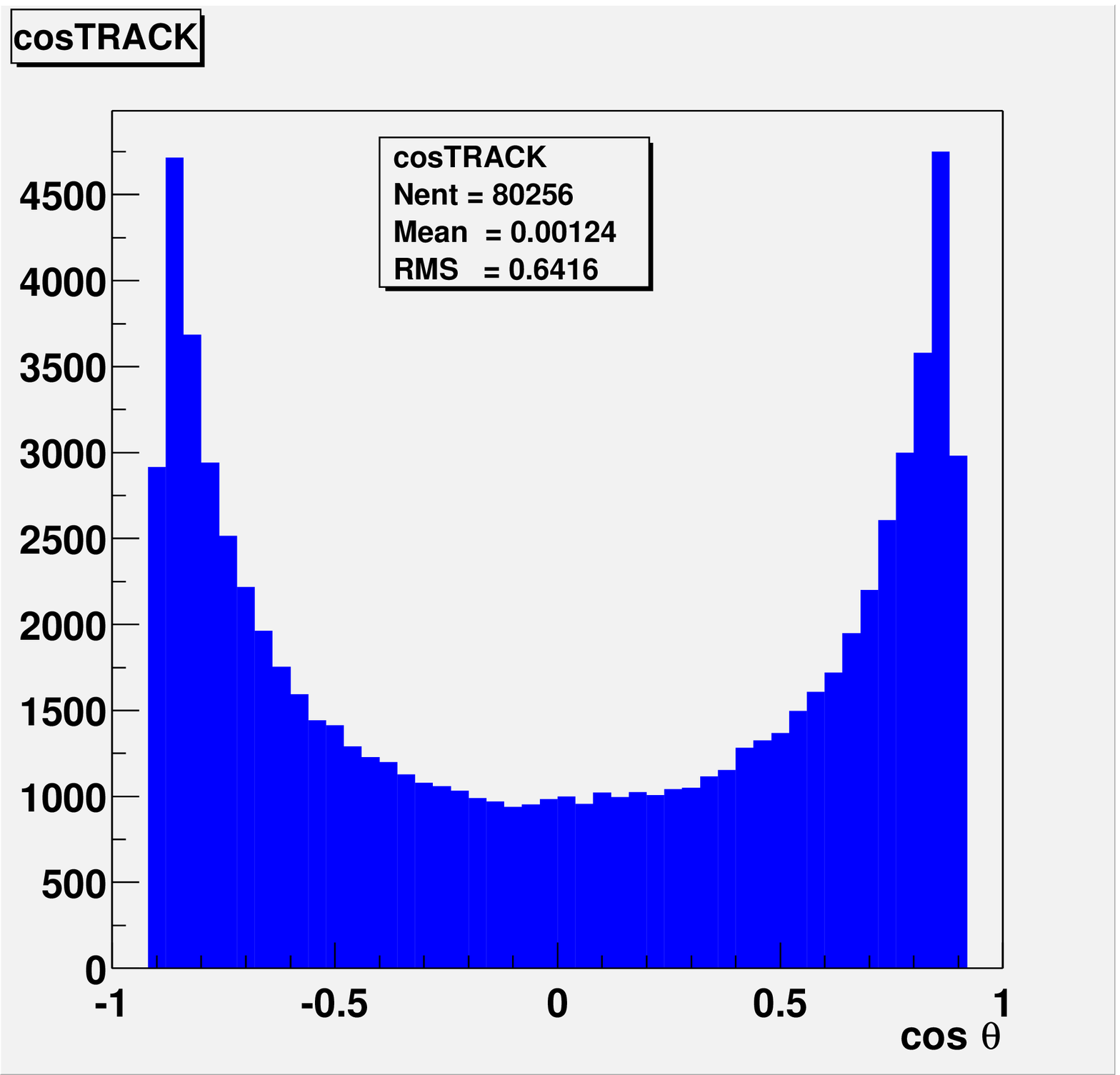}}
\caption{\label{fig:500track}
Tracks contributing to the resolved photon background
for $\rtsee=500$~GeV.
Energy and    $\cos\theta$ distribution for tracks with $p>0.2$~GeV.
The plots corresponds to 20,500 beam crossings.}
\end{center}\end{figure}

\begin{figure}\begin{center}
\resizebox{.9\textwidth}{!}{
\includegraphics{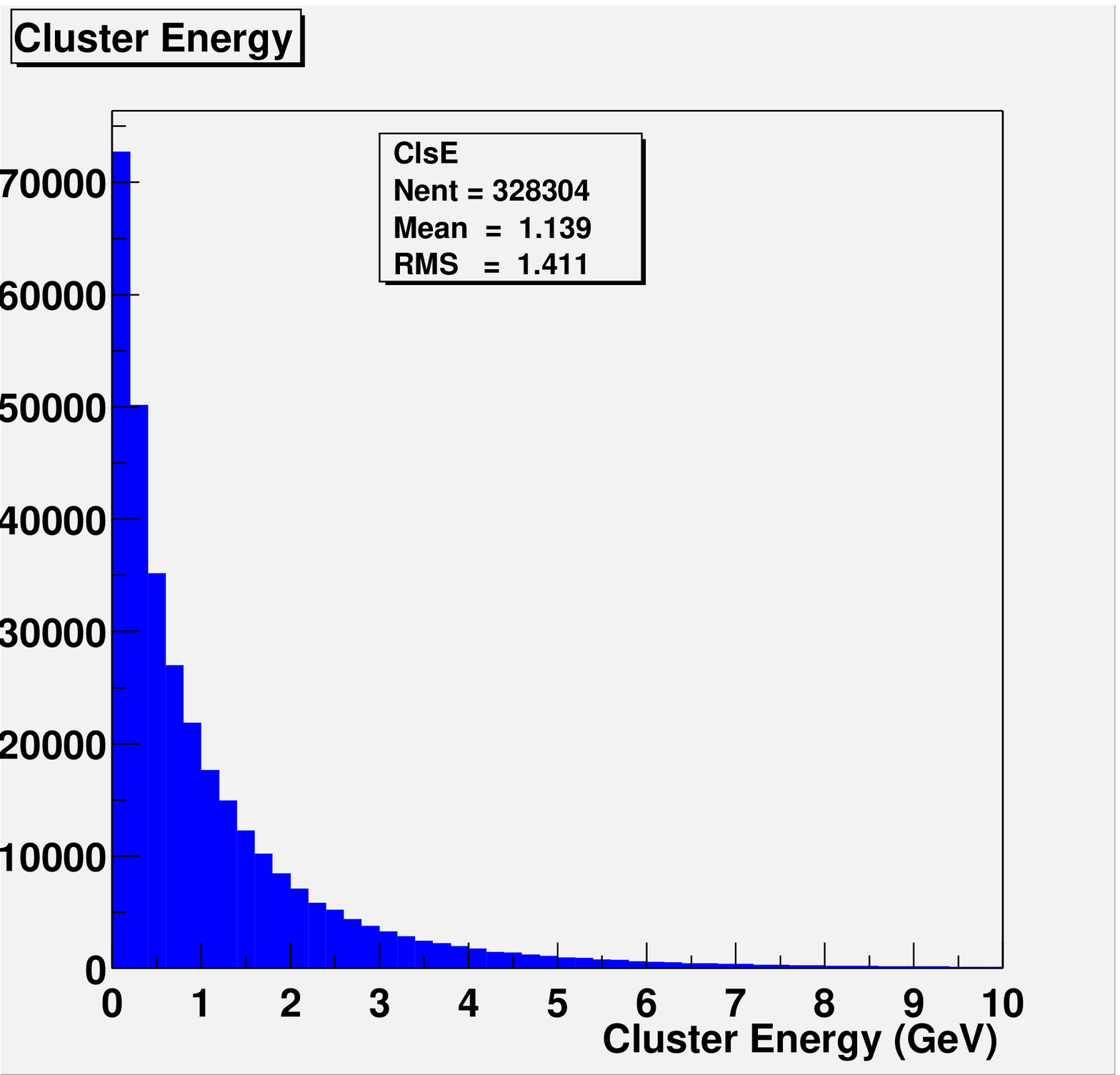} 
\includegraphics{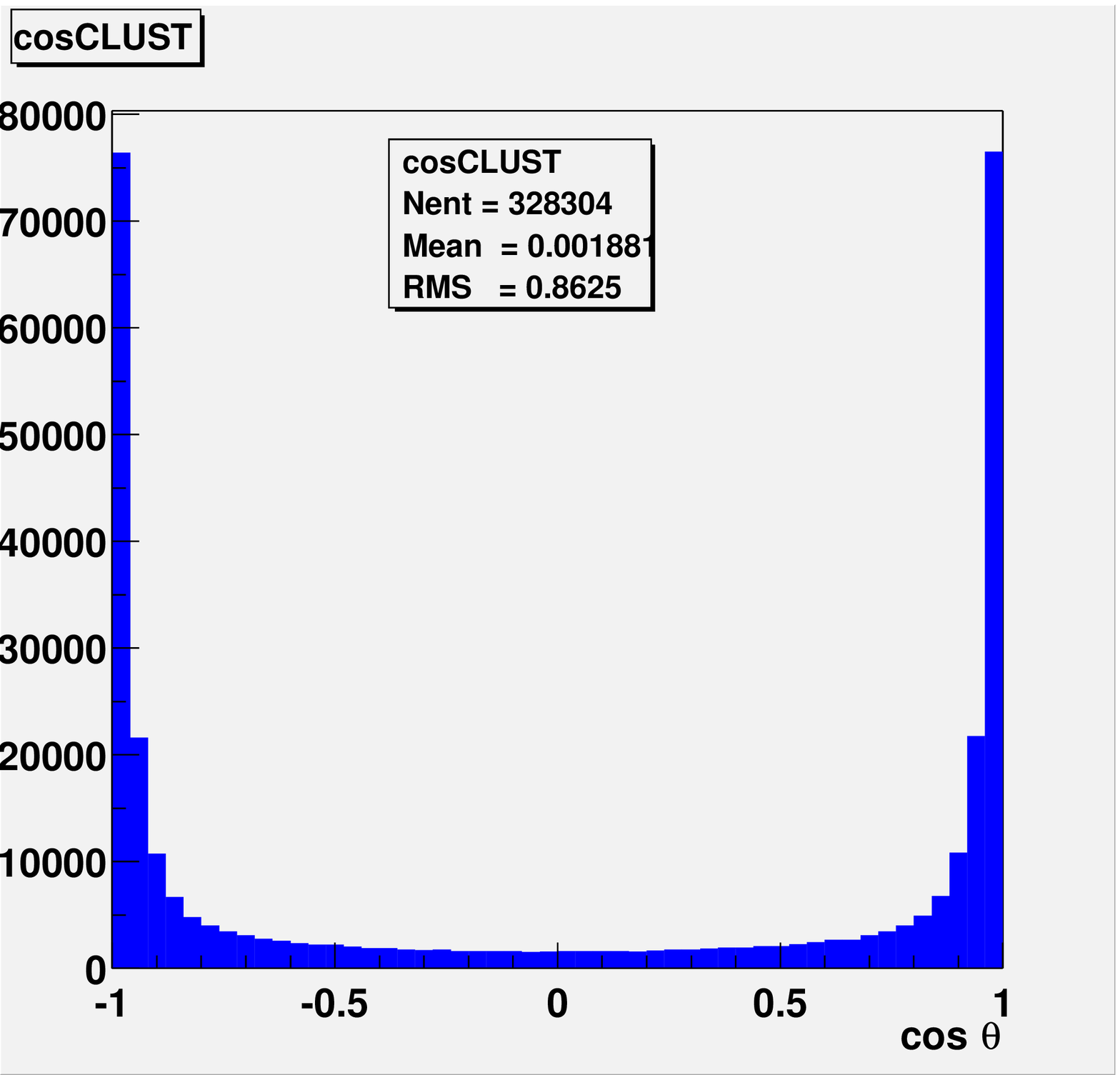}}
\caption{\label{fig:500clust}
Showers contributing to the resolved photon background
for $\rtsee=500$~GeV.
Energy and    $\cos\theta$ distribution for showers with $E>0.1$~GeV.
The plots corresponds to 20,500 beam crossings.}
\end{center}\end{figure}

We process the events through the LC Fast MC detector simulation within the 
ROOT\cite{Brun:1997pa} framework.  The simulation includes calorimeter 
smearing and the detector configuration 
described in Section 4.1 of Chapter 15 of Ref.~\cite{Abe:2001nr}.
From our preliminary studies, we  find that the two-photon cross section
is dominated by the twice resolved process at both 160~GeV and 500~GeV.
For the beam parameters given in Table~\ref{tab:beampar}, 
we expect a  background event rate of 6,700 and 20,500 events/second for the
160~GeV and 500~GeV machines, respectively. This corresponds to 0.6
and 1.8 overlay events per beam crossing. For the TBA machine beam parameters
discussed in Ref.\cite{cliche}, 0.1 overlay events per beam crossing
are expected.
Most of the products of the $\gamma\gamma \to$~hadrons will be produced
at small angles relative to the photon beam and will escape  undetected
 down the beam pipe. We are interested in the decay products that
enter the detector. For  this reason we are only considering 
tracks and showers with $|\cos\theta|<0.9$ in the laboratory
frame, and we require  
tracks to have momentum greater than 200~MeV, while 
the showers must have energy greater than 100~MeV.
The resulting track and shower energy distributions integrated over 
17,000 beam crossings for $\rtsee=160$~GeV and 5,500 beam crossings for 
$\rtsee=500$~GeV are shown in
Fig.~\ref{fig:160track}-\ref{fig:500clust},
 and summarized in Table~\ref{tab:2resmult}.
Experimental, theoretical and modeling errors have not yet been evaluated for
these distributions.

\begin{table}
\caption{\label{tab:2resmult} Event Multiplicity Due to Resolved Photon Backgrounds.}
\begin{center}
\begin{tabular}{lcc}
\hline
\hline
 & 160 GeV & 500 GeV\\
\hline
Events/Crossing & 0.6 & 1.8 \\
Tracks/Crossing ($p>0.2$ GeV, $|\cos\theta|< 0.9$) & 3.7 & 14.6 \\
Energy/Track ($p>0.2$ GeV, $|\cos\theta|< 0.9$) & 0.70 GeV & 0.74 GeV \\
Clusters/Crossing ($E>0.1$ GeV, $|\cos\theta|<0.9$) & 5.5 & 21.8 \\
Energy/Cluster ($E>0.2$ GeV, $|\cos\theta|<0.9$) & 0.45 GeV & 0.49 GeV \\
\hline 
\hline
\end{tabular}
\end{center}
\end{table}

\par
Future studies of the physics possibilities of a \gamc\ should
include the impact of the resolved photons on the event reconstruction, and
we need to determine the appropriate number of beam crossing that 
we need to integrate over.
It is generally assumed that the \gamc\ and $e^+e^-$
detectors will be the same or at least have comparable performance.
At $e^+e^-$, the plan is to integrate over 100's to 1000's of beam crossings.
This depends, of course, on the choice of detector technology as well as 
the bunch structure of the electron beam. At a \gamc\
experiment, the desire to minimize the resolved photon backgrounds 
may drive the detector design to be able to minimize the 
number of crossings that are being readout together.
In preliminary studies, integrating over about 10 crossings appears
tractable. The required detector readout rate due to backgrounds from
resolved photons call into question the assumptions 
that the $e^+e^-$ and $\gamma\gamma$ detectors will be based on the same
technology and/or have comparable performance. An important
distinction between the TESLA and NLC/TBA machine designs 
is the time between bunch crossings, 337~ns and $\le$2.8~ns\footnote{The NLC-$e^+e^-$design has 1.4~ns bunch spacing -- see the caption of Table~\ref{tab:beampar}
for discussion.}
, respectively.
The authors plan to study the difference between
the detector design and performance.

The impact of resolved photon backgrounds on the physics reach of a 
\gamc\ has yet to be determined. 
Previously, we reported that including the resolved photon
contribution to the background in an analysis of $\gamma\gamma \to h \to
\gamma\gamma$~\cite{korea} cause a  $S/B\sim 3$, and that
the background was dominated by the resolved photons. However,
now we know that the 
resolved photon background was overestimated by a factor 
of six.


\section{Physics studies}

\begin{figure}[t]
\begin{center}
\mbox{\epsfig{file=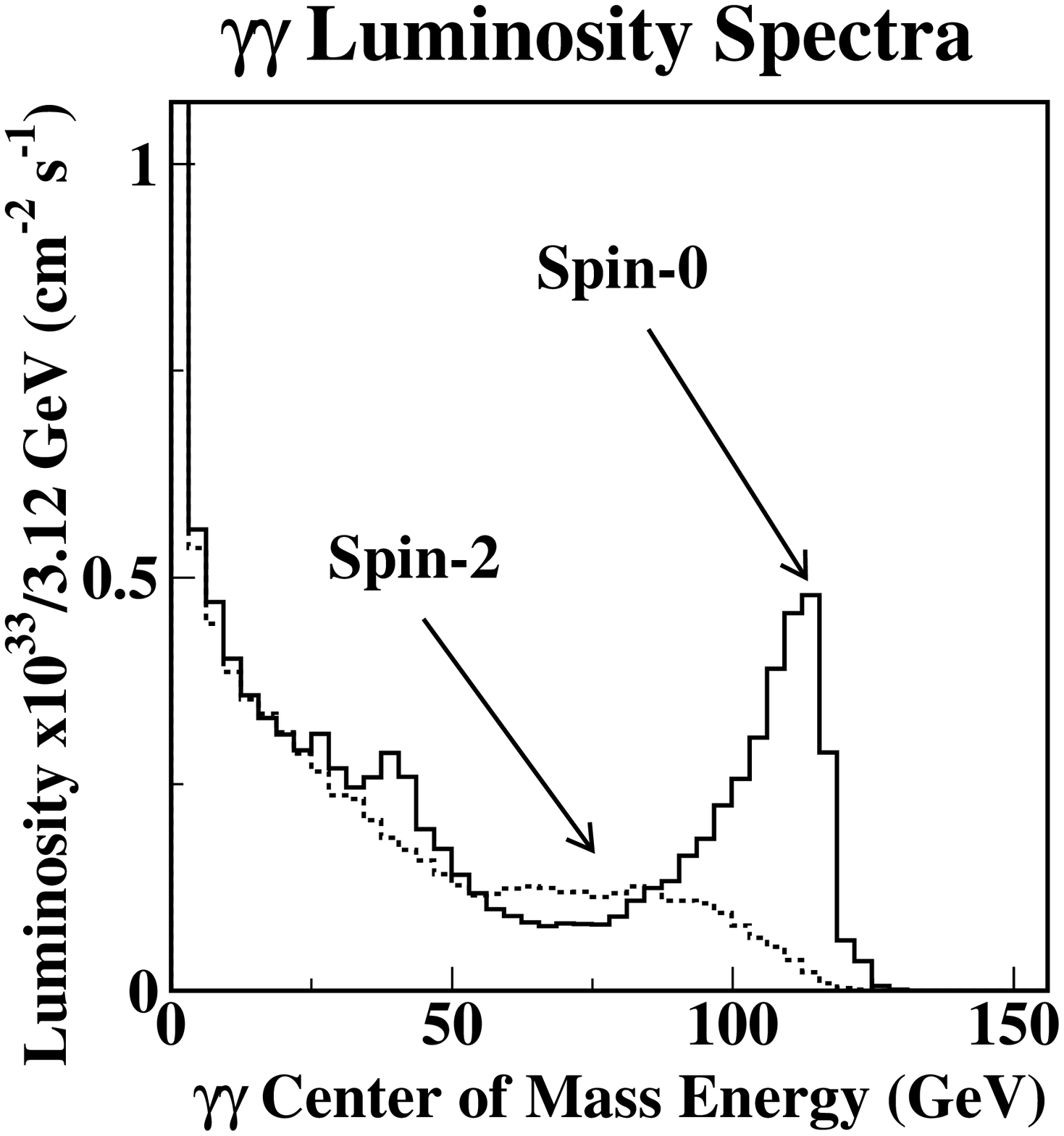,height=8cm}}
\mbox{\epsfig{file=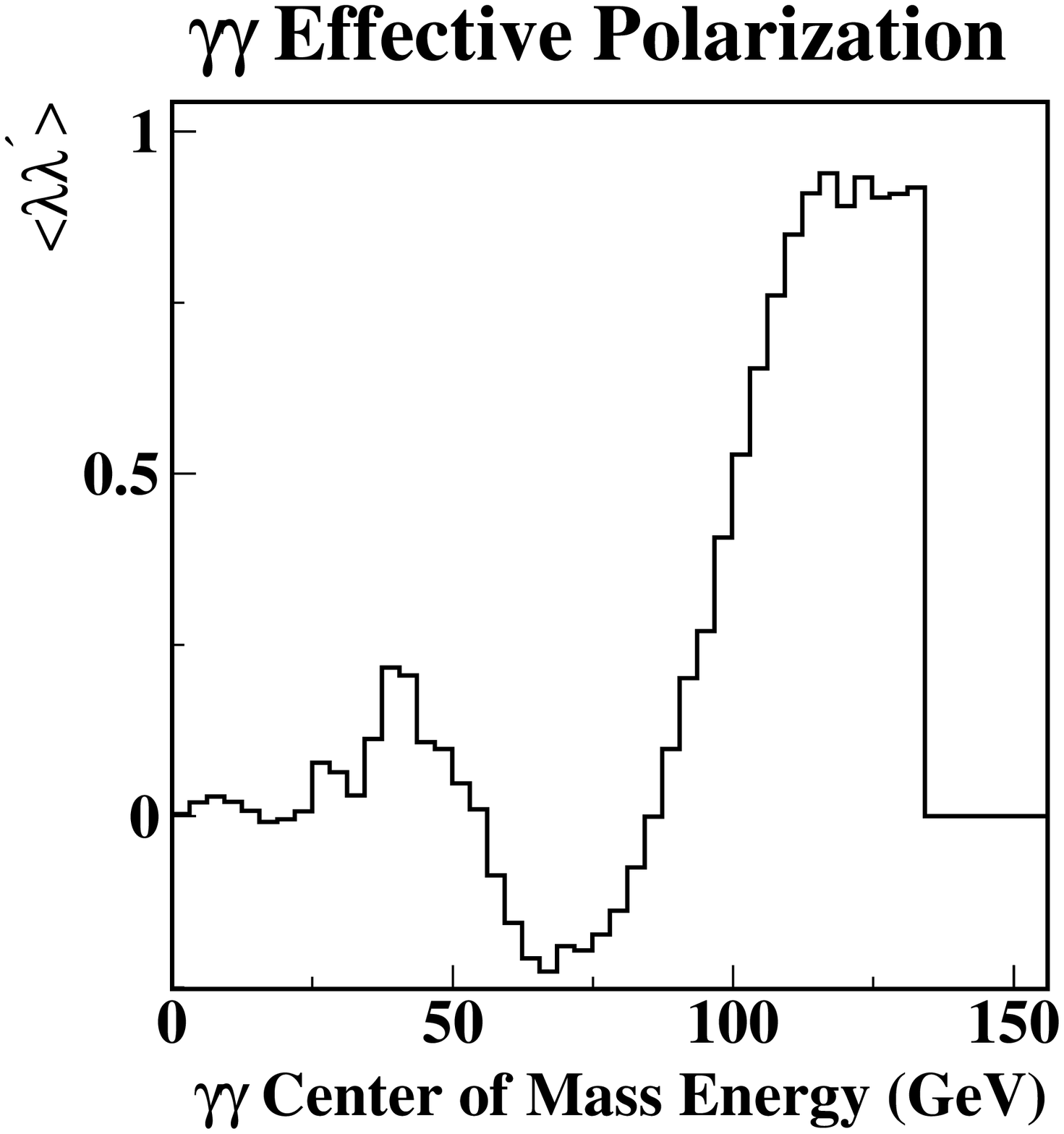,height=8cm}}
\end{center}
\caption[.]{\label{lum_cliche}\it
Luminosity spectra and beam polarization as functions of
$E_{CM}$($\gamma\gamma$) for the CLIC~1 parameters for 75~GeV electrons
obtained with {\tt DIMAD}~\cite{dimad} and {\tt CAIN}~\cite{cainref} for
 ${{\cal L}_{ee}=4.8 \times10^{34} \rm cm^{-2}s^{-1}}$.
}
\end{figure}

All the studies presented here are based on the CLICHE 
parameters given in~\cite{cliche}.  This \gamc\ was tunned
for a 115~GeV Standard Model Higgs, and it is designed to be 
a Higgs factory capable of producing
around 20,000 light Higgs bosons per year. 
The CLIC-1 beams and a laser backscattering system are
expected to be capable of   producing
a $\gamma\gamma$ total~(peak) luminosity of $2.0~(0.36) \times
10^{34}$~cm$^{-2}$s$^{-1}$, see Fig.~\ref{lum_cliche}. 
In the MSSM and some of the other models
under consideration,  the Higgs could have a mass 
as low as 90~GeV, and as high 
as 130~GeV. For the low mass case, the CLIC-1 energy could be lower,
but     an energy upgrade would be required  in the high mass case.

\subsection{Standard Model expactations at the  \gamc\ Higgs factory}
In the Standard Model, the branching ratios for ${\cal
B}r(H \to b\bar{b})$, ${\cal B}r(H \to WW)$, ${\cal B}r(H \to ZZ)$
 and ${\cal B}r(H \to\gamma\gamma)$ for a Higgs mass of 115~GeV are:  
73.7\%, 8.8\%  0.9\% and 0.2\%, respectively. 
In~\cite{cliche}, it was  shown that  the most promising
reaction for a  115~GeV Higgs, is  $\gamma \gamma \rightarrow H
\rightarrow {\bar b} b$, but the expectations for 
 other decay channels like $H \rightarrow W W$ and $H \rightarrow
\gamma\gamma$ were also given. Their results are summarized in 
Table~\ref{table:decay}.

However, since one of the objectives of a \gamc\ Higgs factory
would be to test the 
prediction for these branching ratios, 
and use their measurements to distinguish between the
Standard Model and its possible extensions, 
we are making an effort to also look at the  $\gamma \gamma \rightarrow H
\rightarrow ZZ$.  In order to evaluate the signal to background ratio 
in this decay mode,  we need the   cross sections for
$\gamma \gamma \rightarrow Z Z$ and
$\gamma \gamma \rightarrow four-fermions$.

We have made progress adding the processes $\gamma \gamma \rightarrow
\gamma \gamma$, $\gamma \gamma \rightarrow Z Z$ and $\gamma \gamma
\rightarrow \gamma Z$ to Pandora~\cite{PeskinPandora}. To date, these
processes have not been included in Pandora as they are 1-loop processes,
and as such the cross section is difficult to calculate. As it is the
most interest to us, we have focused on the $\gamma \gamma \rightarrow Z Z$
process.

The FormCalc and LoopTools packages~\cite{HahnLoopTools} can be used to
calculate the cross sections for various loop processes.  We use code
generated from these packages for us by Thomas Hahn, called \texttt{AAAA},
which can be used to calculate one loop integrals for general 2~to~2
processes, and 2~to~3 process.  Given the mass, charge, polarization and
nature (scalar, fermion, vector or photon) of the initial and final
particles \texttt{AAAA} calculates the cross section for a given center of
mass energy.  We have modified the \texttt{AAAA} code to create a
subroutine that returns the cross section for the process $\gamma \gamma
\rightarrow Z Z$ given the masses of the $Z$'s, the initial and final
polarizations, $\cos(\theta)$ and center of mass energy.

We have based the Pandora class for the $\gamma \gamma \rightarrow Z Z$
process on the $\gamma \gamma \rightarrow W W$ class.  When the cross
section is required by Pandora we call the subroutine discussed above.


The tools that we have develop will allows us to use the CAIN prediction
of the full $\gamma\gamma$ energy spectra and their corresponding
polarization. The expected cross section for $\gamma \gamma \rightarrow Z Z$
is smaller than 0.01~fb, for the CLICHE luminosity spectra and polarization 
shown in Fig.~\ref{lum_cliche}.

In order to determine the impact of $\gamma\gamma \rightarrow 
four - fermions$, background samples were generated with WHIZARD 1.24.
For details concerning the usage of WHIZARD for four-fermion processes,
please refer to point {\bf B}{\it 3} below.

At the moment, we have only 
studied the prospects for the detection of $\gamma\gamma \rightarrow
ZZ$ based on the search for the heaviest decay $b\bar{b}b\bar{b}$, 
and compare them to the light decays like $e^+e^-e^+e^-$,
$e^+e^-\mu^+\mu^-$ and $\mu^+\mu^-\mu^+\mu^-$ final states.

The signal sample for $\gamma\gamma \rightarrow H \rightarrow ZZ \rightarrow
b\bar{b}b\bar{b}$ was generated using Pythia 6.158 with an interface
to CAIN to get the correct CLICHE spectrum.  Event reconstruction and
analysis were done in the framework of the FastMC program and PAW.
About 75\% of generated events had four or more reconstructed
jets.  We required
$|cos\Theta_j|<0.9$ for $j=1,2,3,4$, where $\Theta_j$ is the polar
angle of the jet with respect to the beam direction, measured in the
lab frame.  The efficiency of this cut for the four-fermion background
processes depends strongly on the mass of the particle involved and for
$\gamma\gamma \rightarrow b\bar{b}b\bar{b}$ it produces a 20-fold
reduction of the initial
sample.  Meanwhile, signal events are distributed nearly isotropically,
so that $\sim$70\% of them will survive the cut.

Necessary signature to identify the $ZZ$ intermediate state is the
appearance of the $Z$ mass peak.  We therefore assume here that one $Z$
must be on the mass shell, leaving at most a 24 GeV mass for the
other $Z$.  Consequently, we selected only events for which one finds
two jets with 85 GeV $< M_{j_1j_2} <$ 97 GeV, and the remaining two with
12 GeV $< M_{j_3j_4} <$ 30 GeV.  It is to be noted that
barely half of the Pythia-generated signal events do actually satisfy
the above criteria.  Final selection is done by looking
at the four-jet invariant mass and taking events for which
$M_{4j} >$ 100 GeV, which should include nearly all signal events which
survived previous selection criteria.  The final selection efficiency
for signal events is around 25\%.  

No $b$ tagging was simulated.
We assume an additioanl 70\% efficiency for tagging a single $b\bar{b}$ pair.
Taking $\sigma(\gamma\gamma\rightarrow H)$ = 112 fb,
$BR(H\rightarrow ZZ)$ = 0.009 and $BR(Z\rightarrow b\bar{b})$ = 0.15,
we end up at 0.6  reconstructed $\gamma\gamma \rightarrow H \rightarrow ZZ
\rightarrow b\bar{b}b\bar{b}$ events per canonical year of $10^7 s$.
For the background, we have considered the direct $b\bar{b}b\bar{b}$
production,
as well as $\gamma\gamma \rightarrow b\bar{b}c\bar{c}$ with a $c\bar{c}$
pair mistagged as $b\bar{b}$ and $\gamma\gamma \rightarrow c\bar{c}c\bar{c}$
with a double $c\bar{c}$ mistagging.  We assume
a 3.5\% probability of mistagging $c\bar{c}$ as $b\bar{b}$.  Total
selection acceptance for background processes as a function of energy
was found to vary from less than 0.01\% for all three considered processes
at 100 GeV and below to 0.4\% at
120 GeV for $\gamma\gamma \rightarrow b\bar{b}b\bar{b}$, to 0.2\%
at 120 GeV for $\gamma\gamma \rightarrow b\bar{b}c\bar{c}$ and to 0.04\%
at 120 GeV for $\gamma\gamma \rightarrow c\bar{c}c\bar{c}$.  Total cross
sections in this energy region, calculated by WHIZARD 1.24, were found to
be around 300 fb for $\gamma\gamma \rightarrow
b\bar{b}b\bar{b}$, 8 pb for $\gamma\gamma \rightarrow b\bar{b}c\bar{c}$
and 90 pb for $\gamma\gamma \rightarrow c\bar{c}c\bar{c}$, and slowly
varying with energy.  From all this input, we arrive at a final number
of 4.4 $b\bar{b}b\bar{b}$ events, 4.7 $b\bar{b}c\bar{c}$ events and 0.9
$c\bar{c}c\bar{c}$ events in the signal window per $10^7 s$.
Therefore, the signal to background is not optimal in this channel.

However, similar analysis for $Z\rightarrow ee, \mu\mu$  seem to give a 
better signal to background ratio, but higher monte carlo  statistics is
needed to be able to  confirm. The reasons are the following: the reconstruction
efficiency of the signal is much higher, and even though the 
$\sigma(\gamma\gamma\rightarrow 4l, l=e~or~\mu)$ are higher that for $b\bar{b}b\bar{b}$, their 
stronger $cos\Theta$ dependence causes most of the events to go down the 
beampipe.  For example,   $\sigma(\gamma\gamma\rightarrow 4\mu)$=0.16~nb, but 
only 1.7~fb have at least all four $\mu$'s in the detector.  

Further work is needed before we can conclude, but the required tools are
now available.

\begin{table}[t]
\caption{\it The statistical errors on selected decay modes of
a 115~GeV Higgs boson in the Standard Model. The $\gamma\gamma\to h$
cross section for the full~(peak)  ${\cal L_{\gamma\gamma}}$ 
is  assumed to be 112~(624)~fb (see Ref.\cite{cliche}).  
The expected yield for
200~(36)~fb$^{-1}$ is 22,400  Higgs particles.}
\label{table:decay}
\begin{center}
\begin{tabular}{lccccc}
\hline
\hline
decay mode\,\,\,\,\,  &\,\,\,\,\, raw events/year\,\,\,\,\, &\,\,\,\,\, S/B\,\,\,\,\,  & \,\,\,\,\,\,\,\,\,\,$\epsilon_{sel}$\,\,\,\,\, &\,\,\,\,\,\,\,\,\,\, ${\cal B}r$\,\,\,\,\, &\,\,\,\,\, $\Delta \Gamma_{\gamma\gamma} {\cal B}r/\Gamma_{\gamma\gamma}{\cal B}r$ \\
\hline
${\bar b} b$      & 16509         & 2.0  &   0.20 &  73.7\% & 2\% \\
\hline
$W^+ W^-$         & 1971            & 1.2  &  0.32  &  8.8\% & 5\%\\
\hline
$\gamma\gamma$    &  47             & 1.3  &  0.85  &  0.23\% & 22\%\\
\hline
$ZZ$              & 201             &      &        & 0.9\% & 11\%\\
\hline
\hline
\end{tabular}
\end{center}
\end{table}

%
%
\subsection{NMSSM Scenario with $h\rightarrow aa$}


In this section, we demonstrate that a \gamc\ add-on to the CLIC
test machines could provide invaluable complementarity to the LHC
when it comes to studying and verifying difficult Higgs signals 
that can arise at the LHC in the context of
the Next to Minimal Supersymmetric Model or other Higgs sectors
in which a SM-like Higgs boson can decay to two lighter Higgs bosons.

\subsubsection{Introduction}

As repeatedly stressed, 
one of the key goals of the next generation of colliders is the
discovery of Higgs boson(s) \cite{hhg}. 
 Assuming the absence
of CP violation in the Higgs sector, the Higgs bosons of the Minimal
Supersymmetric Model (MSSM) comprise the
CP-even $\hl$ and $\hh$, the CP-odd $\ha$ and the charged Higgs, $\hpm$.
Recent reviews of the prospects for Higgs discovery and
study at different colliders include \cite{Carena:2002es,Gunion:2003fd}.
It has been established that the LHC with $L>100\fbi$ will discover
at least one of these Higgs bosons. A LC or
a \gamc~will be able to perform detailed
precision measurements of great importance to testing the details
of the MSSM Higgs sector and are very likely to discover those
Higgs bosons of the MSSM that can not be seen at the LHC.
For example, the \gamc\ can detect the $\hh$ and $\ha$
in the large-$\mha$, moderate-$\tanb$ wedge region where
the LHC will not be able to detect them.  

The results for the CP-conserving (CPC) MSSM do not generalize to
supersymmetric models with more complicated Higgs sectors.  One highly
motivated extension of the MSSM is the Next-to-Minimal Supersymmetric
Model (NMSSM), in which one additional singlet Higgs superfield is
introduced in order to naturally explain the poorly understood $\mu$
parameter \cite{Ellis:1988er,hhg}.  The NMSSM Higgs sector comprises
three (mixed) CP-even states ($h_{1,2,3}$) two (mixed) CP-odd states
($a_{1,2}$) and a charged Higgs pair ($\hpm$).  Guaranteeing discovery
of at least one NMSSM Higgs boson is a considerable challenge. In
particular, one of the key ingredients in the no-lose theorem for CPC
MSSM Higgs boson discovery is the fact that the SM-like Higgs boson
(the $h$ when $m_A\gsim 125\gev$ or $H$ when $m_A\lsim 115\gev$) never
has significant decays to other Higgs bosons ($h\to AA$ or $H\to
AA,hh$, respectively).  In the NMSSM, Higgs boson masses are not very
strongly correlated, and $h_1\to a_1a_1$ or $h_2\to a_1a_1$ decays can
be prominent \cite{Gunion:1996fb,Dobrescu:2000jt}.  (If
one-loop-induced CP violation is substantial in the MSSM Higgs sector,
decays of one CP-mixed MSSM Higgs boson to two others are also
possible \cite{Carena:2002bb}.)  Such decays fall outside the scope of
the usual detection modes for the SM-like MSSM $h$ on which the MSSM
no-lose LHC theorem largely relies.  The first question is whether
this makes an absolute LHC no-lose theorem for the NMSSM impossible.
Second, if there are regions of parameter space in which the LHC
signal is marginal or of uncertain interpretation, could a LC 
or (our particular interest here) a  \gamc~
{\it alone} (\ie\ in the absence of a LC) guarantee Higgs
discovery or help verify the nature of the signal for such regions.
The purpose of this note is to remark on the complementarity of a
\gamc\ to the LHC in such regions.  We will find that it could be very
crucial.

In earlier work \cite{Ellwanger:2001iw}, a partial no-lose
theorem for NMSSM Higgs boson discovery at the LHC was established.
In particular, it was shown that the LHC would be able to detect at
least one of the NMSSM Higgs bosons (typically, one of the lighter
CP-even Higgs states) throughout the full parameter space of the model, 
excluding only those
parameter choices for which there is sensitivity to the
model-dependent decays of Higgs bosons to other Higgs bosons and/or
superparticles.  

In more recent work, the assumption
of a heavy superparticle spectrum has been retained and
the focus was on the question of whether or not this
no-lose theorem can be extended to those regions of NMSSM parameter
space for which Higgs bosons can decay to other Higgs bosons.
It is found \cite{Ellwanger:2003jt} that the parameter choices such
that the ``standard'' discovery modes fail are such that 
either the $h_1$ or $h_2$ (numerically ordered according to increasing mass)
is very SM-like in its couplings, but
mainly decays to $a_1a_1$. 
Detection of $h_{1,2}\to a_1a_1$ will be difficult since each $a_1$ will decay
primarily to either $b\anti b$ (or 2 jets if $m_{a_1}<2m_b$) 
and $\tau^+\tau^-$, yielding final
states that will typically have large backgrounds at the LHC.
 In the end, there is a signal at the LHC 
for those cases in which the $a_1$ decays to $b\anti b$
with a substantial branching ratio even for these
most difficult cases, but it will be hard to be sure 
if the signal really corresponds to a Higgs boson.
We will show that a \gamc\  would be very important for clarifying the
nature of the signal.

\subsubsection{Details Regarding Earlier Results}

In the earlier work, the simplest version of the NMSSM is considered, 
in which the term
$\mu \widehat H_1 \widehat H_2$ in the superpotential of the MSSM is replaced
by (using the notation $\widehat A$ for the superfield and $A$ for its scalar
component field)
\vspace*{-.1in}
\begin{equation}\label{2.1r}
\lambda \widehat H_1 \widehat H_2 \widehat S\ + \ \frac{\kappa}{3} \widehat S^3
\ \ ,
\end{equation}
\vskip -.1in \noindent
so that the superpotential is scale invariant. 
No assumptions regarding the
``universal'' soft terms are made. 
Hence, the five soft supersymmetry breaking terms
\begin{equation}\label{2.2r}
m_{H_1}^2 H_1^2\ +\ m_{H_2}^2 H_2^2\ +\ m_S^2 S^2\ +\ \lambda
A_{\lambda}H_1 H_2 S\ +\ \frac{\kappa}{3} A_{\kappa}S^3
\end{equation}
are considered as independent. 
The masses and/or couplings of
sparticles are assumed to be such that their contributions to the loop diagrams
for $gg\to h$ and $ \gamma\gamma\to h$ couplings
are negligible. 
In the stop sector, which appears in the radiative corrections
to the Higgs potential, the soft masses $m_Q = m_T \equiv M_{susy}= 1$
TeV are chosen. 
Scans are perfomred over the stop mixing parameter, related to $M_{susy}$ 
and the soft mixing parameter $A_t$ by 
$
X_t \equiv 2 \ \frac{A_t^2}{M_{susy}^2+m_t^2} \left ( 1 -
\frac{A_t^2}{12(M_{susy}^2+m_t^2)} \right ) \ .
$
As in the MSSM, the value $X_t = \sqrt{6}$ -- so called maximal
mixing -- maximizes the radiative corrections to the Higgs boson
masses, and it is found
that it leads to the most challenging points in NMSSM parameter space.

In the studies of \cite{Ellwanger:2003jt}, 
a numerical scan over the free parameters is performed.
For each point,  the masses and mixings of the
 CP-even and CP-odd Higgs bosons, $h_i$ ($i=1,2,3$) and $a_j$ ($j=1,2$)
are computed,
taking into account radiative corrections up to the dominant two
loop terms. 
Parameter choices excluded by LEP constraints
on $e^+ e^- \to Z h_i$ 
 and $e^+ e^- \to h_i a_j$ are eliminated and   
$m_{h^\pm} > 155$~GeV is required, so that $t
\to h^\pm b$ would not be seen.
LHC discovery modes for a Higgs boson considered were
(with $\ell=e,\mu$):

1) $g g \to h/a \to \gamma \gamma$;\par
2) associated $W h/a$ or $t \bar{t} h/a$ production with 
$\gamma \gamma\ell^{\pm}$ in the final state;\par
3) associated $t \bar{t} h/a$ production with $h/a \to b \bar{b}$;\par
4) associated $b \bar{b} h/a$ production with $h/a \to \tau^+\tau^-$;\par
5) $g g \to h \to Z Z^{(*)} \to$ 4 leptons;\par
6) $g g \to h \to W W^{(*)} \to \ell^+ \ell^- \nu \bar{\nu}$;\par
7) $W W \to h \to \tau^+ \tau^-$;\par
8) $W W \to h\to W W^{(*)}$.\par
\noindent The expected statistical
significances at the LHC in all Higgs boson detection modes 1) -- 8) 
was estimated by
rescaling results for the SM Higgs boson and/or the 
the MSSM $h, H$ and/or $A$. 
Among these modes,
the $t\anti t h \to t\anti t b\anti b$ mode is
quite important.  The experimentalists extrapolated
this beyond the usual SM mass range of interest.  
Results for $\nsd=S/\sqrt B$ (where $S$ and $B$
are the signal and background, respectively) 
assuming $K_S=K_B=1$ were employed,
awaiting the time when all $K$ factors are known.
(For all cases where both $K_S$ and $K_B$ are known, their
inclusion improves the $\nsd$ value.)
For each mode, the procedure was to
use the results for the ``best detector'' (e.g. CMS
for the $t\anti t h$ channel), assuming
$L=300\fbi$ for that {\it one} detector.

Some things that changed between the 1st study of \cite{Ellwanger:2001iw}
and the 2nd study of \cite{Ellwanger:2003jt} were the following.
1) The $gg\to\hsm\to \gam\gam$ $\nsd$ values from CMS
were reduced by the inclusion of detector cracks
and other such detector details.
2) The CMS $t\anti t\hsm\to t\anti t b\anti b$ $\nsd$
vales came in substantially larger than the ATLAS values.
3) The experimental evaluations of the $WW$ fusion
channels yield lower $\nsd$ values than the original
theoretical estimates.

The results from \cite{Ellwanger:2001iw} can be summarized as follows.
For parameter space regions where none of the ``higgs-to-higgs'' decays
\begin{eqnarray}
& i) \ h \to h' h' \; , \quad ii) \ h \to a a \; , \quad iii) \ h \to h^\pm
h^\mp \; , \quad iv) \ h \to a Z \; , \nonumber \\
& v) \ h \to h^\pm W^\mp \; , \quad vi) \ a' \to h a \; , \quad vii) \ a \to h
Z \; , \quad viii) \ a \to h^\pm W^\mp \; .\nonumber
\end{eqnarray}
is kinematically allowed it is possible for the LHC signals
to be much weaker than SM Higgs signals.
In particular, one can find parameters such that
the $gg\to h_i \to \gam\gam$ rates are all greatly suppressed and
all the $h_i\to WW$ branching fractions and couplings are suppressed.
  The result is greatly decreased $N_{SD}$ values for
all the channels 1) -- 8), and a not very wonderful
combined statistical significance  after summing over 
various sets of channels.
Nonetheless, the very worst parameter choices for the no-higgs-to-higgs
decay part of parameter space does predict a $\geq 5\sigma$ signal
for at least one Higgs boson in at least one channel --- in particular,
in the $t\anti t \h\to t\anti t b\anti b$ channel or the $WW\to \h\to
\tauptaum$ channel.

In order to probe the complementary part of the parameter space, 
in \cite{Ellwanger:2003jt}, it was required that at least one of
the decay modes $i) - viii)$ is allowed. 
In the set of randomly scanned points,
those for which all the
statistical significances in modes 1) -- 8) are below
$5\sigma$ were selected. 
There were a lot of points, all with similar
characteristics. Namely, in the Higgs spectrum, there is a very
SM-like CP-even Higgs boson with a mass between 115 and 135~GeV ({\it
  i.e.} above the LEP limit), which can be either $h_1$ or $h_2$, with
essentially full strength coupling to the gauge bosons.
This state
decays dominantly to a pair of (very) light CP-odd states, $a_1a_1$,
with $m_{a_1}$ between 5 and 65~GeV.  
(The singlet component of $a_1$
has to be small in order to have a large $h_1 \to a_1 a_1$ or $h_2\to a_1a_1$
branching ratio when the
$h_1$ or $h_2$, respectively, is the SM-like Higgs boson.) 
Since the $h_1$ or $h_2$ is very SM-like, 
the $e^+ e^- \to h_1 a_1$ or $e^+e^-\to h_2 a_1$
associated production processes have very low rate and place no constraint 
on the light CP-odd state at LEP. 
For illustration, six
difficult benchmark points were selected. The features of
these points are displayed in Table~\ref{tpoints}.
For points 1 -- 3, $h_1$ is the SM-like
CP-even state, while for points 4 -- 6 it is $h_2$. 
One should note the large $\br(h\to a_1 a_1)$
of the SM-like $h$ ($h=h_1$ for points 1 -- 3 and $h=h_2$
for points 4 --6). 
 For points 4 -- 6, with $m_{h_1}<100\gev$,
the $h_1$ is mainly singlet and therefore evades LEP constraints.
Further, in the case of the points 1 -- 3, the $h_2$ would not
be detectable either at the LHC or the LC. For points 4 -- 6,
the $h_1$, though light, is singlet in nature and would
not be detectable.
Further, the $h_3$ or $a_2$
will only be detectable for points 1 -- 6 if a super
high energy LC is eventually built so that
$e^+e^-\to Z\to h_3 a_2$ is possible.
Thus, it was necessary to focus on searching for the SM-like $h_1$ ($h_2$)
for points 1 -- 3 (4 -- 6) using
the  dominant $h_1(h_2)\to a_1a_1$ decay mode.

\begin{table}[ht]
\caption{\label{tpoints} \sl Properties of selected scenarios that could escape detection
at the LHC. The production rate
for $gg\to h_i$ fusion
relative to the $gg$ fusion rate for
a SM Higgs boson with the same mass is given. Important absolute branching
ratios are displayed. For points 2 and 6, $\br(a_1\to jj)\simeq 
1-\br(a_1\to \tau^+\tau^-)$.
For the heavy $h_3$ and $a_2$, only their masses are shown.
For all points 1 -- 6, the statistical
significances for the detection of any Higgs boson in any of the channels 1) --
8) (as listed in the introduction) are tiny; their maximum is indicated in the
last row, together with the process number and the corresponding Higgs state.}
\begin{center}
{\small
\begin{tabular} {|l|l|l|l|l|l|l|} 
\hline
Point Number & 1 & 2 & 3 & 4 & 5 & 6  \\
\hline \hline
Bare Parameters &\multicolumn{6}{c|}{} \\
\hline
$\lambda$            & 0.2872 & 0.2124 & 0.3373 & 0.3340 & 0.4744 & 0.5212 \\
\hline
$\kappa$             & 0.5332 & 0.5647 & 0.5204 & 0.0574 & 0.0844 & 0.0010 \\
\hline
$\tan\beta$          &   2.5  &   3.5  &   5.5  &    2.5 &    2.5 & 2.5 \\
\hline
$\mu_{\rm eff}~({\rm GeV})$&    200 &    200 &    200 &    200 &    200 & 200 \\
\hline
$A_{\lambda}~({\rm GeV})$  &    100 &      0 &     50 &    500 &    500 & 500 \\
\hline
$A_{\kappa}~({\rm GeV})$   &      0 &      0 &      0 &      0 &      0 & 0 \\
\hline \hline
CP-even Higgs Boson Masses and Couplings &\multicolumn{6}{c|}{} \\
\hline \hline
$m_{h_1}$~(GeV)      &    115 &    119 &    123 &     76 &     85 &  51\\
\hline
Relative 
gg Production Rate   &   0.97 &   0.99 &   0.99 &   0.00 &   0.01 &  0.08\\
\hline
$\br(h_1\to 
b\anti b)$           &   0.02 &   0.01 &   0.01 &   0.91 &   0.91 &  0.00\\
\hline
$\br(h_1\to 
\tau^+\tau^-)$      &   0.00 &   0.00 &   0.00 &   0.08 &   0.08 &  0.00\\
\hline
$\br(h_1\to a_1 a_1)$&   0.98 &   0.99 &   0.98 &   0.00 &   0.00 &  1.00\\
\hline \hline

$m_{h_2}$~(GeV)      &    516 &    626 &    594 &    118 &    124 &  130\\
\hline
Relative
gg Production Rate   &   0.18 &   0.09 &   0.01 &   0.98 &   0.99 &  0.90\\
\hline
$\br(h_2\to 
b\anti b)$           &   0.01 &   0.04 &   0.04 &   0.02 &   0.01 &  0.00\\
\hline
$\br(h_2\to 
\tau^+\tau^-)$      &   0.00 &   0.01 &   0.00 &   0.00 &   0.00 &  0.00\\
\hline
$\br(h_2\to a_1 a_1)$&   0.04 &   0.02 &   0.83 &   0.97 &   0.98 &  0.96\\
\hline \hline

$m_{h_3}$~(GeV)      &    745 &   1064 &    653 &    553 &    554 &  535\\
\hline \hline
CP-odd Higgs Boson Masses and Couplings &\multicolumn{6}{c|}{} \\
\hline \hline
$m_{a_1}$~(GeV)      &     56 &      7 &     35 &     41 &     59 &  7\\
\hline
Relative
gg Production Rate   &   0.01 &   0.03 &   0.05 &   0.01 &   0.01 &  0.05\\
\hline
$\br(a_1\to 
b\anti b)$           &   0.92 &   0.00 &   0.93 &   0.92 &   0.92 &  0.00\\
\hline
$\br(a_1\to 
\tau^+\tau^-)$      &   0.08 &   0.94 &   0.07 &   0.07 &   0.08 &  0.90\\
\hline \hline

$m_{a_2}$~(GeV)      &    528 &    639 &    643 &    560 &    563 &  547\\
\hline 
Charged Higgs  
Mass (GeV)           &    528 &    640 &    643 &    561 &    559 &  539\\
\hline\hline
Most Visible Process No. &  2 ($h_1$) &  2 ($h_1$) &  8
                  ($h_1$) &  2 ($h_2$) &  8 ($h_2$)  &  8 ($h_2$)\\
\hline            
Significance at 300~${\rm fb}^{-1}$
                     &   0.48 &   0.26 &   0.55 &   0.62 &  0.53  & 0.16\\
\hline
\end{tabular}
}
\end{center}
\end{table}

\begin{figure}[h!]
\centerline{LHC, $\sqrt{s_{pp}}=14$ TeV}
\begin{center}
\includegraphics[width=8cm,angle=90]{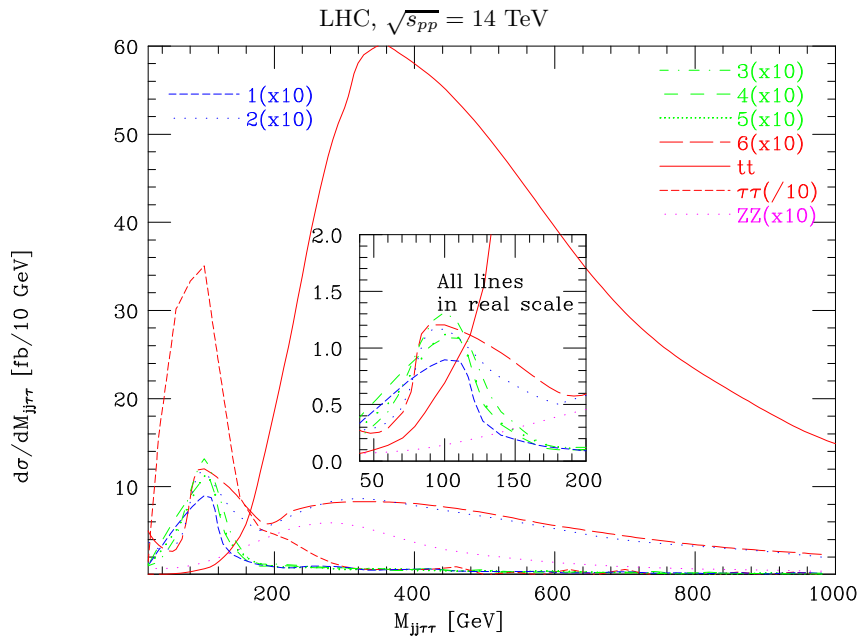}
\end{center}
\caption{\footnotesize \sl Reconstructed mass of the $jj\tau^+\tau^-$ system for signals and backgrounds after the selections described, at the LHC. 
The distribution $d\sigma/dM_{jj\tau^+\tau^-}$ [fb/10~GeV] vs. $M_{jj\tau^+\tau^-}$~[GeV] is plotted. 
The lines corresponding to points 4 and 5
are visually indistinguishable. No $K$ factors are included. 
\label{MHLHC}
}
\end{figure}

For points 1 and 3 -- 5, $a_1\to b\anti b$ decays are dominant
with $a_1\to \tauptaum$ making up the rest.
In the case of points 2 and 6, $m_{a_1}<2m_b$ so that
$a_1\to \tau^+\tau^-$ decays are dominant, 
with $a_1\to jj$ decays making up most of the rest.
For points 1 and 3 -- 5,
the $b$ jets will not be that energetic and $b$-tagging will
be somewhat inefficient. However, because of a large $jj\tauptaum$
background from Drell-Yan $\tauptaum$ pair production, $b$-tagging will be
needed.  The situation is illustrated in Fig.~\ref{MHLHC}, where
the cross section vs. the reconstructed $M_{jj\tauptaum}$ is plotted. 
The signals and backgrounds are plotted prior to the implementation of
$b$-tagging. For all six NMSSM setups, the Higgs resonance produces a 
somewhat amorphous bump in the very end of the
low mass tail of the $t\bar t$ background (see the insert in the top
frame of Fig.~\ref{MHLHC}).   The large size of the $\tauptaum$ background is
apparent.  After implementing single $b$-tagging, assuming 
criteria for which $\eps_b=0.5$ and
$\eps_{mis-tag}=0.01$, the signal and $t\anti t$ background rates
will be reduced by a factor of about 2 and the $\tau$ background will
be negligible for points 1 and 3 -- 5.  
(Viable techniques for observation of points 2 and 6 have not
yet been developed.)
Although the signal is somewhat amorphous in nature, 
statistics are significant. 
To estimate $S/\sqrt B$, the following procedure
was employed. First, assume $L=300~{\rm fb}^{-1}$,
a $K$ factor of 1.1 for $WW$ fusion
and a $K$ factor of 1.6 for the $t\anti t$ background.
(These $K$ factors are not included in the plots of Fig.~\ref{MHLHC}.)
Then, sum events over the region $60\leq M_{jj\tau^+\tau^-}\leq 90$~GeV.
Assuming a net efficiency of 50\% for single $b$-tagging,
the $t\anti t$ background rate is $B\sim 160$.
For points 1 and 3 -- 5, one finds signal rates of about $S=445$,  $375$,
$515$, $460$, corresponding to $N_{SD}=S/\sqrt B$ of 35, 30, 41 and 36,
respectively. However, given the broad distribution of
the signal, it is clear that a crucial question will be the accuracy
with which the background shape can be predicted from theory.  
(The background {\it normalization} after the cuts imposed in the analysis
would be very well known from the higher $M_{jj\tau^+\tau^-}$ regions.) 
Even more important is the question of how certain the interpretation
of this signal will be, given that it would be the {\it only } signal
for Higgs boson production. In this regard, detection of the $a_1a_1\to b\anti b b\anti b$ final state would be very valuable as it would allow
us to determine if $\br(a_1\to \tauptaum)/\br(a_1\to b\anti b)$ was
in the ratio predicted (by the ratio of fermion masses). 
However, there are large backgrounds in the $b\anti b b\anti b$ final state.
A study is in progress.

\subsubsection{The role of a \gamc}

While further examination of and refinements in the LHC analysis may ultimately
lead one to have good confidence in the viability of 
the NMSSM Higgs boson signals discussed above, 
an enhancement at low $M_{jj\tau^+\tau^-}$ of the type
shown (for some choice of $m_{a_1}$) will nonetheless be the only evidence
on which a claim of LHC observation of Higgs bosons can be based.
Ultimately, confirmation and further study at another collider
will be critical.

One possibility would be an LC, with energy up to 800~GeV.
(In the following, 
$h=h_1$ for points 1--3 and $h=h_2$ for points 4--6 in
Table~\ref{tpoints}.)
Because the $ZZh$ coupling is nearly full strength in all cases, 
and because the $h$ mass is of order 100~GeV,
discovery of the $h$ would be very straightforward via $e^+e^-\to Z h$ 
using the $e^+e^-\to ZX$ reconstructed $M_X$ technique which
is independent of the ``unexpected'' complexity of the $h$ decay
to $a_1a_1$. 
This would immediately provide a direct measurement
of the $ZZh$ coupling with very small error.
The next stage would be
to look at rates for the various $h$ decay final states, $F$,
and extract $BR(h\to F)=\sigma(e^+e^-\to Zh\to ZF)/\sigma(e^+e^-\to Zh)$.
For the NMSSM points considered here, the main channels would
be $F=b\anti b b\anti b$, $F=b\anti b \tau^+\tau^-$ and
$F=\tau^+\tau^-\tau^+\tau^-$. 
At the LC, a fairly accurate
determination of $BR(h\to F)$ should be possible in all three cases.
This would allow us to determine $BR(h\to a_1 a_1)$ independently.  
As demonstrated in \cite{Ellwanger:2003jt}, the $WW$ fusion production
mode would also yield an excellent signal and provide a second
means for studying the $h$ and its decays.
Indeed, at 800~GeV or above, $WW$ fusion
is the dominant Higgs boson production channel for
CP-even Higgs bosons in the intermediate mass range. 

Here, however, we wish to focus on the \gamc.
In scenarios (1), (3), (4) and (5) outlined in Table~\ref{nmssmtab}, 
the main signal channel to look for
experimentally is $\gamma\gamma \rightarrow h \rightarrow 
aa \rightarrow b\bar{b}b\bar{b}$.
A sizable signal would also be present for the $b\bar{b}\tau^+\tau^-$
final state.  In scenarios (2) and (6), the most promising channel to look at
is $\gamma\gamma \rightarrow h \rightarrow aa \rightarrow
\tau^+\tau^-\tau^+\tau^-$.
In what follows, we will demonstrate the photon collider potential for
observing the $h_{NMSSM} \rightarrow aa$ decay based on
the detection of its $b\bar{b}b\bar{b}$ final state.

Samples of signal and background events were generated and respective
detector responses simulated.
Signal samples were generated with Pythia 6.158, using a full $\gamma\gamma$
luminosity spectrum as obtained from CAIN.  We assume here the standard CLICHE
spectrum as employed in \cite{cliche}, 
which peaks at $E_{CM} = 115$ GeV and falls off quickly at larger
energies.  Note that in most scenarios under consideration here, $m_h > 115$
GeV, and so this particular spectrum is unlikely to be the optimum in the
general case.  The Pythia process
$\gamma\gamma \rightarrow H \rightarrow AA$ was
considered, where $H$ and $A$ are usually taken to be the ``heavy" scalar
and pseudoscalar Higgs bosons, respectively (e.g., in the MSSM).  Their
masses were set accordingly to the respective $h$ and $a$ masses in the
four relevant NMSSM scenarios.  The $H$ couplings were changed by hand
to ensure a close to unity branching fraction for $H \rightarrow AA$.
Overall normalization was given by the total cross section for $\gamma\gamma
\rightarrow h_{SM}$, where $h_{SM}$ is the Standard Model Higgs boson of
an appropriate mass.  Higgs production cross sections in each of the
four relevant
scenarios are listed in Table~\ref{nmssmtab} (fourth column).

The main background sources are the direct four-quark production processes:
$\gamma\gamma \rightarrow b\bar{b}b\bar{b}$, $\gamma\gamma \rightarrow
b\bar{b}c\bar{c}$ with a $c\bar{c}$ pair mistagged as $b\bar{b}$, and
$\gamma\gamma \rightarrow c\bar{c}c\bar{c}$ with a double $c\bar{c}$
mistagging.  Background samples were generated with WHIZARD 1.24.
Several cross checks of WHIZARD have been done, which are relevant for
this analysis.
The four-fermion generator in WHIZARD was cross checked for the
processes $\gamma\gamma \rightarrow e^+e^-e^+e^-$,
$e^+e^-\mu^+\mu^-$ and $\mu^+\mu^-\mu^+\mu^-$, and output cross
sections were found to be consistent with existing theoretical computations 
\cite{4lept}.  Typical cross sections
in this energy region were found to be around 300 fb for
$\gamma\gamma \rightarrow
b\bar{b}b\bar{b}$, 8 pb for $\gamma\gamma \rightarrow b\bar{b}c\bar{c}$
and 90 pb for $\gamma\gamma \rightarrow c\bar{c}c\bar{c}$, and slowly
varying with energy.
The effects of beam polarization were cross checked between WHIZARD
and Pythia for
$\gamma\gamma \rightarrow b\bar{b}$ and $c\bar{c}$ and found to be consistent.
For the four fermion processes, beam polarization plays a minor role.
Events were generated at several fixed center-of-mass energies, analysis
was consistently repeated for every sample.  Correct $\gamma\gamma$
luminosity spectra were obtained by interpolation of the results 
(cross sections and selection efficiencies) to arbitrary
intermediate energies and applying appropriate weights to events.

Event reconstruction and analysis was done in the framework of the
FastMC program.  We required exactly 4 reconstructed jets and
$|cos\Theta_j|<0.9$ for $j=1,2,3,4$, where $\Theta_j$ is the polar
angle of the jet with respect to the beam direction, measured in the
lab frame.  Four-fermion processes have angular distributions strongly
peaked at small angles, so that this cut reduces the amount of
surviving background by up to two orders of magnitude.  Meanwhile,
signal events are distributed nearly isotropically, with $\sim$80\% of
them surviving the cut.

The remaining events were checked for the two-jet invariant masses.
From the three possible combinations of two two-jet invariant masses
in a four-jet event, the one yielding two values the closest to
each other was selected.  We required consistency of the two values
within 10 GeV.  The acceptance of this cut is close to 100\% for signal
events and around 60\% for background events.

No $b$ tagging was simulated.  We assume a 50\% $b$ tagging efficiency
and a 3.5\% probability of mistagging a $c\bar{c}$ pair as $b\bar{b}$.

\begin{table}[h!]
\caption{NMSSM points 1,3,4,5.  We give cross sections, acceptance
after all cuts other than $b$-tagging, and the number of events
(assuming a $b$-tagging efficiency of $\eps_b=0.5$)
in a canonical $10^6$ second year.
\label{nmssmtab}
}
\begin{center}
\begin{tabular}{ c|c|c|c|c|c }
\hline
\hline
Scenario & $m_h$ (GeV) & $m_a$ (GeV) & $\sigma(\gamma\gamma \rightarrow h)$ (fb) & Acceptance &
No. events / 10$^6 s$\\
\hline
(1) & 115 & 56 & 112 & 0.26 & 139 \\
(3) & 123 & 35 & 9.1 & 0.33 & 14.7 \\
(4) & 118 & 41 & 46 & 0.28 & 63 \\
(5) & 124 & 59 & 6.0 & 0.24 & 7.1 \\
\hline
\hline
\end{tabular}
\end{center}
\end{table}

Signal samples remaining at this point in each of the four scenarios
considered, normalized to $10^6 s$ of data taking,
are given in Table~\ref{nmssmtab} (quoted acceptances do not include the
$b$ tagging efficiency).
Obviously, the final amount of signal strongly depends
on the $h$ mass.  This dependence is, however, almost entirely
due to the luminosity spectrum assumed and the resulting sensitivity
of the total $h$ production cross section to $m_h$.

\begin{figure}[here]
\includegraphics[width=11cm]{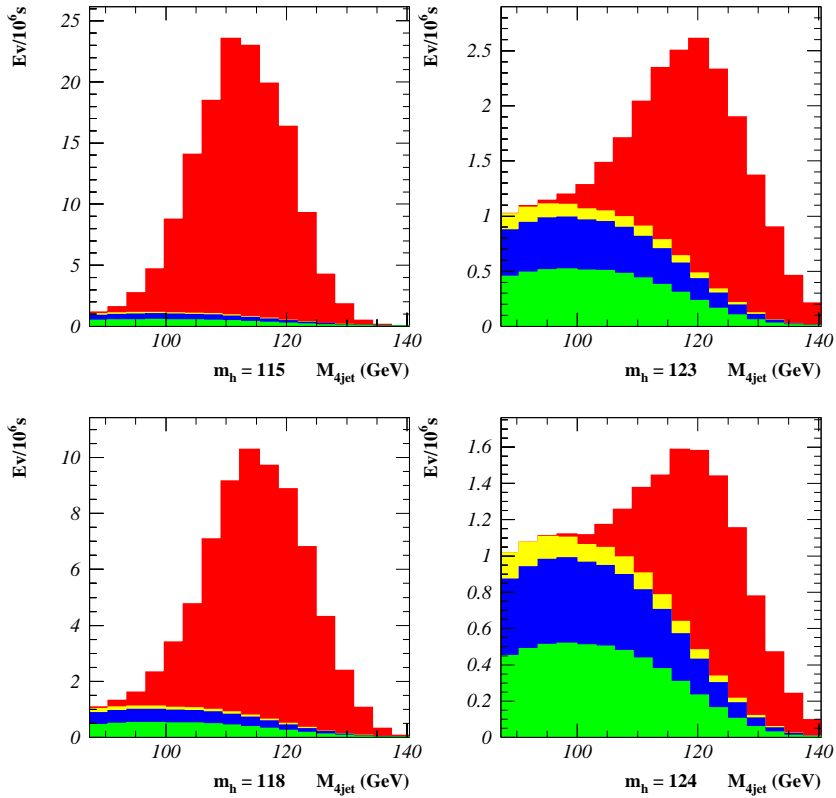}
\caption{\label{4jet}Four-jet invariant mass distributions after $10^6 s$ of
running, and after including $b$-tagging with
$\eps_b=0.5$, in scenarios (1), (3), (4) and (5).  Signal $\gamma\gamma
\rightarrow h \rightarrow aa \rightarrow b\bar{b}b\bar{b}$ is shown in red,
backgrounds: $\gamma\gamma \rightarrow b\bar{b}b\bar{b}$ in green,
$\gamma\gamma \rightarrow b\bar{b}c\bar{c}$ in blue and
$\gamma\gamma \rightarrow c\bar{c}c\bar{c}$ in yellow.}
\end{figure}

The expected four-jet invariant mass distributions of all surviving
signal and background events are depicted in Fig.~\ref{4jet} (the
background samples are the same in each case).  Absolute numbers are
again normalized to $10^6$ seconds.  We find a clear signal, visible
over the background after a short time of running, even in the most
disfavorable scenario (5).  It is also clear that the particular
choice of $E_{CM}=115$ GeV was not the most fortunate for this study
and that an upgrade of $E_{CM}$ to 120-125 GeV would very probably
 have the effect of
producing signals for cases (3) and (5) that look much more like those
shown here for cases (1) and (4).  If the beam energy happens to be
well tuned to $m_h$ (as was the case here for scenario (1)), a
prominent signal peak in the four-jet invariant mass is bound to
appear immediately.

For the two scenarios not considered here, (2) and (6), the prospects
for observing $\gamma\gamma \rightarrow
h \rightarrow aa \rightarrow \tau^+\tau^-\tau^+\tau^-$
are potentially just as good as found here for the $b\anti b b\anti b$
channel for scenarios (1) and (3 -- 5).  With the cross
section for direct 4$\tau$ production being of the order of 500 fb,
the final signal to background ratio should, in fact, be very similar
to the one obtained here in the 4$b$ channel, except for the slightly
different $h$ masses; the overall normalization will depend
on the $\tau$ reconstruction efficiency. 
The main difficulty will be the impossibility of
fully reconstructing the mass peak in the $\tauptaum\tauptaum$
channel. A full study is needed.

Returning to scenarios (1), (3), (4) and (5) given Table~\ref{nmssmtab}, 
we note that $BR(a \rightarrow \tau^+\tau^-)
\approx 0.1$ so that signal detection seems
plausible also in the channel $b\bar{b}\tau^+\tau^-$, provided $m_h$ is
not too far away from $E_{CM}$ (a slight upgrade of $E_{CM}$ should be
enough for this purpose).  Combination of data obtained from the two
channels will provide invaluable information for physics studies.
In particular,
the very important ratio $\br(a_1\to \tauptaum)/\br(a_1\to b\anti b)$
can be measured and the consistency with predictions for this ratio
in the case of the NMSSM (or more generally with expectations for
a pseudoscalar Higgs boson) can be checked.  In addition,
we can expect that other possible decay modes for the $h$ can
be strictly limited, allowing a determination of the absolute rate
for $\gam\gam \to h$ production and hence of $\Gamma(h\to \gam\gam)$.
The magnitude of this partial width is determined 
primarily by the $h$'s couplings
to $WW$, $t\anti t$ and $b\anti b$ (assuming a heavy sparticle spectrum
in the case of the NMSSM).  Thus, a measurement of $\Gamma(h\to \gam\gam)$
provides invaluable constraints on these couplings and would,
in particular, allow us to check for consistency with their being
approximately SM-like.

\subsubsection{Conclusions for the NMSSM}

We have demonstrated that a \gamc\ would provide an invaluable 
confirmation of the very amorphous and difficult to interpret
LHC signals for $h\to aa$ production that would be present
in the $aa\to b\anti b\tauptaum$ channel.  In particular, we have
demonstrated a clear signal in the important complementary channel
$aa\to b\anti b b\anti b$. We have also argued that the other two
final state channels of $aa\to b\anti b \tauptaum$ and $aa\to \tauptaum\tauptaum$ will also yield clear signals at a \gamc\ that will allow us
to confirm the Higgs boson nature of the signals.  In contrast,
the viability of a LHC signal in the $aa\to b\anti b b\anti b$
final state is quite uncertain. Finally, in cases
where $m_a$ is so small that $a\to b\anti b$ decays are not allowed,
we believe that the \gamc\ will be able to detect the 
$aa\to \tauptaum\tauptaum$
and possibly the $aa\to jj\tauptaum$ final states that have not yet
been shown to be detectable at the LHC.

\subsection{Higgs boson physics in the  Complex MSSM}

The MSSM with complex parameters (cMSSM) offers an interesting
extension of the ``real MSSM'' (rMSSM), especially concerning Higgs
boson physics. 
At the tree-level, $\cp$-violation in the Higgs boson sector is
absent. However, complex phases of the trilinear couplings, $\At$,
$\Ab$,~\ldots, of the gluino mass and of the gaugino masses, $\mgl$,
$M_2$ and $M_1$ and possibly also from the Higgs mixing parameter
$\mu$ can induce $\cp$-violation at the loop-level. As a consequence
all three neutral Higgs bosons, $h$, $H$ and $A$ are no longer $\cp$
eigenstates, but 
can mix with each other~\cite{mhiggsCPXgen}, 
\begin{equation}
\KL h, H, A \KR \to \KL \He, \Hz, \Hd \KR {\rm ~~~with~~} 
\mHe \le \mHz \le \mHd~.
\end{equation}
Higher order corrections to the Higgs boson masses and couplings have
been evaluated in several 
approaches~\cite{mhiggsCPXEP,mhiggsCPXRG1,mhiggsCPXRG2,mhiggsCPXFD1}. 

Due to loop corrections, the lightest Higgs boson can decouple from
the gauge bosons. In this case, at $e^+ e^-$~colliders, the main
production modes for Higgs bosons can be
\BEA
e^+ e^- &\to& Z \,\to\, Z\, \Hz \non \\
e^+ e^- &\to& Z \,\to\, \Hz\, \He ~.
\EEA
Higgs boson searches at LEP in the context of the
cMSSM~\cite{mhexpOPAL} have shown that in this case the decay 
$\Hz \to \He\He$ poses special experimental problems. 
This decay mode, which is often dominant in this case, results in a
six jet final state topology (with the Higgs bosons decaying to
$b$~quarks). Such a final state can be quite complicated to handle,
e.g.\ due to jet pairing problems. 

In a \gamc\ the situation is more favorable. Here the Higgs
bosons are produced singly. Thus the decay $\Hz \to \He\He$ results in
a four jet final state that is easy to handle. (In a similar fashion
the problem at an $e^+ e^-$~collider could be overcome if the Higgs
bosons are produced in the $WW$~fusion mode at high energies.)
The topology of the decay $\Hz \to \He\He$ resembles strongly the
topology that arises for $h \to AA$ in the rMSSM, the NMSSM
 or the THDM. Thus the
analysis methods can (nearly) directly be taken over from these
cases. 

\subsection{The Little Higgs boson at a Photon Collider}
%




The Standard Model (SM) of the strong and electroweak interactions has
passed stringent tests up to the highest energies accessible today.
The precision electroweak data \cite{Hagiwara:fs} point to the existence of
a light Higgs boson in the SM, with mass $m_H \lsim 200$ GeV.
The Standard Model with such a light Higgs boson can be viewed as an effective 
theory valid up to a much higher energy scale $\Lambda$, possibly all the way
up to the Planck scale. 
In particular, the precision electroweak data exclude the presence of
dimension-six operators arising from strongly coupled new physics 
below a scale $\Lambda$ of order 10 TeV \cite{Barbieri};
if new physics is to appear
below this scale, it must be weakly coupled.
However, without protection by a symmetry, the Higgs mass is quadratically
sensitive to the cutoff scale $\Lambda$ via quantum corrections, 
rendering the theory with $m_H \ll \Lambda$ rather unnatural.
For example, for $\Lambda = 10$ TeV, the ``bare'' Higgs mass-squared 
parameter must be tuned against the quadratically divergent radiative
corrections at the 1\% level.
This gap between the electroweak scale $m_H$ and the cutoff scale
$\Lambda$ is called the ``little hierarchy''.

Little Higgs models \cite{littleh,Littlest} revive an old idea to keep the
Higgs boson naturally light: they make the Higgs particle a
pseudo-Goldstone boson \cite{PNGBhiggs} of some broken global symmetry.
The new ingredient of little Higgs models is that they
are constructed in such a way that at least two
interactions are needed to explicitly break all of the global symmetry
that protects the Higgs mass.  This forbids quadratic divergences in the
Higgs mass at one-loop; the Higgs mass is then smaller than the cutoff 
scale $\Lambda$ by {\it two} loop factors, making the cutoff scale
$\Lambda \sim 10$ TeV natural and solving the little hierarchy problem.

From the bottom-up point of view, in little Higgs models
the most important quadratic divergences 
in the Higgs mass due to the top quark, gauge boson, and Higgs boson loops
are canceled by loops of new weakly-coupled 
fermions, gauge bosons, and scalars with masses around a TeV.  In 
contrast to supersymmetry, the cancellations in little Higgs models
occur between loops of particles with the {\it same} statistics.
Electroweak symmetry breaking is triggered by a
Coleman-Weinberg \cite{coleman-weinberg} potential, generated by
integrating out the heavy degrees of freedom, which also gives the Higgs
boson a mass at the electroweak scale.

The Littlest Higgs model \cite{Littlest} is a minimal model of this type.  
It consists of a nonlinear sigma model with a global
$SU(5)$ symmetry which is broken down to $SO(5)$ by a vacuum condensate 
$f \sim \Lambda/4\pi \sim$ TeV.
The gauged subgroup $[SU(2)\times U(1)]^2$ is broken at the
same time to its diagonal subgroup $SU(2)\times U(1)$, identified
as the SM electroweak gauge group. 
The breaking of the global symmetry leads to 14 Goldstone bosons, four of
which are eaten by the broken gauge generators, leaving 10 states that
transform under the SM gauge group as a doublet $h$ (which becomes
the SM Higgs doublet) and a triplet $\phi$.
A vector-like pair of colored Weyl fermions is also needed to cancel 
the divergence from the top quark loop.  The particle content and interactions
are laid out in detail in Ref.~\cite{LHpheno}.

In the Littlest Higgs model, the tree-level couplings of the Higgs 
boson to SM particles are modified from those of the SM Higgs boson 
by corrections of order $v^2/f^2$ 
(where $v \simeq 246$ GeV is the SM Higgs vacuum expectation value)
that arise due to the nonlinear sigma
model structure of the Higgs sector and, for the couplings to gauge bosons
and the top quark, due to mixing between the SM particles and the new
heavy states.
The experimental sensitivity to the Higgs couplings to $W$ and $Z$
bosons in this model has been considered in Ref.~\cite{Debajyoti}.
Here we focus on the loop-induced Higgs coupling to photon pairs,
which receives corrections from the modifications of the Higgs couplings
to the SM top quark and $W$ boson, as well as from the new heavy 
particles running in the loop \cite{LHloop}.  
Like the tree-level couplings, the loop-induced Higgs couplings receive
corrections of order $v^2/f^2$.

Any charged particle that couples to the Higgs boson will contribute
to $H \to \gamma\gamma$.
In the Littlest Higgs model, those states include the heavy charged $SU(2)$
gauge boson $W^{\pm}_H$, the vector-like top quark partner $T$, and the
charged scalars $\Phi^{\pm}$, $\Phi^{\pm\pm}$.
Besides the condensate $f$ as the most important scale parameter,
the mass and couplings of each new state depend upon additional dimensionless
parameters.  
The mixing between the two gauge groups
$SU(2)_1$ and $SU(2)_2$, with couplings $g_1$ and $g_2$ respectively,
is parameterized by $c$.
The mixing between the top quark and
the heavy vector-like quark $T$ is parameterized by $c_t$.
In the Higgs sector, the ratio of the triplet and
doublet vacuum expectation values $(v^{\prime}/v)$ is parameterized by $x$. 
More explicitly, we have (for additional details, see 
Refs.~\cite{LHpheno,LHloop}):
\begin{equation}
      0<  c  = \frac{g_1}{\sqrt{g_1^2+g_2^2}}<1, \qquad
      0<c_t = \frac{\lambda_1}{\sqrt{\lambda_1^2 + \lambda_2^2}}<1, \qquad
       0\le x = \frac{4fv^\prime}{v^2}<1.
\end{equation}
We also define $s \equiv \sqrt{1-c^2}$ and $s_t \equiv \sqrt{1 - c_t^2}$.
The electroweak data prefers a small value for $c$ \cite{ewprecision}, 
while the positivity of the heavy scalar mass requires $x < 1$.

The partial width of the Higgs boson into two photons is 
given in the Littlest Higgs model by \cite{hhg,LHloop}
\begin{equation}
	\Gamma(H \to \gamma\gamma) = 
	\frac{\sqrt{2} G_F \alpha^2 m_H^3 y^2_{G_F}}{256 \pi^3}
	\left| \sum_i y_i N_{ci} Q_i^2 F_i \right|^2,
\end{equation}
where $N_{ci}$ and $Q_i$ are the color factor and electric charge, 
respectively,
for each particle $i$ running in the loop.  The standard
dimensionless loop factors
$F_i$ for particles of spin 1, 1/2, and 0 are given in Ref.~\cite{hhg}.
The factor $y^2_{G_F}$ contains the order $v^2/f^2$ correction to the 
relation between the Higgs vacuum expectation value $v \simeq 246$ GeV 
and the Fermi constant $G_F$ in the Littlest Higgs model:
$v^{-2} = \sqrt{2} G_F y^2_{G_F}$, where \cite{LHloop}
\begin{equation}
	y^2_{G_F} = 1 + \frac{v^2}{f^2} 
	\left[ -\frac{5}{12} + \frac{1}{4} x^2 \right].
\end{equation}
The remaining factors $y_i$ incorporate the couplings and mass suppression
factors of the particles running in the loop.  For the top quark and 
$W$ boson, whose couplings to the Higgs boson are proportional to their masses,
the $y_i$ factors are equal to one up to a correction of order 
$v^2/f^2$ \cite{LHloop}:
\begin{eqnarray}
	y_t &=& 1 + \frac{v^2}{f^2} \left[ -\frac{2}{3} + \frac{1}{2} x
	- \frac{1}{4} x^2 + c_t^2 (1 + c_t^2) \right], \nonumber \\
	y_W &=& 1 + \frac{v^2}{f^2} \left[ -\frac{1}{6} 
	+ \frac{3}{4} (c^2-s^2)^2 - x^2 \right].
\end{eqnarray}
For the heavy particles in the loop, on the other hand, the $y_i$ 
factors are of order $v^2/f^2$.  This reflects the fact that the masses
of the heavy particles are not generated by their couplings to the 
Higgs boson; rather, they are generated by the $f$ condensate.
This behavior naturally respects the 
decoupling limit for physics at the scale $f \gg v$.
The couplings are \cite{LHloop}:
\begin{eqnarray}
	y_T &=& - c_t^2 (1 + c_t^2) \frac{v^2}{f^2}, \nonumber \\
	y_{W_H} &=& -s^2 c^2 \frac{v^2}{f^2}, \nonumber \\
	y_{\Phi^+} &=& \frac{v^2}{f^2} 
	\left[ -\frac{1}{3} + \frac{1}{4} x^2 \right].
\end{eqnarray}
The $\Phi^{++}\Phi^{--}H$ coupling is zero at leading order, so the
corresponding $y_{\Phi^{++}}$ is suppressed by an extra factor of 
$v^2/f^2$ and we thus ignore it.

\begin{figure}[h]
    \resizebox{\textwidth}{!}{
        \rotatebox{270}{\includegraphics{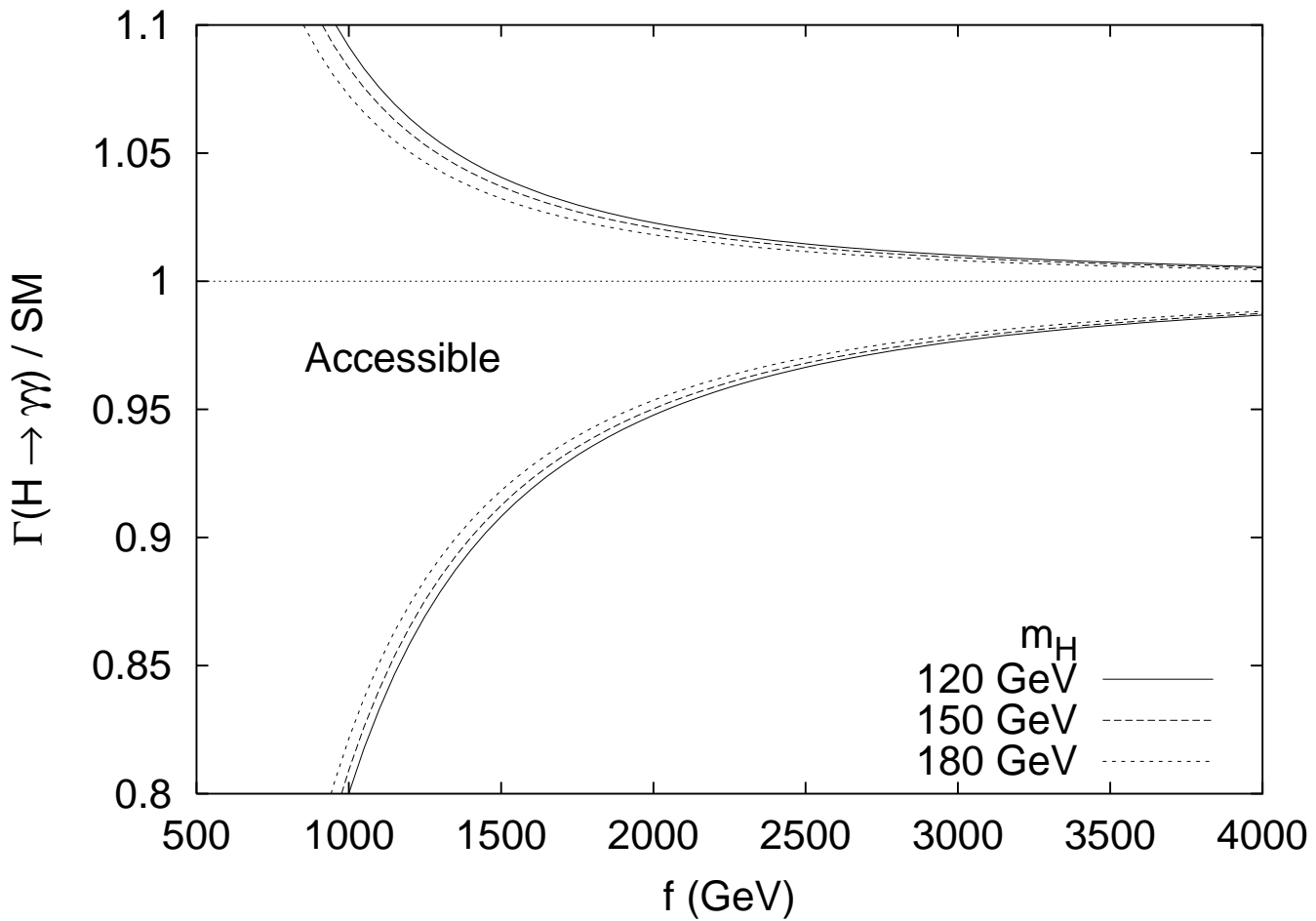}}}
    \caption{Range of values of
        $\Gamma(H \to \gamma\gamma)$ accessible in the Littlest
	Higgs model as a
        function of $f$, normalized to the SM value, for $m_H = 120$,
        150 and 180 GeV.  From Ref.~\cite{LHloop}.}
    \label{fig:range}
\end{figure}

For a fixed value of the scale $f$, only three parameters of the 
Littlest Higgs model affect the loop-induced partial width of 
$H \to \gamma\gamma$: $c$, $c_t$, and $x$.  Varying these parameters
within their allowed ranges, we obtain the possible range of 
$\Gamma(H \to \gamma\gamma)$, which is shown normalized to its SM value
as a function of $f$ in Fig.~\ref{fig:range}.
The effect can be quite significant: for instance,
for $f=1$ TeV the deviation from the SM prediction ranges between 
$+10\%$ and $-20\%$.

A \gamc\ can produce the Higgs resonance in the $s$-channel
with a cross section proportional to $\Gamma(H \to \gamma\gamma)$.
For a light Higgs boson with mass around $115-120$ GeV,
the most precisely measured process will be
$\gamma\gamma \to H \to b \bar b$, with an uncertainty of about
2\% on the rate \cite{cliche,gammagamma}.
(The uncertainty rises with increasing Higgs mass to 
about 10\% for $m_H = 160$ GeV.)
This can be combined with the $\sim 1.5-2$\% measurement of the branching ratio
of $H \to b \bar b$ from the $e^+e^-$
collider \cite{LCbook,TESLATDR,ACFA} to extract
$\Gamma(H \to \gamma\gamma)$ with a precision of about 3\%.
Such a measurement could probe $f < 2700$ GeV at the
$1 \sigma$ level, or $f < 1800$ GeV at the $2 \sigma$ level.
A $5 \sigma$ deviation is possible for $f < 1200$ GeV.
For comparison, the electroweak precision constraints \cite{ewprecision}
on the Littlest Higgs model require $f \gsim 1$ TeV, while naturalness
considerations prefer as low a value of $f$ as possible.

In the absence of an $e^+e^-$ collider, a similar analysis could be done
combining LHC and $\gamma$C data on Higgs boson production and decay rates
to look for deviations of the various Higgs couplings from their SM values
that would indicate the little Higgs nature of the Higgs boson.  Such an
analysis has not yet been done.

\subsection{Higgs-Radion Mixing}


In this section, we demonstrate the important complementarity of
a \gamc\ for probing the Higgs-radion sector of the Randall-Sundrum
(RS) model \cite{Randall:1999ee}.

\subsubsection{Introduction}

In the RS model,
there are two branes, separated in the 5th dimension ($y$) and
$y\to -y$ symmetry is imposed.  With appropriate boundary conditions,
the metric
\begin{equation}
ds^2=e^{-2\sigma(y)}\eta_{\mu\nu}dx^\mu dx^\nu-b_0^2dy^2, 
\label{metricz}
\end{equation}
where $\sigma(y)\sim m_0b_0|y|$, is an exact solution of Einsteins equations.
The factor $e^{-2\sigma(y)}$ is called the warp factor; 
scales at $y=0$ of order $\mpl$
on the hidden brane are reduced to scales at $y=1/2$ of order TeV on the
visible brane, thereby solving the hierarchy problem.

Fluctuations of $g_{\mu\nu}$ relative to the flat space metric $\eta_{\mu\nu}$ 
are the KK excitations $h^n_{\mu\nu}$.
Fluctuations of $b(x)$ relative to $b_0$ define, after
rescaling to obtain correct kinetic energy normalization, 
the radion field, $\phio$.
In addition, we place a Higgs doublet $\Hhat$ on the visible brane.
After various rescalings, the properly normalized quantum fluctuation
field is called $h_0$.

It is extremely natural (and certainly not forbidden) for
there to be Higgs-radion mixing via a term in the action of the form
\begin{equation}
S_\xi=\xi \int d^4 x \sqrt{g_{\rm vis}}R(g_{\rm vis})\Hhat^\dagger \Hhat\,,
\end{equation}
where $R(g_{\rm vis})$ is the Ricci scalar 
for the metric induced on the visible brane. The $\xi\neq 0$ kind
of phenomenology has been studied in \cite{wellsmix,csakimix,Han:2001xs,Chaichian:2001rq,Hewett:2002nk,Csaki:1999mp,Dominici:2002jv,Battaglia:2003gb}.
The ensuing discussion in this introductory
section is a summary of results found in \cite{Dominici:2002jv}.

A crucial parameter is the ratio
$\gamma\equiv \vo/\lphi\,$,
where $\lphi$ is vacuum expectation value of the radion field.
After writing out the full quadratic structure of the Lagrangian,
including $\xi\neq 0$ mixing, we obtain a form in which
the $\ho$ and $\phio$ fields for $\xi=0$ are mixed and have complicated
kinetic energy normalization.
We must diagonalize the kinetic energy and rescale to get canonical
normalization.
\bea
\ho&=&\left (\cos\theta -{6\xi\gam\over Z}\sin\theta\right)\h
+\left(\sin\theta+{6\xi\gam\over Z}\cos\theta\right)\phi
\equiv d\h+c\phi
\label{hform}
\\
\phio&=&-\cos\theta {\phi\over Z}+\sin\theta {\h\over Z}
\equiv a\phi+b\h\,. \label{phiform}
\eea
In the above equations
\begin{equation}
Z^2\equiv 1+6\xi\gam^2(1-6\xi)\,,
\label{z2}
\end{equation}
and the procedure for computing $\theta$ is given in \cite{Dominici:2002jv}.
To avoid a tachyonic situation, $Z^2>0$ is required.
This leads to a constraint on the maximum negative and positive $\xi$ values:
\begin{equation}
\frac{1}{12}\left(1-\sqrt{1+\frac{4}{\gamma^2}}\right)
\leq \xi \leq
\frac{1}{12}\left(1+\sqrt{1+\frac{4}{\gamma^2}}\right)\,.
\label{xilim}
\end{equation}
The process of inversion is very critical to the phenomenology
and somewhat delicate and leads to further constraints on the $\xi$ range.
The result found is that the physical mass eigenstates $h$ and $\phi$ 
cannot be too close to being degenerate
in mass, depending on the precise values of $\xi$ and $\gam$;
extreme degeneracy is allowed only for small $\xi$ and/or $\gam$.
For fixed $\mh$ and $\lphi$, the resulting theoretically allowed
region in $(\mphi,\xi)$ space has an hourglass shape that
can be seen in Fig.~\ref{fig:compl120}.
In the theoretically allowed region of parameter space,  for given choices of 
$\xi$, $\gam$, $\mh$ and $\mphi$ we compute $Z^2$
and perform the inversion to obtain $\mho^2$ and $\mphio^2$, 
and $\theta$ from which we compute
$a,b,c,d$ in Eqs.~(\ref{hform}) and (\ref{phiform}). 

In all it takes four 
independent parameters to completely fix the mass diagonalization
of the scalar sector when $\xi\neq 0$.
These are:
\begin{equation}
\xi\,,\quad \gam\,,\quad \mh\,,\quad \mphi\,,
\end{equation}
where we recall that $\gam\equiv \vo/\lphi$ with $\vo=246\gev$.
The quantity $\lwh={1\over \sqrt 3}\lphi$ fixes
the KK-graviton couplings to the $h$ and $\phi$
and 
\begin{equation}
m_1=x_1 {m_0\over\mpl} {\lphi\over\sqrt 6}
\label{m1form}
\end{equation}
is the mass of the first KK graviton excitation, where $x_1$ is the first
zero of the Bessel function $J_1$ ($x_1\sim 3.8$). Here,
$m_0/\mpl$ is related to the curvature of the brane
and should be a relatively small number for consistency
of the RS scenario.
Sample parameters that are safe from precision EW data
and RunI Tevatron constraints are
$\lphi=5\tev$ (corresponding to $\lwh\sim 3\tev$)
and $m_0/\mpl=0.1$. We will focus on these parameter choices for the
studies discussed here.
The latter implies $m_1\sim 780\gev$; i.e. $m_1$ is typically too large
for KK graviton excitations 
to be present, or if present, important, in $h,\phi$ decays.
But, KK excitations in this mass range (and much higher)
will be observed and well measured at the LHC.
This will provide important information. In particular,
the mass gives $m_1$ in the above notation, while 
the excitation spectrum as a function of $m_{jj}$ 
determines $m_0/\mpl$. These can be combined
ala Eq.~(\ref{m1form}) to get $\lphi$.
Knowledge of $\lphi$ will really help in any study of the Higgs sector.

The $\ho$ has SM-like couplings to $VV$ and $f\anti f$.
The $\phio$ has $VV$ and $f\anti f$ couplings deriving
from the interaction $-{\phi_0\over \lphi}T_\mu^\mu$.
The result, after introducing the Higgs-radion mixing, is that
\begin{equation}
\gfvh\equiv{g_{VVh}\over g_{VV\hsm}}={g_{f\bar{f}h}\over g_{f\bar f \hsm}}=d+\gamma b,,\quad 
\gfvphi\equiv {g_{VV\phi}\over g_{VV\hsm}}={g_{f\bar{f}\phi}\over g_{f\bar f \hsm}}=c+\gamma a\,.
\label{coups}
\end{equation}
An important point to note is that the factors for $WW$ and $f\bar f$ couplings are the same.
The $gg$ and $\gam\gam$ couplings to the $h$ and $\phi$ are not 
related to those of the $\hsm$ simply by the same universal rescaling factors.
In addition to the standard one-loop contributions, that {\it are}
simply rescaled by $\gfvh$ or $\gfvphi$, there are anomalous contributions.
These are expressed in terms of 
the SU(3)$\times$SU(2)$\times$U(1) $\beta$ function coefficients  $b_3=7$,
$b_2=19/6$ and $b_Y=-41/6$ and enter only through
their radion admixtures, $g_h=\gamb$ for the $h$, and $g_\phi=\gama$
for the $\phi$.
Finally, there are cubic $\h\to \phi\phi$ and $\phi\to \h\h$ couplings,
the former being particularly interesting since the coupling is only
present for $\xi\neq 0$. These couplings depend somewhat on
the $\phio^3$ coupling associated with the radion stabilization mechanism.
Graphs presented below assume small direct $\phio^3$ coupling from
this source.

\begin{figure}[h]
\includegraphics[width=2.5in,angle=90]{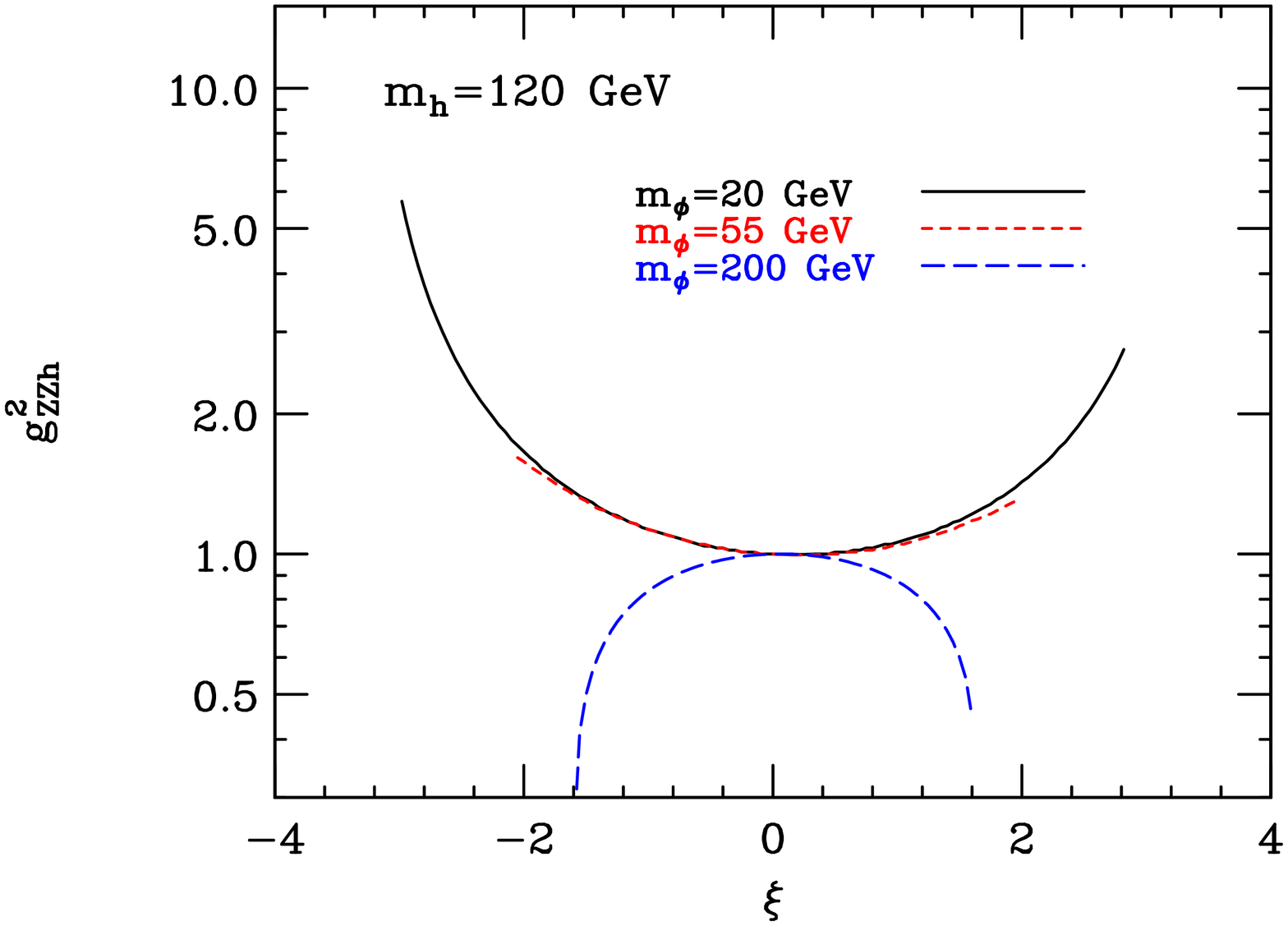}
\includegraphics[width=2.5in,angle=90]{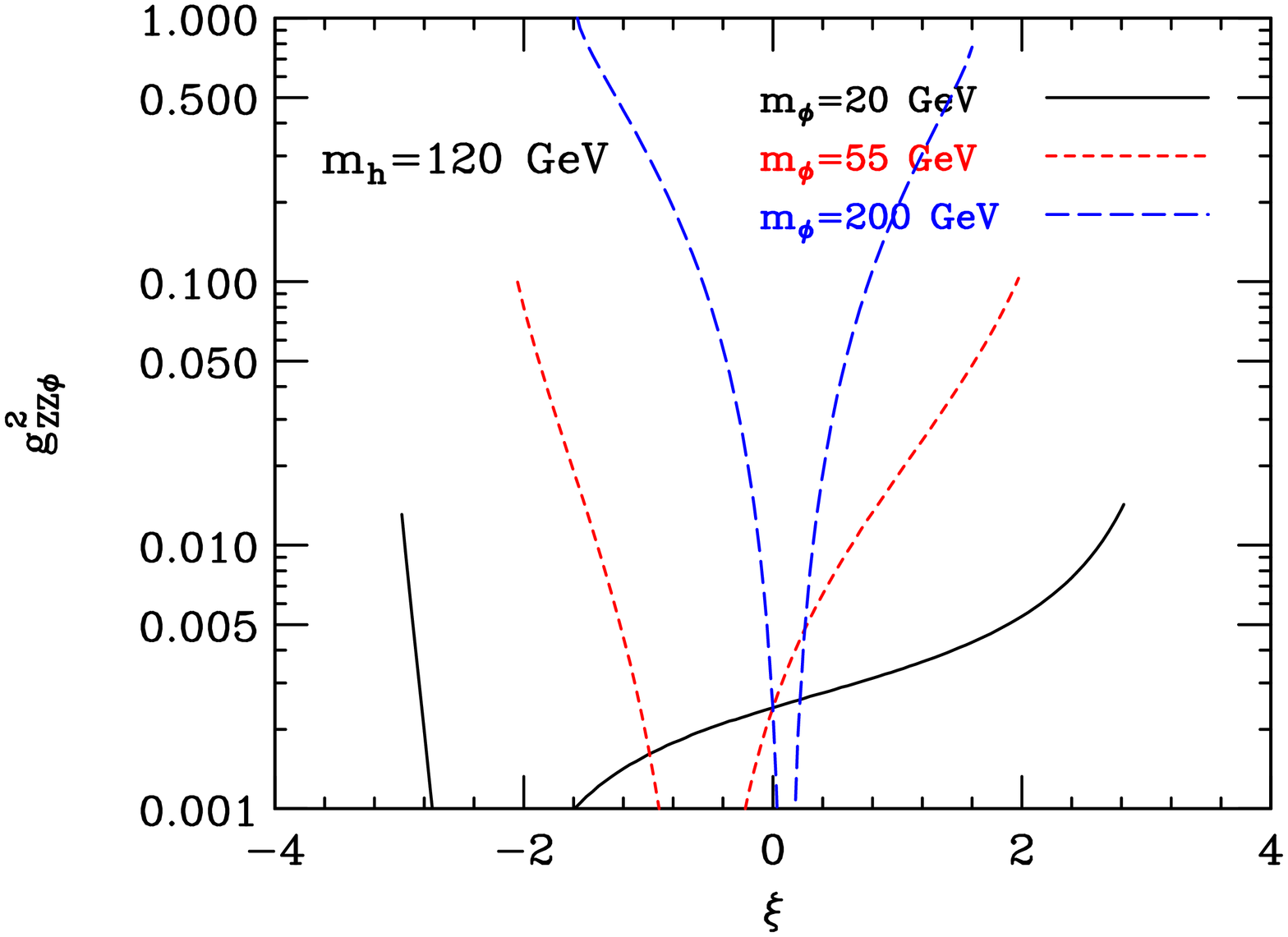}
\caption{\sl In the left figure, we plot
$g_{ZZ\h}^2/g_{ZZ\hsm}^2 = g_{f\anti f \h}^2 / g_{f\anti f\hsm}^2$ as a function of $\xi$ for
several $\mphi$ values. The right figure shows
$g_{ZZ\phi}^2/g_{ZZ\hsm}^2 = g_{f\anti f \phi}^2/ g_{f\anti f\hsm}^2$. 
\label{gzzsqmh120}
}
\end{figure}

The behavior of $\gfvh^2$ and $\gfvphi^2$ is illustrated in Fig.~\ref{gzzsqmh120}.
Some important general features of these couplings are the following.
First, if $\gfvh^2<1$ ($\gfvh^2>1$) is observed then $\mphi>\mh$
($\mphi<\mh$), respectively, except for a small region near $\xi=0$.
For $\mphi>\mh$, the suppression is most severe for large $|\xi|$ where 
$\gfvh^2\sim 0.1\div 0.2$ is possible.   
Second, for any given $\mh$, $\gfvphi$ can be quite small and even has
zeroes as one scans over the allowed $\xi$ range at a given $\mphi$.
However, at large $|\xi|$, if $\mphi>\mh$ then $\gfvphi$ can be a
substantial fraction of the SM strength, $\gfvphi=1$, implying SM type
discovery modes could become relevant.  The most important result of
this is that $gg\to \phi \to ZZ\to 4\ell$ can be a viable discovery
mode when $\mphi\gsim 2\mz$ and $|\xi|$ is near the maximum allowed.

The universal scaling of the $f\anti f$ and $VV$ couplings of the $h$
and $\phi$ relative to the SM $\hsm$ implies that, unless these
couplings are very suppressed, the $h$ and $\phi$ will have branching
ratios that are quite SM-like.  The only exceptions are the following.
(1) Decays of the type $h\to \phi\phi$ can be significant at larger
$\xi$ when $\mphi$ is modestly below the $\mh/2$ threshold ---
$\br(h\to\phi\phi)_{\rm max}<0.2$ is typical for $\lphi>5\tev$
and large $|\xi|$.  (2) Where
$\gfvphi$ is near a zero, $\phi\to gg$ decays, induced by the
anomalous coupling, will dominate.  (3) For $\mphi>2\mh$, $\br(\phi\to
hh)\sim 0.3\div0.5$ is typical. Of course, the total width of the $h$
or $\phi$ will be scaled down relative to an $\hsm$ of the same mass when
$\gfvh^2<1$ or $\gfvphi^2<1$ (respectively).

\subsubsection{Numerical Results: LHC and \gamc}

In order to illustrate the complementarity of the LHC and \gamc\
we will focus on the case $\mh=120\gev$, $\lphi=5\tev$
and $m_0/\mpl=0.1$. Once these parameters are fixed,
the remaining free parameters are $\mphi$ and $\xi$.
of the results found there.

\begin{figure}
\hspace*{.3in} 
\includegraphics[height=5.5in,width=3in,angle=90]{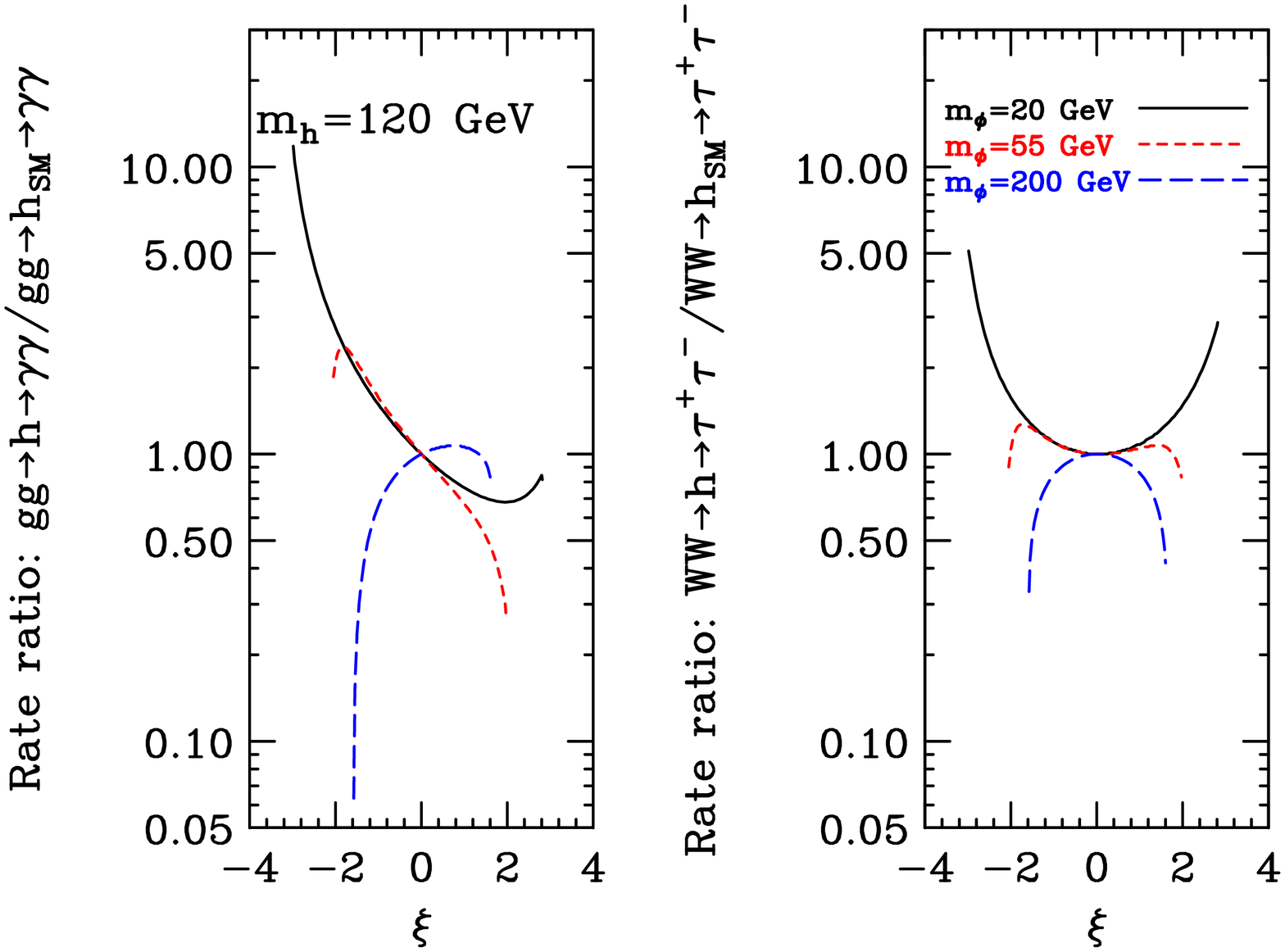}
\caption{\sl \small The ratios $gg\to h \to \gam\gam/gg\to\hsm\to\gam\gam$
and $WW\to h \to \tau^+\tau^-/WW\to \hsm \to \tau^+\tau^-$ (the latter being
the same as for
$gg\to t\anti t h\to t\anti t b\anti b$) for $\mhsm=\mh=120\gev$ and
$\lphi=5\tev$.}
\label{prodh_mh120}
\end{figure}

\begin{figure}[h!]
\begin{center}
\includegraphics[height=5.5in,width=3in,angle=90]{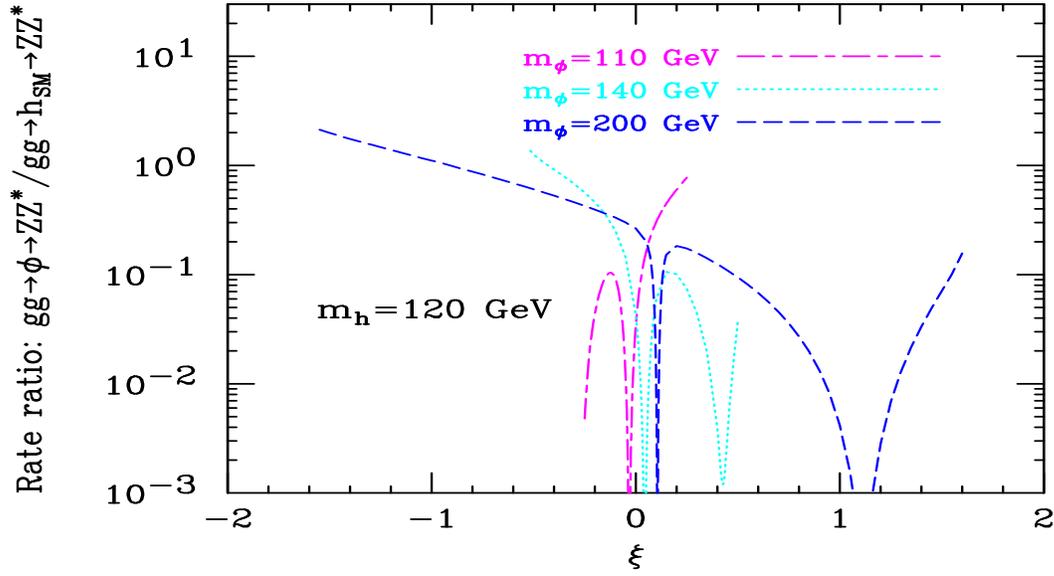}
\end{center}
\caption{\sl \small The ratio of the rate for $gg\to \phi\to ZZ$
to the corresponding rate for a SM Higgs boson with mass $\mphi$
assuming $\mh=120\gev$ and $\lphi=5\tev$ as a function of $\xi$
for $\mphi=110$, $140$ and $200\gev$. Recall that the $\xi$ range
is increasingly restricted as $\mphi$ becomes more degenerate
with $\mh$. {Note:} for $\mphi>\mh$ the mode approaches SM strength
if $\xi<0$ and is nearing SM strength if $\xi>0$ and near maximal.
\label{prodphiggtozz_mh120}
}
\end{figure}

To illustrate the impact of the couplings on the standard LHC discovery
modes, we give two plots. In Fig.~\ref{prodh_mh120} we
see that the $\h\to \gam\gam$
mode can be greatly suppressed if $\mphi>\mh$, especially if $\xi<0$.  
Also, if $\mphi<\mh$, $\h\to\gam\gam$
will be strong if $\xi<0$, but can be considerably weakened if $\xi>0$.
From Fig.~\ref{prodphiggtozz_mh120}, we observe that the discovery
mode $gg\to\phi\to ZZ\to 4\ell$ can approach SM level viability
at the largest allowed $|\xi|$ values, especially for $\xi<0$.

\begin{figure}[h]
\includegraphics[height=3in]{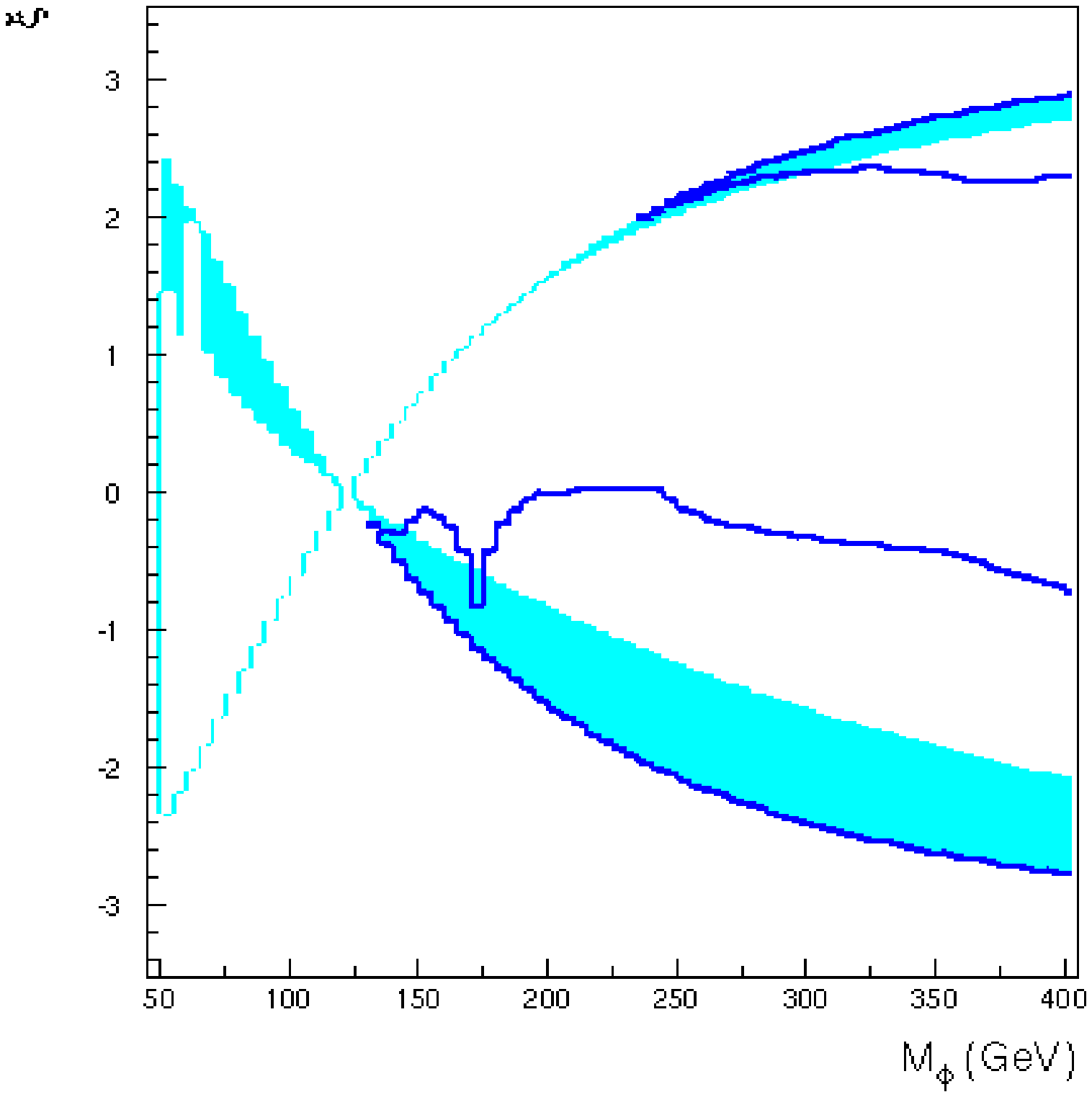}
\includegraphics[height=3in]{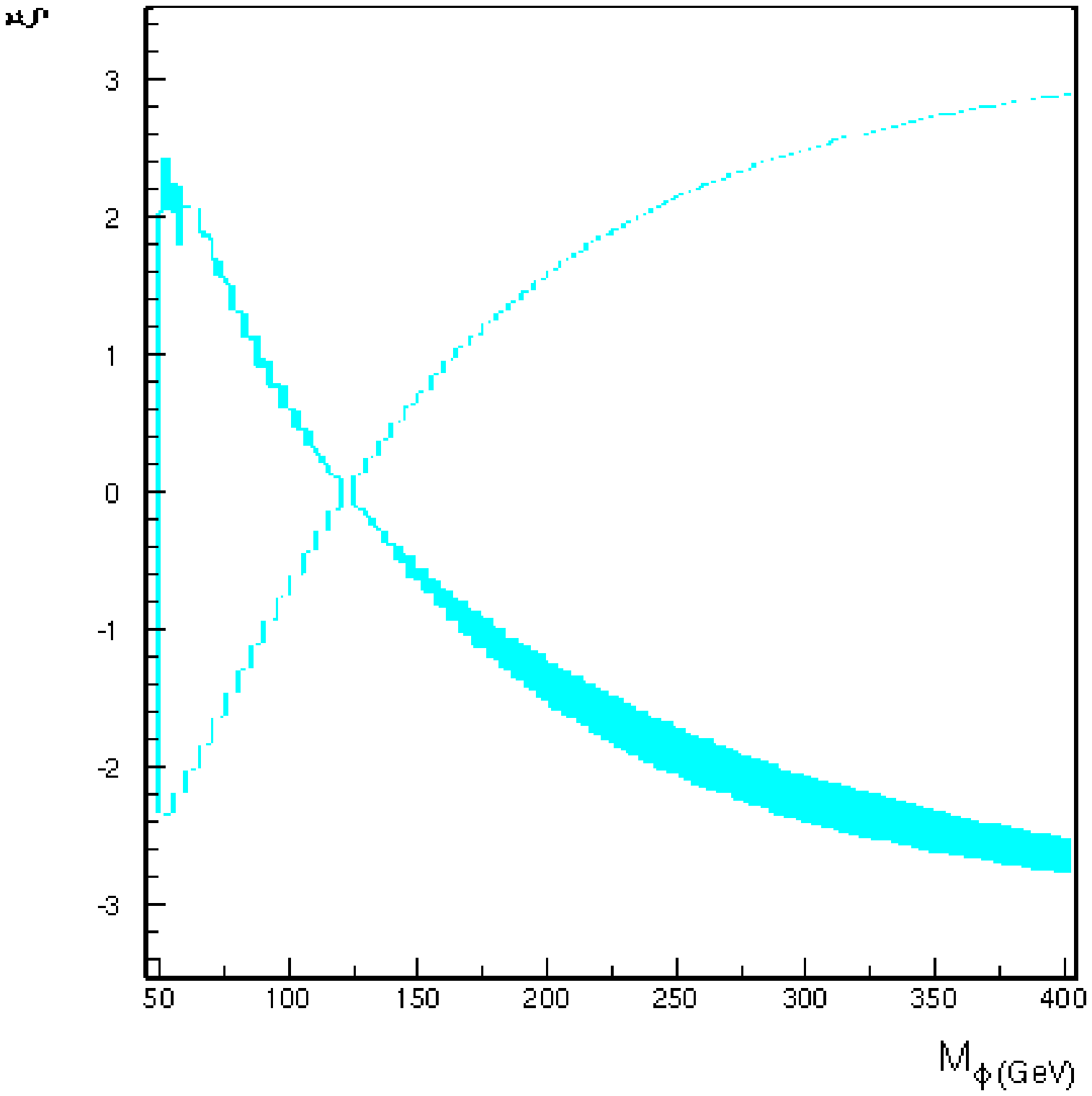}
\caption{\sl \small Illustration of mode complementarity at the LHC for $\mh=120\gev$. The outer hourglass shape defines the theoretically
allowed region in the $(\mphi,\xi)$ parameter space.
The cyan regions show where neither the $gg\to\h\to\gam\gam$ mode
nor the (not very important at this $\mh$ value)
$gg\to\h\to 4\ell$ mode yields a $>5\sig$ signal. LHC discovery
of the $\h$ with $>5\sigma$ is possible throughout the white region
within the hourglass-shaped boundary.  
The graphs are for $\lphi=5\tev$ and $L=30\fbi$ (left) and $L=100\fbi$ (right).
In the left-hand, $L=30\fbi$ plot, 
the regions between the dark blue curves are those 
where $gg\to\phi\to 4\ell$ is $>5\sig$. These regions expand
somewhat as $L$ increases. These figures are taken 
from \cite{Battaglia:2003gb}. 
\label{fig:compl120}
}
\end{figure}

A study of the resulting LHC prospects for $h$ and $\phi$ detection was
performed in \cite{Battaglia:2003gb}. Based on the preceding
discussion,  it should be possible to understand the features
of Fig.~\ref{fig:compl120} where we show the regions of
$(\mphi,\xi)$ parameter space in which $h$ and/or $\phi$ discovery
using LHC modes, designed for SM Higgs boson discovery, will be possible.
The results are obtained by simply rescaling the signal rates for
a SM Higgs boson of the same mass as the $h$ or $\phi$ using the
type of rescaling factors plotted in Figs.~\ref{prodh_mh120} and \ref{prodphiggtozz_mh120}.

Fig.~\ref{fig:compl120} exhibits regions
of $(\mphi,\xi)$ parameter space in which {\it both} the $h$ and $\phi$
mass eigenstates will be detectable.
In these regions, the LHC will observe two scalar bosons
somewhat separated in mass, with the lighter (heavier) having a non-SM-like 
rate for the $gg$-induced $\gamma\gamma$ ($Z^0Z^0$) final state.
Additional information will be required to ascertain whether these two Higgs
bosons derive from a multi-doublet or other type of extended
Higgs sector or from the present type of model with Higgs-radion mixing. 

If a linear collider is constructed, there is no region of $(\mphi,\xi)$
parameter space for which the $h$ (assuming modest $\mh$) will not be detected.
This is because the $ZZ\h$ coupling strength is always such that
$\gfvh^2>0.01$ (a value easily probed at the LC in the $\epem\to ZX$
missing mass discovery mode). Depending on $\mphi$, the $\phi$
will also be detectable in the $ZX$ final state if $\gfvphi^2>0.01$.
In particular, LC observation of the $\phi$ should be possible
in the region of low $\mphi$, large $\xi>0$  within which detection of either
at the LHC might be difficult. This is because the $ZZ\phi$
coupling-squared is $\gsim 0.01$ relative to the SM for most of
this region.

However, let us imagine that the LC has not yet been built but
that a photon-collider add-on to the CLIC test machine has been constructed.
Let's remind ourselves about the results for the 
SM Higgs boson obtained in the CLIC study of~\cite{cliche}.
There, a SM Higgs boson with $\mhsm=115\gev$ was examined.
After the cuts, one obtains about $S=3280$ and $B=1660$
in the $\gam\gam\to \hsm\to b\anti b$ channel, corresponding to
$S/\sqrt B \sim 80$!

\begin{figure}[h!]
\begin{center}
\includegraphics[height=5.5in,width=3.5in,angle=90]{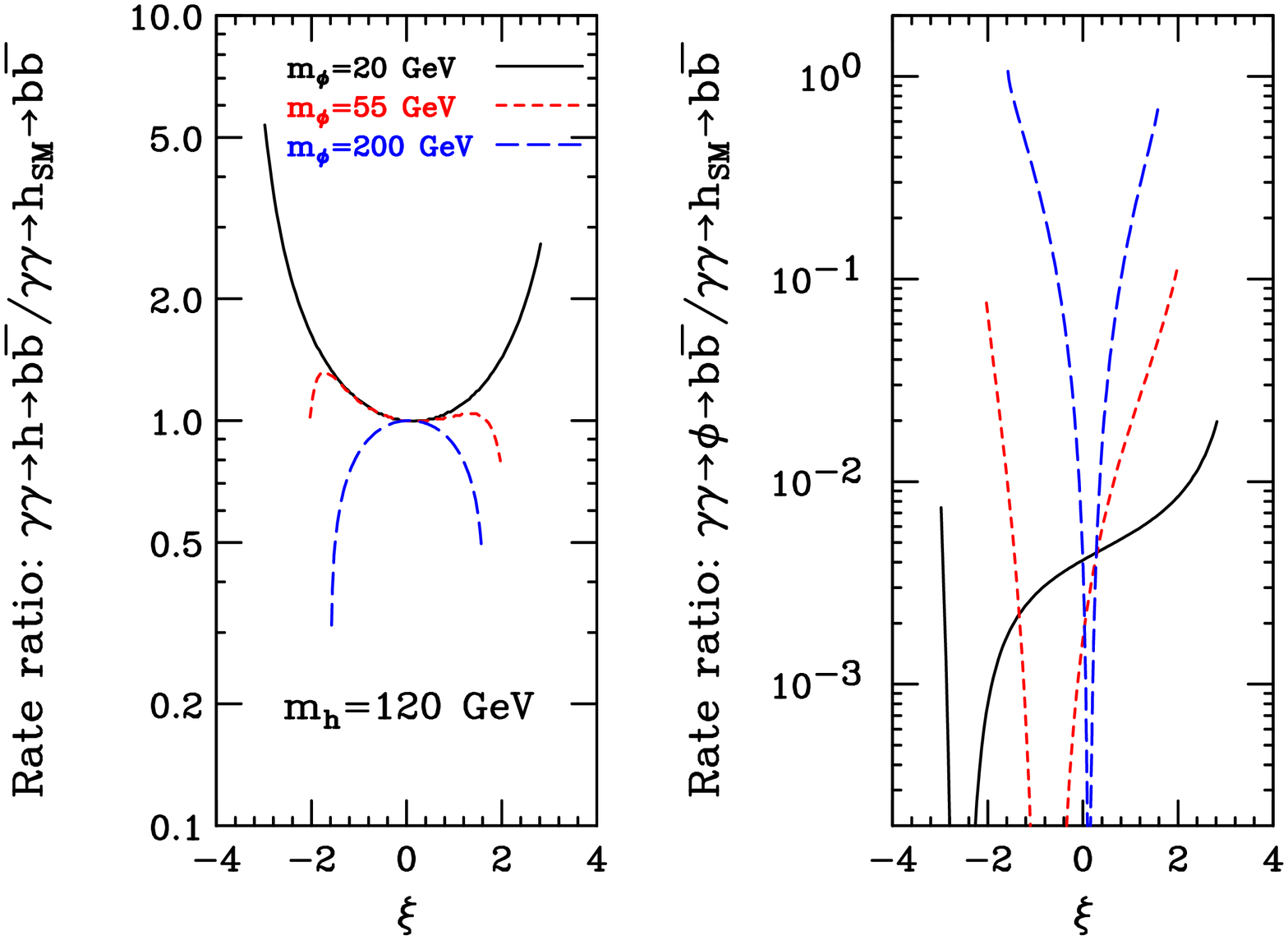}
\end{center}
\caption{The rates for $\gam\gam\to h \to b\anti b$
and $\gam\gam\to\phi\to b\anti b$ relative
to the corresponding rate for a SM Higgs boson of the same mass.
Results are shown
for $m_h=120\gev$  and $\lphi=5\tev$ as functions of $\xi$
for $\mphi=20$, $55$ and $200\gev$.
\label{gagatobb_mh120}
}
\end{figure}

We will assume that these numbers do not change significantly
for a Higgs mass of $120\gev$.
After mixing, the $S$ rate for the $h$ 
will be rescaled relative to that for the $\hsm$.  Of course, $B$
will not change.
The rescaling is shown in Fig.~\ref{gagatobb_mh120}.
The $S$ for the $\phi$ can also be obtained by rescaling
if $\mphi\sim 115\gev$.
For $\mphi<120\gev$, the $\phi\to b\anti b$ channel will
continue to be the most relevant for $\phi$ discovery,
but studies have not yet been performed to obtain the $S$
and $B$ rates for low masses.

Observe that for $\mphi<\mh$ we have either little change or enhancement, whereas
significant suppression of the $gg\to \h \to \gam\gam$ rate
was possible in this case for positive $\xi$.
  Also note that for $\mphi>\mh$ and large $\xi<0$
(where the LHC signal for the $h$ is marginal) there is much less suppression 
of $\gam\gam\to\h\to b\anti b$ than for $gg\to \h\to\gam\gam$ ---
at most a factor of 2 vs a factor of 8 (at $\mphi=200\gev$).
This is no problem for the \gamc\ since $S/\sqrt B\sim \half 80 \sim 40$
is still a very strong signal.
{In fact, we can afford a reduction by a factor
of $16$ before we hit the $5\sig$ level!}  
Thus, {\it the $\gam\gam$
collider will allow $h$ discovery (for $\mh=120$) 
throughout the entire hourglass},
which is something the LHC cannot absolutely do.

Using the factor of $16$ mentioned above
and referring to the rescaling factors plotted in 
Fig.~\ref{prodphiggtozz_mh120},
it is apparent that the $\phi$ with $\mphi<120\gev$ is very likely to
elude discovery at the $\gam\gam$ collider. (Recall that
it also eludes discovery at the LHC for this region.)
The only exceptions to this statement occur at the very largest $|\xi|$
values for $\mphi\geq 55\gev$ where $S_\phi>S_{\hsm}/16$.

Of course, we need to have signal and background results
after cuts for these lower masses to know if the factor of 16
is actually the correct factor to use.
To get the best signal to background ratio we would want to lower
the machine energy (as would be relatively easy at a CLIC
test facility) and readjust cuts and so forth.  
This study should be done.
For the $\mphi>\mh$ region, we will need results for
the $WW$ and $ZZ$ modes that are being worked on.

\subsubsection{Conclusions and discussion for Higgs-Radion Mixing}

\begin{figure}[h!]
\includegraphics[height=5.5in,width=3in,angle=90]{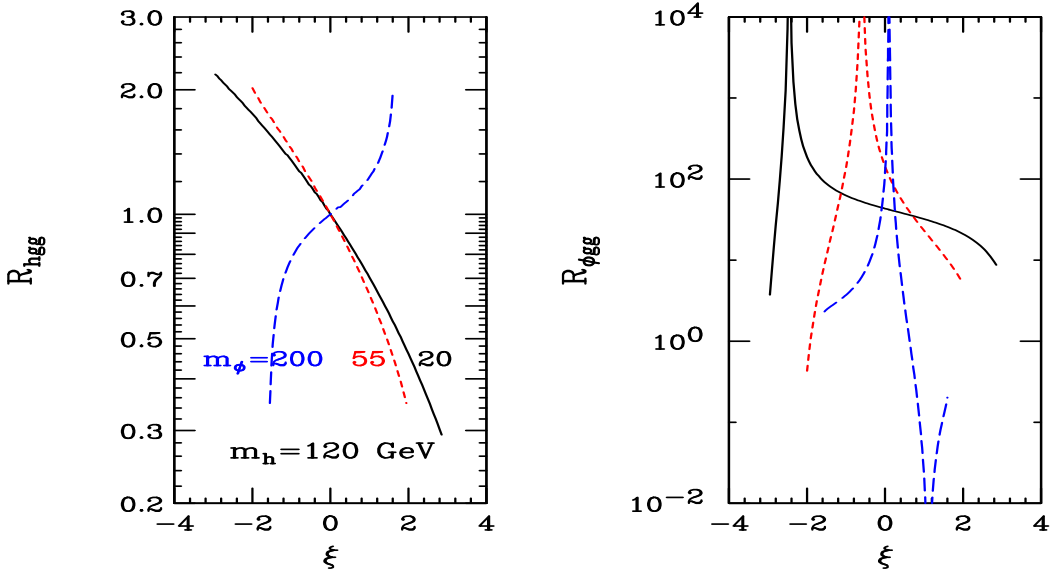}
\caption{\sl \small We plot
the ratios $R_{\h gg}$ and $R_{\phi gg}$
of the $\h gg$ and $\phi gg$ couplings-squared
including the anomalous contribution
to the corresponding values expected in its absence.
Results are shown
for $m_h=120\gev$  and $\lphi=5\tev$ as functions of $\xi$
for $\mphi=20$, $55$ and $200\gev$.
(The same type of line is used for a given $\mphi$ in 
the right-hand figure as is used in the left-hand figure.) 
\label{gggagaanomalouscoup_mh120}
}
\end{figure}

Overall, the $\gam C$ is more than competitive with the LHC for $h$ discovery.
In particular,
the $\gam C$ can see the $h$ where the LHC signal will be marginal (\ie\
at the largest theoretically allowed $\xi$ values). Of course,
the marginal LHC regions are not very big for full $L$.
Perhaps even more interesting is the fact that
there is a big part of the hourglass where the $h$ will be seen 
at both colliders. When the LHC achieves $L>100\fbi$, this comprises
most of the hourglass. Simultaneous observation of the $h$ at
the two different colliders will greatly increase our knowledge about the $h$
since the two rates measure different things.
The LHC rate in the $\gam\gam$ final state measures
$\Gamma(h\to gg)\Gamma(h\to \gam\gam)/\Gamma_{\rm tot}^h$ while
the \gamc\ rate in the $b\anti b$ final state determines
$\Gamma(h\to \gam\gam)\Gamma(h\to b\anti b)/\Gamma_{\rm tot}^h$.
Consequently, the ratio of the rates gives us
$
 {\Gamma(\h\to gg)\over \Gamma(\h\to b\anti b)}\,,
$
in terms of which we may compute
\begin{equation}
R_{hgg}\equiv \left[{\Gamma(\h\to gg)\over \Gamma(\h\to b\anti b)}\right]
 \left[{\Gamma(\h\to gg)\over \Gamma(\h\to b\anti b)}\right]^{-1}_{SM}\,.
\end{equation}
This is a {\it very} interesting number since it directly probes
for the presence of the anomalous $gg h$ coupling. 
In particular, 
$R_{hgg}=1$ if the only contributions to $\Gamma(\h\to gg)$ come
from quark loops and all quark couplings scale in the same way.
A plot of $R_{hgg}$ as a function of $\xi$ for $\mh=120\gev$, $\lphi=5\tev$
and $\mphi=20$, $55$ and $200\gev$ appears in Fig.~\ref{gggagaanomalouscoup_mh120}.

We can estimate the accuracy with which $R_{hgg}$ can be measured
as follows.  Assuming the maximal reduction of 1/2 for the
signal rate ($S$) rescaling at the $\gam\gam$ CLIC collider, we find that
$\Gamma(\h\to \gam\gam)\Gamma(\h\to b\anti b)/
\Gamma_{tot}^h$ can be measured
with an accuracy of about $\sqrt{S+B}/S\sim \sqrt{3200}/1600\sim 0.035$.
The dominant error will then be from the LHC which will typically
measure $\Gamma(\h\to gg)\Gamma(\h\to \gam\gam)/\Gamma_{tot}^h$
with an accuracy of between $0.1$ and $0.2$ (depending on
parameter choices and available $L$).  From Fig.~\ref{gggagaanomalouscoup_mh120}, we see that $0.2$ fractional accuracy will reveal deviations
of $R_{hgg}$ from $1$ for all but the smallest $\xi$ values.
The ability to measure $R_{hgg}$ with good accuracy 
may be the strongest reason in the Higgs context
for having the $\gam C$ as well as the LHC.
Almost all non-SM Higgs theories predict $R_{hgg}\neq 1$
for one reason another, unless one is in the decoupling limit.

Depending on $L$ at the LHC, there might be a small part
of the hourglass 
(large $|\xi|$ with $\mphi>\mh$) where {\it only} the $\phi$
will be seen at the LHC and the $h$ will only be seen at the $\gam C$.
This is a nice example of complementarity between the two machines.
By having both machines we maximize the
chance of seeing both the $h$ and $\phi$.

As regards the $\phi$, we have already noted from
Fig.~\ref{gagatobb_mh120} that the $b\anti b$ final state rate
(relevant for the $\mphi=20$ and $55\gev$ cases) will only be
detectable in the latter case (more generally for $55\gev <
\mphi<2\mw$), and then only if $|\xi|$ is as large as theoretically
allowed.  If $\gam\gam\to \phi\to b\anti b$ can be observed,
Fig.~\ref{gggagaanomalouscoup_mh120} shows that a large deviation for
$R_{\phi gg}$ relative to the value predicted for an $\hsm$ of the
same mass is typical (but not guaranteed).  
For $\mphi>2\mw$, $\br(\phi\to b\anti b)$ will
be very small and detection of $\gam\gam\to \phi \to b\anti b$ will
not be possible.  We are currently studying $\gam\gam\to \phi\to
WW,ZZ$ final states in order to assess possibilities at larger
$\mphi$.

Overall, there is a strong  case for the $\gam C$ in the
RS model context, especially if a Higgs boson is seen
at the LHC that has non-SM-like rates and other properties.

\section{Outlook}

Important progress in establishing the TBA technique
has been made over the last 7 years with the various CLIC Test
facilities, and  
a Technical Design Report for a machine based on TBA  technology 
could be available by 2008.  The next technological step
toward a multi-TeV collider could be the construction and operation of
several full CLIC modules, each of them providing acceleration by about 70~GeV
in 600\,m, or even think of a staged program with different physics
capablities  that will eventually take us to a multi-TeV machine. Some of
this could be running at the $Z$, $WW$ and/or $t\bar{t}$ threshold, etc.

By the time one would have to decide on the
physics program of the initial stages of  CLIC, it will be clear from the
LHC if the Higgs exists, and if it is within the mass reach of CLICHE. 
Under this circumtances, CLICHE will have a strong motivation because
it could provide 
complementary information on the Higgs boson from that obtained at the LHC, 
and help to distinguish among models. 

In addition, the experience on a \gamc\ that could be gained at a dedicated 
facility such as CLICHE will be very useful  in order to learn how to
operate what eventually will be a multi-TeV collider based on TBA technology 
just like CLIC~\cite{MultiTeVGG}. 

It is true that CLICHE is only one of several options for doing
physics with  a limit number of CLIC modules.
 However, we consider CLICHE to be a very
attractive option  for  an early stage of a TBA machine because  it 
could simultaneously test all components for accelerating high-energy
beams, and in addition  it can give important scientific output. CLICHE   
could provide  unique information on the properties of 
the Higgs boson, whose study will be central to physics at
the high-energy physics frontier over the next decade or two.

\begin{acknowledgments}

We would like to thank John Ellis and Daniel Schulte for important
discussion of the physics and machine issues of CLICHE.
We would  also like to thank Thomas Hahn and Michael Peskin for providing
some of the tools used in our simulations.

This research was partially supported by the Illinois Consortium for
Accelerator Research, agreement number~228-1001 and 
the U.S. Department of
Energy by the University of California, Lawrence Livermore National
Laboratory under Contract No.W-7405-Eng.48.
H.E.L. is supported in part by the U.S.~Department of Energy
 under grant DE-FG02-95ER40896
 and in part by the Wisconsin Alumni Research Foundation.
J.F.G. thanks U. Ellwanger, C. Hugonie, and S. Moretti for their
contributions to their joint
 NMSSM studies and he thanks M. Battaglia, S. de Curtis,
D. Dominici, B. Grzadkowski, and M. Toharia,
for their collaboration on the Higgs-radion mixing phenomenology.
 J.F.G. is supported
by the U.S. Department of Energy and by the Davis Institute for High
Energy Physics.

\end{acknowledgments}


\begin{thebibliography}{99}
\bibitem{ourproposal}  We plan to demonstrate the technical feasibility of the
NLC $\gamma\gamma$ interaction region design.
We propose to integrate a 1/2 scale NLC $\gamma\gamma$ interaction region 
into the SLC and generate two-photon luminosity by Compton scattering  30~GeV
beams from the SLC off a 0.1 J short pulse laser.
\end{thebibliography}
\end{document}